\documentclass[preprints,article,accept,moreauthors,pdftex]{Definitions/mdpi}
\firstpage{1} 
\pubvolume{1}
\issuenum{1}
\articlenumber{0}
\pubyear{2021}
\copyrightyear{2021}
\externaleditor{{Academic Editor: Andrea Giachero }} 
\datereceived{24 February 2021} 
\dateaccepted{19 April 2021} 
\datepublished{} 
\hreflink{https://doi.org/} % If needed use \linebreak

\usepackage{empheq,etoolbox,amssymb}
\patchcmd{\subequations}% <cmd>
  {\theparentequation\alph{equation}}% <search>
  {\theparentequation.\alph{equation}}% <replace>
  {}{}% <success><failure>

\newcommand{\Eqref}[1]{Equation~(\ref{#1})}

\newcommand{\figref}[1]{Figure~\ref{#1}}

\newcommand{\beq}[1]{\begin{equation}\label{#1}}
\newcommand{\eeq}{\end{equation}}

\Title{A Review of X-ray Microcalorimeters Based on Superconducting Transition Edge Sensors for Astrophysics and Particle Physics}
\TitleCitation{A Review of X-ray Microcalorimeters Based on Superconducting Transition Edge Sensors for Astrophysics and Particle Physics}%Please check if anything is missing in the main text and in order to process this paper smoothly, please reply item by item. Thank you very much for your cooperation.

\Author{Luciano Gottardi* and Kenichiro Nagayoshi }

\AuthorNames{Luciano Gottardi and Kenichio Nagayoshi}
\AuthorCitation{Gottardi, L.; Nagayoshi, K.}
\address[1]{%
 NWO-I/SRON Netherlands Institute for Space Research, Niels Bohrweg 4, \mbox{2333 CA  Leiden, The Netherlands}; 
}
\corres{\hangafter=1 \hangindent=1.0em \hspace{-.82em} Correspondence: l.gottardi@sron.nl}

\abstract{The state-of-the-art technology of  X-ray microcalorimeters based on  superconducting transition-edge sensors (TESs),  for applications in astrophysics and particle physics, is reviewed. We will show the advance in understanding the detector physics and describe the recent breakthroughs in the TES design that are opening the way towards the fabrication and the read-out of very large arrays of pixels with unprecedented energy resolution. The most challenging low temperature instruments for space- and ground-base experiments will be described.}

  % List three to ten pertinent keywords specific to the article, yet reasonably common within the subject discipline.

\begin{document}
%%%%%%%%%%%%%%%%%%%%%%%%%%%%%%%%%%%%%%%%%%

%%%%%%%%%%%%%%%%%%%%%%%%%%%%%%%%%%%%%%%%%%

\section{Introduction}\label{sec:Intro}

Many space-based observatory and  ground-based experiments in the field of astrophysics and particle-physics are improving dramatically their sensitivity and overall capabilities thanks to the use of large arrays of superconducting transition-edge sensor (TES) microcalorimeters~\cite{Adams2021,Barret20,LiteBIRD2020,Simons2021,ACTPol16,QUIBC2021,Spica14,Nucciotti2018}. 

A TES is a superconducting thin film that, due to its sharp superconducting-to-normal  transition, can be used as an extremely sensitive thermometer. It is weakly coupled to a thermal bath at temperature lower than the TES critical temperature. When operating as a calorimeter, a~TES is  well coupled to a radiation absorber. 
 By means of an ac or dc voltage bias circuit, it can be self-heated and kept stable within the transition. The~current flowing in the TES provides the signal, which is amplified by inductively coupled superconducting quantum interference devices (SQUIDs).
Researchers have attained a high level of understanding of the underlying detector physics. Thanks to this, several groups in the world have recently succeeded in optimally designing the detectors to enable the multiplexing, with~exquisite energy resolution, of~a large number of pixels both in dc-bias multiplexing schemes, like Time/Code Division Multiplexing (TDM/CDM) \cite{Doriese2016,Morgan2016} and GHz-Frequency Division Multiplexing (GHz-FDM) \cite{Mates2017}, and~ac-bias multiplexing schemes like MHz-Frequency Division Multiplexing (MHz-FDM) \cite{vdKuur2016}. 

In this review, we will take a snapshot of the state-of-the-art of 
the TES microcalorimeter technology for soft X-ray instrumentations. Our work is built upon existing reviews~\cite{McCammon2005,IrwinHilton,Ullom_2015}, and~it will focus more on the applications of TES in future X-ray space observatories and for experiments aiming to explore the new frontiers of~physics. 

The TES X-ray calorimeter technology is entering a new era where the fabrication of arrays with more than  1000 pixels is becoming a routine and instruments with almost one hundred of low temperature read-out channels are being build.  The~field  is moving from  experimental research towards fine and sophisticated cryogenic engineering to fabricate cutting-edge instruments for fundamental research, with~the potential to build detectors with 100 kilo-pixels  in the coming 20 years.
To achieve the goals  set by these ambitious projects, it is essential to further optimize the pixel design  and to improve the efficiency of the available read-out technology. The~sensor and the read-out are influencing each other and, when aiming to the highest level of performance, they cannot be developed separately. 
For these reasons, it is important to give an overview of the fundamental physical processes affecting the TES resistive transition  and the  noise, with~the  focus on the interaction between sensors and the read-out system. We will describe the major breakthrough, after~the last review~\cite{Ullom_2015}, on~the TES design optimization, which opened the way to the read-out of large arrays of~pixels.

 This paper is organized as follows. In~Section \ref{sec:TEStheory}, we present the basic theory of TES microcalorimeters and discuss the physics of the resistive transition and the major noise sources. We will describe as well the generalized system of equations that can be solved to perform realistic simulations of the dynamical response of the TES and its read-out circuit. Thanks to the available computational power of modern computers, the~end-to-end simulation  will become an essential tool to guide the development and the calibration of the very complex instruments under construction. 
 
  In Section \ref{sec:LAFab}, the~challenges in the fabrication of the sensors, absorbers, and thermal links in a large array are described. The~vivid research work of the past few years in the improvement of  the performance and uniformity of the TES X-ray calorimeters, both for the dc and ac voltage biased read-out,  is covered in Section \ref{sec:opt}. Most of the development in this area is driven by the stringent requirements from the  X-ray Integral Field Unit (X-IFU) \cite{XIFU2018}, which is one of the two instruments of the Athena astrophysics space mission~\cite{Barret20} approved by ESA in the Cosmic Vision 2015--2025 Science~Programme. 

 In Section \ref{sec:mux}, we summarize the recent progress in the multiplexing demonstration of a large number of pixels with high energy resolution.
 Finally, in~Section \ref{sec:future} we will describe several ground- and space-based applications where the core technology is based on large arrays of TES X-ray~microcalorimeters.

%%%%%%%%%%%%%%%%%%%%%%%%%%%%%%%%%%%%%%%%%%
\section{TES Physics and~Models}
\label{sec:TEStheory}
In this section, we will focus on the theory of TESs. We will discuss primarily the case of X-ray microcalorimeters. However, the~theory presented is general enough to be adapted to other TES-based energy or power~detectors. 

 A figure of merit that characterizes single photons detectors is the resolving power $E/\Delta E$ for photons with  energy $E$. A~fundamental limit for the minimum energy resolution $\Delta E$ achievable with a calorimeter is given by the random exchange of energy between the detector and the thermal bath~\cite{McCammon2005}. This thermodynamic limit is given by $<\Delta E_{TD}^2>=k_BT^2C$. It depends quadratically on the temperature $T$ of the calorimeter, linearly on the detector heat capacity $C$, and it is independent on the thermal conductance $G$ of the thermal link. This value however does not set a limit to how accurately the thermal fluctuations can be measured. This can be understood by looking at the signal and noise power spectrum of the thermodynamic fluctuations. Each frequency bin provides an uncorrelated estimation of the signal amplitude and the signal-to-noise ratio can be improved as the square root of the number of bins averaged~\cite{McCammon2005}.   
The two most important parameters used to estimate the detector signal-to-noise ratio and the energy resolution are the unit-less logarithmic  temperature and current sensitivity $\alpha=\partial\log R/\partial\log T$ and $\beta=\partial\log R/\partial\log I$, calculated, respectively, at a constant current $I$ and temperature $T$. They have been conveniently  introduced~\cite{Lind2004} to parametrize the TES resistive transition 
\beq{Ralphabeta} 
R(T,I)\approxeq R_0+\alpha\frac{R_0}{T_0}\delta T +\beta\frac{R_0}{I_0}\delta I,
\eeq 
and they can be estimated experimentally at the quiescent operating point ($R_0,T_0,I_0$) in the transition.
 A detailed calculation of the minimum energy resolution achievable with a TES calorimeter is possible. After~a careful analysis of the thermometer sensitivity, the~detector noise and the signal bandwidth, gives
\begin{equation}\label{eq:dE}
\Delta E_{FWHM}\simeq 2\sqrt{2\ln 2}\sqrt{4k_BT^2_0C\frac{\sqrt{\zeta(I)}}{\alpha}\sqrt{\frac{nF(T_0,T_{bath})}{1-(T_{bath}/T_0)^n}}}.
\end{equation}
The unit-less parameter  $F(T_0,T_{bath})$ depends on the thermal conductance  exponent $n$, which is related to the physical nature (radiative or diffusive) of the thermal link between TES and the heat sink at $T_{bath}$ \cite{McCammon2005}. For~a TES $F(T_0,T_{bath})\simeq 0.5$ \cite{IrwinHilton}.
The parameter $\zeta(I)$ takes into account the non-linear correction terms to the linear equilibrium Johnson noise, which conveniently helps to write the Johnson voltage spectral density  as $S_V=4k_BTR_0\zeta(I)$. For~a linear resistance $\zeta(I)=1$. For~a TES, it is usually defined as $\zeta(I)=(1+2\beta)(1+M^2)$ where the factor $1+2\beta$  is the first order, near~equilibrium, non-linear correction term as discussed in~\cite{Irwin2006} and the unit-less factor $M^2$ parameterizes  any  {\it unexplained} %can the italics be removed?
observed deviation from this approximation. More details on this important noise contribution are given further in~Section \ref{sect_noise}. 

From \Eqref{eq:dE}, it is clear that  low heat capacity devices, operating at very  low temperature $T_0$, could achieve very high energy resolution. The~$C$ value,  however, is not  a free parameter and is typically defined by the dynamic range, $\mathrm{DR}\propto C/\alpha$, requirement of the specific application.  By~developing TES with a large temperature sensitivity $\alpha$, the~photon energy  can be measured with a much higher resolution than the magnitude set by thermodynamic fluctuations. To~achieve then the ultimate sensitivity, for~given $C$ and $T_0$, the~factor $\sqrt{\zeta(I)}/\alpha$ has to be minimized.

The details of the TES electro-thermal response and noise will be discussed here below. 
In~Section \ref{sect_WL}, we will review the recent development in the understanding of the physics of the TES resistive transition. The~differential electrical and thermal equations describing a TES and its bias circuit are derived in Section \ref{sect_ETEq}, while an overview of the fundamental noise sources in a TES-based detector will be presented in Section \ref{sect_noise}, including the recent new insight on this topic.
%The deviations from a small signal, linear analysis will be briefly discussed in Section \ref{sect_LS}. We will conclude the session with a short review of the pulse processing techniques (Section \ref{sect_PP}) and the most recent results on the multiplexing read-out of large arrays of TESs (Section \ref{sect_MUX}). 

A TES operates under voltage-bias condition in the negative electro-thermal feedback (ETF) mode. This guarantees stable operation and self-biasing within its sharp superconducting transition~\cite{Irwin1995}. As~we will see later on in this section, depending on the multiplexing technology chosen to read out a large array of pixels, a~TES is dc voltage biased or ac voltage biased in the MHz range.   
While keeping the discussion as general as possible, we will place the emphasis on  the MHz biased TESs for two reasons. The~first one is due to the fact that it is the least considered case in the literature, since most of the phenomena observed under MHz bias have only been recently studied in the detail and understood. The~second reason is that the MHz read-out (in-phase and quadrature) allows us to directly probe many of the TES physics effects discussed in Section \ref{sect_WL}.

\subsection{The Proximity Effects and the Resistive~Transition}\label{sect_WL}
\begin{figure}[H]
\widefigure
%\center
\includegraphics[width=15cm]{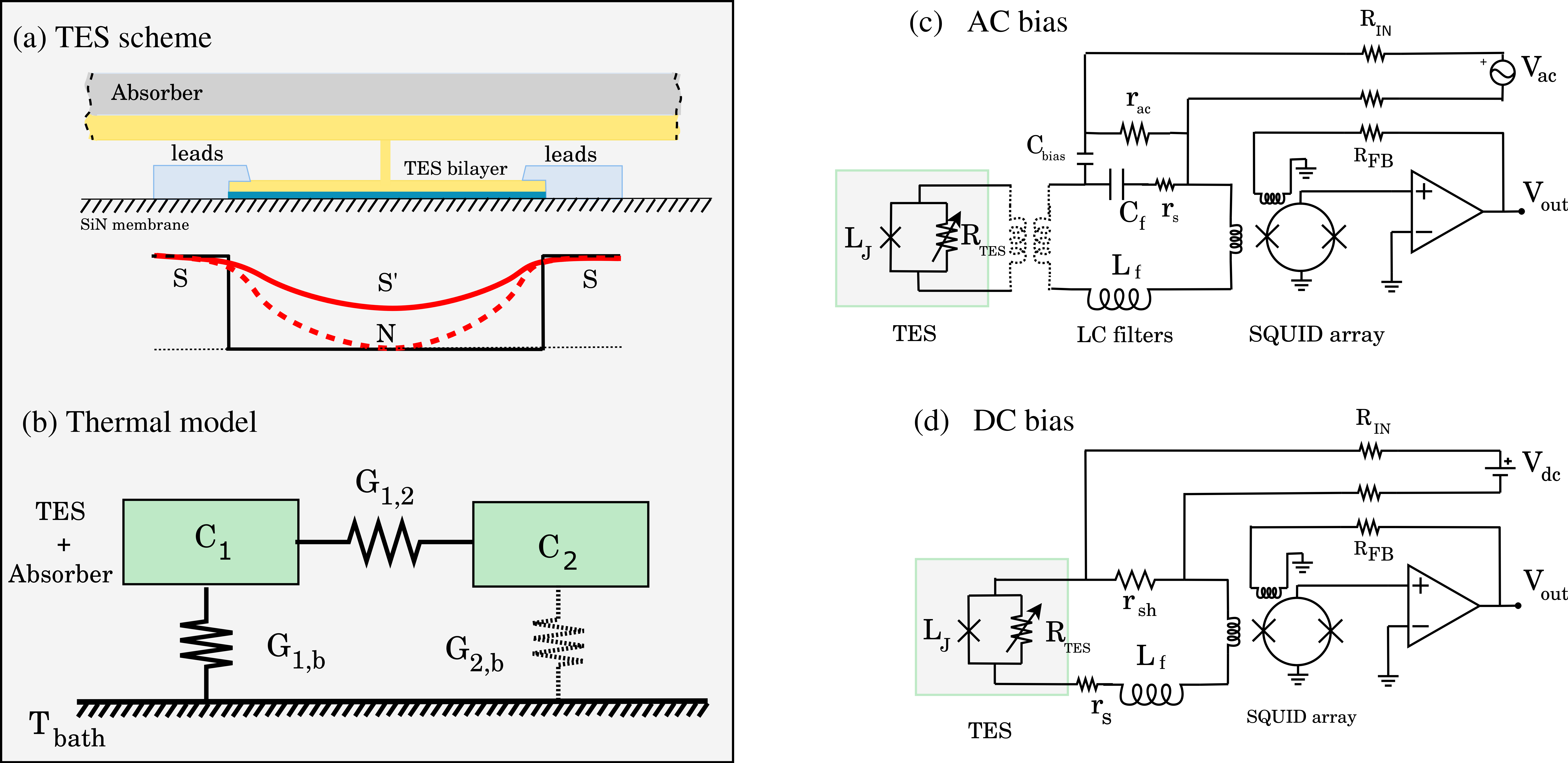}
\caption{\label{fig:ETscheme} (\textbf{a}) Schematic diagram of a TES bilayer connected to the bias leads and the absorber. The~red curves represent the superconducting order parameter for an SS'S (solid) and a SNS (dashed) TES weak-link.  (\textbf{b}) Thermal model of a TES-Absorber system ($C_1$) including a second thermal body $C_2$. $G_{1,b}$ is the main thermal conductance to the bath, $G_{bath}$, and~$G_{2,b}\ll G_{1,b}$. (\textbf{c},\textbf{d}) The electrical circuits for the ac and dc bias read-out, respectively.}
\end{figure}

 A TES consists of a thin film made of a superconducting bilayer with an intrinsic critical temperature $T_{ci}$ typically around $100\, \mathrm{mK}$. It is directly connected to superconducting bias leads with  critical temperature $T_{cL}\gg T_{ci}$. \figref{fig:ETscheme}a shows a schematic of a TES connected to the bias superconducting leads, while more details on the TES structures and their fabrication are reviewed in Section \ref{sec:LAFab}.    
% start a new page without indent 4.6cm
\clearpage
\end{paracol}
\nointerlineskip

\begin{paracol}{2}
%\linenumbers
\switchcolumn

It was first reported by Sadleir~et~al.~\cite{Sadleir10,Sadleir11} that TES structures behave as a superconducting weak-link due to the long-range longitudinal proximity effect (LoPE) originating from the superconducting Nb leads. Their conclusion was based on three solid experimental findings, namely (i) the exponential dependence of the critical current $I_c$ upon the TES length $L$ and the square root of the temperature difference $T-T_{ci}$:
\beq{WLIcT}
I_c(T,L)\propto \frac{L}{\xi(T)}e^{-\frac{L}{\xi(T)}},
\eeq
with $\xi(T)=\xi_i/(T/T_{ci}-1)^{1/2}$ for $T>T_{ci}$, (ii) the scaling of the effective transition temperature $T_{c}$ and the transition width $\Delta T_c$ as $1/L^2$, and~(iii) the Fraunhofer-like oscillations of the critical current as a function of the applied perpendicular magnetic field. The~longitudinal proximity effect was observed over extraordinary long ($L>100\,\upmu \mathrm{m}$) distances
and it is responsible for the enhancement, in~the proximity of the leads, of~the spatially varying superconducting order parameters $\Psi$ of the TES bilayer, as~shown in \figref{fig:ETscheme}a. It has been shown as well~\cite{Sadleir11} that the order parameter is suppressed in proximity to  normal metal structures deposited along or on top of the TES due to the lateral inverse proximity effect (LaiPE). Both the longitudinal and the lateral-inverse proximity effects can be used to fine tune the effective TES critical temperature $T_c$. By~measuring many devices with different geometries and aspect-ratios, fabricated from the same bilayer film, it is possible to separate the contribution of the longitudinal proximity effect from that of the lateral inverse proximity effect . In~\cite{deWit_2020}, for~example, the~effect of the Nb leads has been identified  by measuring $T_c-T_{ci}$ of TESs of equal
width $W$, but~different lengths $L$. The~results are then used to correct for the longitudinal proximity effects and look at the lateral inverse proximity effects in the devices of varying~widths.

Ridder~et~al.~\cite{Ridder2020} have recently  investigated  the effects of the longitudinal proximity effects in Ti/Au TES structures using Ti and Nb superconducting leads material, with~low and high $T_{cL}$ respectively. A~comparison was done with the previous results obtain with Mo/Au~\cite{Sadleir11} and Mo/Cu~\cite{Ullom_2015} TESs.   The~measured characteristic length scale of the proximity effect was the lowest for the Ti/Au TES with Ti leads. The~reported $T_{c}-T_{ci}$ was about 2.5 higher for the Ti/Au TES with Nb leads, and~almost 8 times higher for the MoAu TES with Nb~leads. 

{
The discovery of the proximity effects in TES structures naturally leads to treat these devices as superconducting SNS or SS'S weak-link, with~the bilayer being N  when $T>T_{ci}$ or S' when $T<T_{ci}$. 
Extensive reviews on the Josephson weak-links have been given by   Likharev~\cite{Likharev79} and Golubov~et~al.~\cite{Golubov04}.
The Josephson effect in SS'S junctions at arbitrary temperatures was analysed first by Kupriyanov and Lukichev~\cite{KupLikLuk81,KupLuk82} in the framework of the Usadel equations. They showed that for long junctions and $T\geq T_{ci}$ the $I_c(T)$ has an exponential dependence, as~shown in \Eqref{WLIcT}, with~the effective coherence length $\xi(T)$ larger than the intrinsic coherence length of the material:
\begin{equation}\label{eq:xi_i}
\xi(T)=\xi_N\sqrt{\frac{T_{cL}}{T}\left[1+\frac{\pi^2}{4}\ln^{-1}\left(\frac{T}{T_{ci}}\right)\right]}.
\end{equation}
Here, $\xi_N=\sqrt{\hbar D_N/(2\pi k_B T)}$ is the coherence length  for the normal material N in the dirty limit, with~$D_N=v_Fl_e/3$ the electronic diffusivity for a material with Fermi velocity $v_F$ and mean-free path $l_e$.  
\Eqref{eq:xi_i} can be approximated as
\begin{equation}\label{eq:xi_i_aprox}
\xi(T)=\frac{\pi}{2}\xi_N\sqrt{\frac{T_{cL}}{T}}\frac{1}{(T/T_{ci}-1)^{1/2}}.
\end{equation}
For a 200 nm Au metal film at $T=100\,\mathrm{mK}$, with~$v_F=1.39\cdot10^6$ m/s and thickness limited $l_e\sim200\, \mathrm{nm}$, the~diffusivity $D_N\simeq 0.092\, \mathrm{m}^2\mathrm{s}$ and the coherence length $\xi_N\simeq\, 1 \, \upmu\mathrm{m}$. 
When studying the physics of a TES, one has to keep in mind that its coherence length $\xi(T)$ is enhanced by a factor $\sqrt{T_{cL}/T}$ due to the presence of the leads with a critical temperature $T_{cL}$ and by the fact that the detector operates close to $T_{ci}$. For~typical TESs, the~coherence length can become much larger than $10\,\upmu\mathrm{m}$.
}

\textls[-15]{A  microscopic model of a TES as a weak-link has been developed by Kozorezov~et~al.~\cite{Kozo_ieee2011} using the Usadel approach. Usadel equations have been used later on by Harwin~et~al.~\cite{Harwin_2017,Harwin_2018}  to investigate the effect of lateral normal metal structures  and  to reproduce the TES current-to-voltage~characteristics.}

%, to study the effects of device geometry and material composition.

 The most successful macroscopic model developed to describe Josephson junctions and superconducting weak-links of many different kind is the resistively and capacitively shunted junction (RCSJ) model. The~overdamped limit ($C\sim0$) of the RSCJ model, namely the resistively shunted junction model (RSJ), was formalized by Kozorezov~et~al.~\cite{Kozorezov11} for a dc-biased TES. The~RSJ model can be used to calculate analytically  the TES resistive transition and to generalize the TES response to an alternating current. The~most interesting  prediction in this work is the existence of an intrinsic TES reactance, which can be calculated exactly. As~we will see in Section \ref{sect_ETEq}, this  has important implications on the TES response, in~particular when operating under MHz voltage~biasing.

Using the Smoluchowski equation approach for quantum Brownian motion in a tilted periodic potential in the presence of thermal fluctuations~\cite{Coffey_2008}, an~analytical solution for the dependency of the TES resistance on temperature and current was found~\cite{Kozorezov11}. In~the limit of zero-capacitance and small value of the quantum parameter, we can write:
\begin{equation}\label{RTI}
R(T,I)=R_N\left\{ 1+\frac{1}{x}\mathrm{Im}\left[\frac{I_{1+i\gamma x}(\gamma)}{I_{i\gamma x}(\gamma)}\right]\right\},
\end{equation}
 with $R_N$ the normal state resistance, $\gamma=(\hbar I_c(T,L))/(2ek_BT)$ the ratio of the Josephson coupling to thermal energy, and~ $x=I/I_c(T,L)$. Here, $I_c(T,L)$ is  the critical Josephson current and $I_{1+i\gamma x}(z)$ and $I_{i\gamma x}(z)$ are the modified Bessel function $I_\mu (z)$ of the complex order, $\mu$, and~real variable, $z$. An~example of an $R(T,I)$ curve calculated from \Eqref{RTI} is shown in \figref{fig:RTI}.
By differentiating \Eqref{RTI}, the~parameters $\alpha$ and $\beta$ are obtained:
\begin{equation}\label{alpha_WL}
 \alpha=-\frac{\gamma}{x}\frac{\partial\mathrm{ln}I_c}{\partial\mathrm{ln}T}\frac{R_N}{R}\mathrm{Im}\left\{\frac{I_{-1+i\gamma x}(\gamma)I_{1+i\gamma x}(\gamma)}{I^2_{i\gamma x}(\gamma)}\right\},
 \end{equation}
\begin{equation}\label{beta_WL}
 \beta=-1+\frac{R_N}{R}\left\{ 1-2\mathrm{Re}\left[\frac{1}{I^2_{i\gamma x}(\gamma)}\int_0^\gamma I_{1+i\gamma x}(z)I_{i\gamma x}(z) dz, \right] \right\}.
 \end{equation}
\begin{figure}[H]
\includegraphics[width=8.6cm]{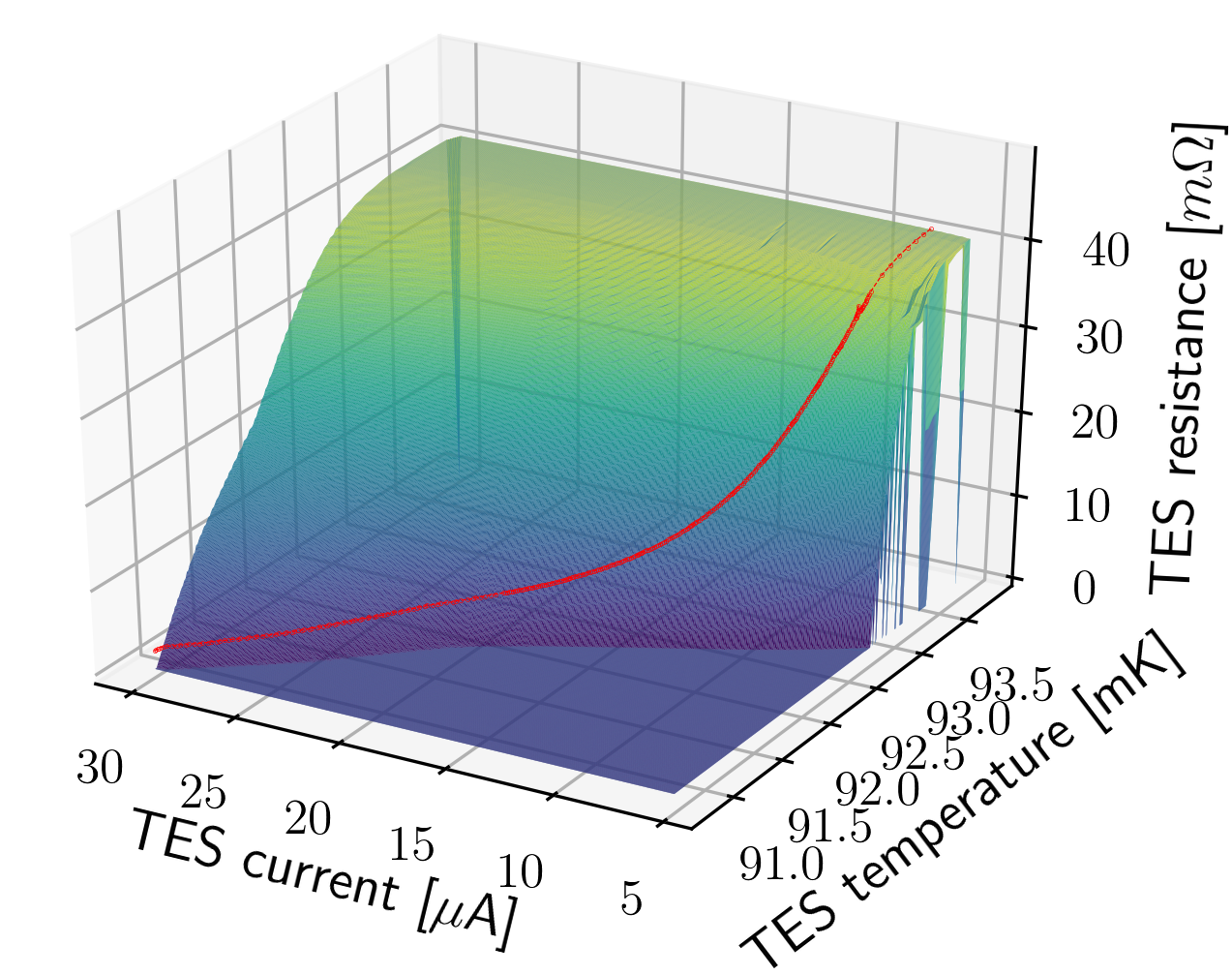}
\caption{\label{fig:RTI} Example of an R(T,I) curve calculated for an X-ray TES calorimeter developed for X-IFU. The~red curve shows the TES bias~trajectory.}
\end{figure} 

 A TES generally operates in the limit of $\gamma\rightarrow \infty$ (Josephson coupling energy larger than the thermal energy)  and $I/I_c>1$. The~device is superconducting ($R(T,I)=0$) for $I/I_c<1$ and normal  ($R(T,I)=R_N$) for $I/I_c\gg 1$. Under~these conditions, the~above equations and the relation between $\alpha$ and $\beta$ can be greatly simplified following the work of Ambegaokar and Halperin~\cite{AmbHal69}.
The  resistive transition $R(T,I)$ can then be written as~\cite{Smith13}
\beq{RTI_simple}
\frac{R(T,I)}{R_N}=\left(1-\left(\frac{I_c(T)}{I}\right)^2 \right)^{1/2},
\eeq
 and $\alpha$ and $\beta$ reduce to
\beq{alpha_simple}
\alpha = -\left(\left(\frac{R_N}{R}\right)^2-1\right)\frac{\partial \mathrm{ln}I_c(T)}{\partial \mathrm{ln}T},
\eeq
\beq{beta_simple}
\beta = \left(\frac{R_N}{R}\right)^2-1.
\eeq
A simple relation for the $\alpha/\beta$ ratio holds:
\beq{alphabeta}
\frac{\alpha}{\beta}=-\frac{\partial \mathrm{ln}I_c(T)}{\partial \mathrm{ln}T}=-\frac{L}{\xi_0}T\left|\frac{T}{T_{ci}}-1\right|^{1/2},
\eeq
which could turn out to be useful for the pixel design optimization~\cite{Smith12}.

The $I_c(T,B)$ curve can be measured rather accurately, very close to $T_c$, for~many orders of magnitude of the current flowing in the TES~\cite{Sadleir10}. By~studying the shape of $I_c(T,B)$ for different TES geometry and as a function of the perpendicular  magnetic field $B_\perp$, in~relation with the TES bias current $I$, one can derive detailed information on the physical mechanism involved in a  TES~\cite{Sadleir10,Sadleir11,Smith12,Smith13,Gottardi14}. A~careful TES current and voltage calibration is however required to accurately calculate the ratio $x=I/I_c(T)$ at each bias point, which is needed to validate experimentally the resistive transition models.
In~\cite{Smith12}, for~example, the~ $I_c(T)$ curve measured for small pitch TES X-ray calorimeters optimized for solar astronomy and with two different design,  has shown a very good agreement with the \Eqref{alphabeta} derived from the RSJ~model. 
 
  There exists a class of devices, where the weak-link effects described above are less dominant. TESs could  for example operate in a stronger superconducting regime, depending on the exact shape of the $I_c(T)$ and the bias current, which in turn depends on the device saturation power required by the applications.  It has been reported~\cite{Ullom_2015}, for~example, that large Mo/Cu TESs, $ 400\times 400\,\upmu\mathrm{m}^2$ in size, with~several normal  metal features on top, operating at high bias current $I\sim200\, \upmu\mathrm{A}$, and~with  the lead $T_{cL}$ closer to the bilayer $T_{ci}$, typically show a $I_c(T)$ curve which is well described by the Ginzburg--Landau (G-L) critical current. The~latter takes the form: 
\beq{GLIcT}
I_c(T)=I_{c0}\left(1-\frac{T}{T_c}\right)^{3/2},
\eeq 
where $I_{c0}$ is the value of $I_c$ at zero temperature.
These devices have typically lower $\alpha$ and $\beta$ values than the ones predicted by the RSJ model (Equations (\ref{alpha_WL}) and (\ref{beta_WL})) \cite{BennetPRB13}.  The~physical mechanism  of phase slip lines (PSL)  \cite{Bennett14} has been proposed, in~the simplified form of a two-fluid model~\cite{Irwin1998,Bennet2012},  as~a way to explain the resistive transition in devices where the ideal weak-link effects were not observed. The~PSL model reproduces the general behaviour of the IV characteristic of these devices as a function of temperature and provides a better  prediction of  $\alpha$ and $\beta$ in the transition.  The~PSL model has also been considered to explain kinks and steps observed along the transition in dc bias TESs~\cite{Bennett14} and is used to guide the optimization of X-ray detectors by studying the dependence of the transition  width on the TES $I/I_c(T)$ \cite{Morgan2017,Morgan2019}.
Bennett~et~al.~\cite{BennetPRB13} made a comparison, updated and expanded later on~\cite{Ullom_2015}, between the RSJ and the PSL models for many TESs with different geometries. The~predictions are not always consistent with all the data presented, and~the influence  of the TES geometry and the effect of the connected normal metal structures is still not clear. The~main result of their comparison is that, over~the range of operation of a TES, for~TESs larger than $80\,\upmu\mathrm{m}$,   the~measured $I_c(T)$ curves are more consistent with \Eqref{GLIcT}. On~the contrary, Smith~et~al.~\cite{Smith12}  have shown, with~pixels developed for solar astronomy application, that TESs as large as $ 140\time 140\, \upmu\mathrm{m}^2$ and with normal metal  noise mitigation structure do follow the weak-link prediction  as in Equations (\ref{WLIcT}) and (\ref{alphabeta}). It is clear that it is not always easy to discriminate between the regime of operation of a TES-based device, in~particular in the presence of complicated normal metal structures affecting the current distribution  in the TES~bilayer.

 As it will be discussed in Section \ref{sec:acbias}, there is a wish to fabricate detectors that operate away from the weak-link regime to minimize non-ideal effects that can deteriorate the detector performance. This is particularly true for ac biased devices. It is then very important to understand the underlying physical mechanisms that shape the $I_c(T,B)$ curve and the TES resistive transition.
To predict the exact $I_c(T,B)$ behaviour in the cross-over region between the RSJ and the phase-slip models, it might be necessary to solve the Usadel equations in 2D for a realistic  TES design (with normal metal stems, bars, and stripes),  as~a follow up of the work done in one dimensional structures~\cite{Kozorezov11,Harwin_2017,Harwin_2018}.

\noindent From the experimental point of view, the~ full characterization  of the new generation of devices, discussed in Section \ref{sec:opt}, might help in the understanding of the resistive~transition.

The theoretical framework reviewed  by Likharev~\cite{Likharev79} and Golubov~et~al.
~\cite{Golubov04} could guide the interpretation of the experimental data. As~described in their work, the~possible physical states in which a TES of different size  likely operate can be illustrated in the diagram of \figref{fig:LxWstate}, simplified and reproduced after~\cite{Likharev79,Kupr75}. 
\begin{figure}[H]
\includegraphics[width=8.6cm]{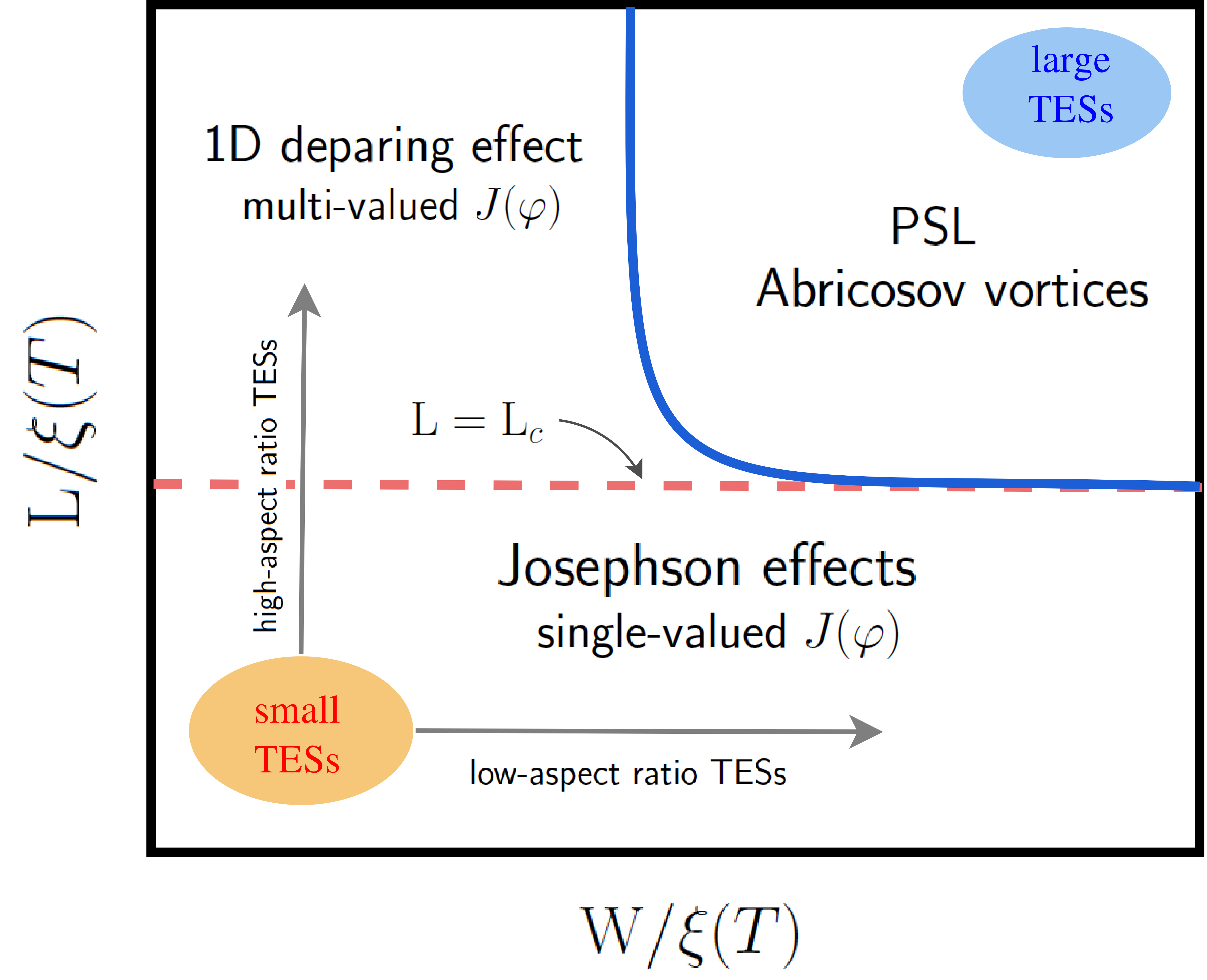}
\caption{\label{fig:LxWstate} Possible physical states in which a TES of different size  likely operates. Reproduced after~\cite{Likharev79,Kupr75}, by~permission from American Physical Society and the~authors.}
\end{figure} 
\noindent The TES length $L$ and width $W$ are normalized with respect to the coherence length $\xi(T)$. The~solid blue line shows the boundary between the regions of large size superconducting structures, where PSL/Abrikosov vortices could be formed, and~the small weak-link where the pure Josephson effect is observed.
The dashed red line corresponds to a TES critical length $L_c$, estimated by Likharev to be $\sim3.49\xi$, above~which, for~small TES width $W$, a~transition from a Josephson effect, with~single valued current-phase relation  $I(\phi)$ to a one-dimensional depairing state (phase slip centers), with~multivalued current-phase relation,   might occur.  
 The exact shape and position  of these boundaries are not trivial to calculate for a TES, since they depend on many parameters, such as the self magnetic field, the~bilayer coherence length and the difference between the bilayer $T_c$ and the  $T_{cL}$ of the leads. The~threshold is temperature dependent as well, so the TES  might switch from a weak to a strong superconducting state depending  on the actual bias conditions. {We have seen before that for a TES with an Au  thickness of $200\, \mathrm{nm}$, the~coherence length close to $T_{ci}\sim\,90\,\mathrm{mK}$ could reach values, enhanced by the electrodes with  $T_{cL}/T_{ci}\simeq 70$,  of~$\xi(T=1.1T_{ci}) \sim\, 50\,\upmu\mathrm{m}$. This leads to values of  $L_c=3.49\xi > 150\, \upmu\mathrm{m}$}.  This estimation, however, is rather approximated, since the cross-over boundary strongly depends on  the current distributions, which in turn depend on the width and the thickness of the bilayer,  as~well as on the presence or not of normal metal structure in a real~TES.
 
The common practice  to discriminate whether a TES behaves as a weakly-linked  or strong superconductor has been so far to look for the exponential behaviour of the $I_c(T)$, its Fraunhofer-like behaviour as a function of  the perpendicular magnetic field, the~oscillatory features  in the TES reactance as a function of the  MHz voltage bias TES~\cite{Gottardi14}, or~the presence of Shapiro-steps in dc bias TES under MHz current or magnetic field excitation~\cite{SmithPoster2014,Sadleir2016,Zhou2018}. The~latter effect, for~example, manifest itself  with a specific current step pattern in the TES $IV$ curves at $V_{m,n}=m/n\hbar\omega/2e$, according to the Josephson law. However, similar steps might be observable in large 2D superconducting films, with~a non-exponential  $I_c(T)$ as in \Eqref{GLIcT}, as~a result of the coherent motion of a large number of Abrikosov vortices~\cite{Likharev79,Bennett14}.  The~characterization of the crossover from weak- to strong superconductivity behaviour is not trivial.
A better way to identify the transition between the possible states, schematically shown in \figref{fig:LxWstate},  would be to  measure experimentally the current-phase relation~\cite{Golubov04}  of  TESs with different geometry in a wide range of biasing condition. The~ideal Josephson effect should take place only when $I(\varphi)$ is single-valued. For~geometries where $L>L_c$, $I(\varphi)$ is multi-valued and   the TES might enter into a non-stationary state where depairing effects, phase slip, and PSL/Abrikosov vortex formation are likely to~occur.

A study that deserves more attention is the modeling of the TES $R(T,I,B)$ transition using the Berezinskii--Kosterlitz--Thouless (BTK) theory, which describes the onset of dissipation due to current assisted vortex pair unbinding. In~a recent work based on a previous investigation done on TESs by Fraser~\cite{Fraser2004}, Fabrega~et~al.~\cite{Fabrega2018} have studied the resistive transition of Mo/Au bare TESs, with~high $T_c\sim$460--640 
$\mathrm{mK}$,  as~a function of temperature, current, and magnetic field. The~observed large-current induced broadening of the transition   at low resistance could be explained  by the BTK mechanism. More investigations in this direction are needed, but~this approach could prove useful to understand non-equilibrium effects in TESs and the transition from a weak-link to a 2D strong superconductivity~regime. 

\subsection{TES Electro-Thermal~Equations}\label{sect_ETEq}
The standard electro-thermal linear model for a dc biased TES calorimeter was originally developed by Lindeman~et~al.~\cite{Lindeman_2011} to describe the noise and the response to photons. No assumption were made on the TES physics, and~the resistance, for~each bias point in the transition, was linearly approximated as in \Eqref{Ralphabeta}. This model has been successfully used to understand the behaviour of TES-based detectors in many  works that followed.
 In this section, we present the Langevin electro-thermal equations  for a TES calorimeter, generalized for the dc and  ac biased case, assuming the detector physics being well described by the Josephson effects, formalized in the RSJ model. A~schematic diagram of the detector thermal model and the electrical ac and dc bias circuit is shown in \figref{fig:ETscheme}.
 
As suggested and observed in many experiments~\cite{Takei08,Kinn12,Maasilta12,Goldie_2009,Wake2019}, a~single-body thermal model for  a TES calorimeter is rarely sufficient to explain the detector response. This is due to the presence of dangling heat capacitance  and parasitic thermal conductance in the TES-absorber structures or to a not sufficiently high thermal conductance between TES and absorber. It can be shown that a two-body model~\cite{Takei08,Goldie_2009,Lindeman_2011}, as~drawn  in \mbox{\figref{fig:ETscheme}b}, is sufficiently general to account for parasitic thermal effects in TES-based detectors. A~more detailed analysis for even more complex thermal structures can be found \mbox{in~\cite{Kinn12,Maasilta12,Goldie_2009,Wake2019}}. However, one has to keep in mind that the model is unconstrained when too many thermal bodies are added into the system of equations. As~a consequence, the~physical interpretation of the results becomes~impossible. 

In \figref{fig:ETscheme}c, the~electrical  circuit for a MHz biased TES is shown. The~TES is placed in series with a high-$Q$ superconducting  $LC$ filter~\cite{Bruijn18,Bruijn14} and the input coil of the SQUID current amplifier~\cite{Kiviranta02}. An~optional superconducting transformer, shown in the picture in dashed lines, could be used to optimize the impedance matching between detector and amplifier. The~ac voltage bias is provided via a capacitive divider in parallel with an ac shunt resistor. From~a simple Thevenin equivalent circuit analysis, the~bias network is equivalent to a shunt resistance  $r_{sh}=r_{ac}C_b/(C_F+C_b)$, in~series with the $LC$ resonator. $r_s$ accounts for  the intrinsic loss in the $LC$-filters~\cite{Gottardi19}. 

The dc bias circuit is shown in \figref{fig:ETscheme}d. It is equivalent to the ac bias one with $C_F=0$, and~$C_b\rightarrow \infty$ and after replacing the ac voltage source with a dc one. In~this case, $L_f$ is the Nyquist inductor added to limit the read-out bandwidth, $r_{sh}$ is a low  shunt resistors needed to provide a stiff dc voltage bias to the TES, and $r_s$ indicates any parasitic resistance in the bias circuit. Both for the ac and dc read-out the following holds: $r_s\ll r_{sh} \ll R_N$.  More details on the ac and dc read-out will be given in Section \ref{sec:mux}.

Within  the RSJ model, the~TES is electrically treated as a Josephson weak-link with a resistor $R_N$ in parallel with a non-linear Josephson inductance. It obeys the standard Josephson equations relating the TES voltage $V_{TES}$ and the Josephson current $I_J$ to the gauge-invariant phase difference $\varphi$ of the superconducting order parameter across the~leads
\beq{Josephson} 
V_{TES}=\frac{\hbar}{2e}\frac{\partial \varphi}{\partial t},\;\;
I_J(t)=I_c(T)\sin \varphi(t),
\eeq 
assuming that a sinusoidal current-phase relation  holds.
The total phase difference across the weak-link depends on the perpendicular magnetic flux coupled into the TES, i.e.,~$\varphi_{tot}=\varphi+2\pi\frac{\Phi(t)}{\Phi_0}$, where $\Phi(t)=A_{eff}(B_{\perp,DC}+B_{\perp,AC}(t))$ and $A_{eff}$ is the effective weak-link area crossed by  the dc ($B_{\perp,DC}$) and ac ($B_{\perp,AC}$) perpendicular magnetic field, respectively. The~$B_\perp$ field can be generated both from an external source or self generated in the TES by the current flowing in the~leads.   

 The total current in the TES $I(t)$ is considered to be the
sum of two components, the~Josephson current $I_{J}(t)$ and a quasi-particle current $I_{qp} = V(t)/R_{N}$. 

 In the general case, $V(t)=V_{dc}+V_{pk}\cos \omega_{0}t$, where $V_{dc}$ is a constant voltage across the TES and the second term account for any potential ac excitation of peak amplitude $V_{pk}$ injected into the bias circuit. In~the pure ac voltage bias case, $V_{dc}=0$,  and~$V_{pk}\cos \omega_{0}t$ is the applied voltage bias with $\omega_{bias}=\omega_{0}$. The~ac voltage across the TES forces the gauge invariant superconducting phase $\varphi$ to oscillate at the bias frequency $\omega_{bias}$,  $\pi/2$ out-of-phase with respect to the voltage. The~$\varphi$ peak value depends on $V_{pk}/\omega_0$. A~rigorous description of the electro-thermal equations for a TES, following the RSJ model,  will be given below. However, the~main features observable in the TES current   can already be understood by a simple {voltage-source} model in the small signal limit, with~$\varphi=2e/\hbar\int V(t)dt$ from the ac Josephson relation in \Eqref{Josephson}. This is valid both for a dc and an ac biased TES.
In the general case, the~Josephson current becomes 
\beq{Ijgeneral}
I_{J}(t)=I_c(T)\sin\left[\frac{2eV_{dc}}{\hbar}t+\frac{2eV_{pk}}{\hbar\omega}\sin \omega_0t\right],
\eeq
which, using standard trigonometric identities and the Bessel function relations, can be written in the form
\beq{IjBessel}
I_{J}(t)=I_c(T)\sum^{\infty}_{n=-\infty}(-1)^nJ_n\left(\frac{2eV_{pk}}{\hbar\omega_{0}}\right)\sin ( \omega_J-n\omega_{0})t.
\eeq
 In the equation above, $J_n$ is the ordinary Bessel function of the first kind and $\omega_J= 2eV_{dc}/\hbar$ is the Josephson oscillation frequency. 
\noindent  \Eqref{IjBessel} says that, when a  dc biased TES is excited by a small ac signal with frequency $\omega_0$, spikes in the TES current can be observed at the TES voltage $V_{dc}=n\hbar\omega_0/2e$, with~$\omega_J=n\omega_0$.
  
 For the ac bias case, thanks to the  high-$Q$ $LC$ filters in the bias circuits, $V_{dc}$=0, $\omega_J=0$ and only the $n=1$ frequency  component  survives in the sum, with~$\omega_0=\omega_{bias}=1/\sqrt{L_FC_F}$. 
\noindent At a fixed bias point and with $V_{dc}=0$ and for small value of  $V_{pk}$, the~non-linear Josephson inductance in parallel to the TES resistance is simply defined by 
 \beq{Lj}
 L_J=\frac{V(t)}{dI/dt}=\frac{\hbar}{2eI_c(T)\cos \varphi}=\frac{\hbar}{2eI_cJ_0\left(\frac{2eV_{pk}}{\hbar \omega_0}\right)}.
 \eeq
In the ac bias read-out, $L_J$ can be derived from the quadrature component of the current measured in the  $I-V$ characteristics. 
 
 The generalized system of  coupled thermal and electrical differential equations for a TES and its bias circuit, which includes the Josephson relations,  becomes 
 \begin{subequations}\label{eq:Langev}	
  \begin{empheq}{align}
%\beq{LangRSJ}  
% \begin{aligned}
   \label{eq:Lang_RSJ}	
  \begin{split}
  \frac{\hbar}{2e}\frac{\partial \varphi}{\partial t}&=I_c(T)\sin \varphi(t)+I(t)R_N,\\
  \end{split} 
  \\[1ex]
  \label{eq:Lang_el}
  \begin{split}
  L_F\frac{\partial I}{\partial t}&=V_{dc}+V_{ac}(t)-\left(\frac{Q(t)}{C_F}+r_{sh}I(t)+\frac{\hbar}{2e}\frac{\partial \varphi}{\partial t}\right)+e_{int}+e_{ext},\\
  \frac{\partial Q}{\partial t}&=I(t),\\
  \end{split}
  \\[1ex]
  \label{eq:Lang_th}
  \begin{split}
  C\frac{\partial T}{\partial t}&=\frac{\hbar}{2e}\frac{\partial \varphi}{\partial t}I(t)-G_{12}T+G_{12}T_{2}-P_{bath}+P_X-I(t)e_{int}+p_{bath}+p_{12}+p_{ext,1},\\
  C_{2}\frac{\partial T_{2}}{\partial t}&=-G_{2,b}T_2+G_{12}T-G_{12}T_2+p_{12}+p_{2,b}+p_{ext,2},
  \end{split}
%  \end{aligned}  
%\eeq
 \end{empheq}
\end{subequations}
where \Eqref{eq:Lang_RSJ} is the superconducting phase relation according to the Jospehson and the RSJ model, Equations (\ref{eq:Lang_el}) and (\ref{eq:Lang_th}) are, respectively, the~electrical bias circuit equations and the thermal equations for the two-body  model shown in \figref{fig:ETscheme}. 
In \Eqref{eq:Lang_el}, $Q$ is the capacitor charge, and~$I$ and $T$ are the TES current and temperature, respectively. For~the ac-bias case, $V_{dc}=0$. For~the sake of simplicity, in~the electrical circuit equations \Eqref{eq:Lang_el}, we have omitted the terms related to the superconducting transformer drawn in \figref{fig:ETscheme}.  These can be easily derived by a standard electrical circuit~analysis.

 In the last two thermal equations, $P_X$ is the input power generated by an absorbed photon, while $P_{bath}$ refers to the power flowing to the thermal bath, given by 
 \beq{eq:Pbath}
 P_{bath}=k(T^n-T^n_{bath}),
 \eeq
where $k=G_{bath}/n(T^{n-1})$, with~$G_{bath}$ the differential thermal conductance to the thermal bath, $n$ the thermal conductance
exponent, and  $T_{bath}$ the bath temperature~\cite{IrwinHilton}. 

The heat capacity $C$ in \Eqref{eq:Langev} is typically the sum of the TES bilayer and the absorber heat capacity, since the two structures are generally thermally well coupled, while $C_2$ is a potential decoupled heat capacitance which could have different physical sources, like a fraction of the TES bilayer, the~supporting $Si_xN_y$ membrane, or the leads~\cite{Takei08, Goldie_2009,Maasilta12,Kinn12,Wake2019}.
The terms $e_{int}$ and $e_{ext}$ indicate the internal and external voltage noise sources, respectively,  while $p_{bath}$ and $p_{12}$, are the power noise sources generated by the finite thermal conductance to the bath ($G_{bath}$ and $G_2$) and between the different internal thermal bodies ($G_{12}$). The~term $p_{ext}$ accounts for potential parasitic external power sources like, for~example, stray light. All these noise sources will be extensively discussed in Section \ref{sect_noise}.

The full equations for a dc-biased TES are retrieved by setting $Q(t)/C_F=0$ in \Eqref{eq:Lang_el}. In~this case, $L_f$ and $r_{sh}$ are, respectively, the~Nyquist inductor and the shunt resistor typically used in a TDM read-out. The~ac voltage term $V_{ac}(t)$ is absent in normal bias conditions. However, it can be used to calculate the TES response to a small, $|V_{ac}|<<V_{dc}$, ac excitation, or~to evaluate the impact on the weak-link TES of electro-magnetic interferences (EMI) coupled to the bias line, by~assuming, for~example, $V_{ac}(t)=V_{pk}\cos(\omega_{EMI} t)$.  Interesting enough, by~adding to the bias line a controlled small signal at a fixed MHz frequency, Shapiro steps can be   observed in the TES IV characteristics, according to \mbox{\Eqref{IjBessel}}, and~the position of the steps can be used to accurately calibrate the TES voltage~\cite{SmithPoster2014,Sadleir2016,Zhou2018}.

The system of coupled equations given in \Eqref{eq:Langev} is typically solved in the small signal regime and in the linear approximation  using standard linear matrix algebra. In~this case, the~RSJ equation \Eqref{eq:Lang_RSJ} is generally ignored  and all the terms including $\partial \varphi/\partial t$ are replaced by $I(t)R$ \cite{IrwinHilton,Lind2004,JvdKuur11,TaralliAIP2019}.
    
The Equations (\ref{eq:Lang_RSJ}) and  (\ref{eq:Lang_el}) can be solved analytically, for~all the equilibrium values $I_0$, $R_0$, and~$T_0$ along the TES transition, by~simultaneously impose that, at~each value of $I(t)$, \Eqref{eq:Pbath} is satisfied. This has been shown in~\cite{Gottardi_APL14} for a TES bolometer ac voltage biased, at~resonance, at~$f_{bias}=1.4$ and $2.4\, \mathrm{MHz}$, with~$V_{ac}(t)=V_{pk}\cos(2\pi f_{bias})$. The~RSJ model  was extended to calculate the stationary non-linear response of a TES to a large ac bias current, following the approached described by Coffey~et~al. in~\cite{Coffey_2009,Coffey00}. A~clear signature of the ac Josephson effect in a TES was observed in the quadrature component of the current. Using the analytic expression for the non-linear admittance of a weak-link, changing in accordance with the power balance variation through the resistive transition, they could well reproduce the measured TES impedance  as a function of bias voltage and frequency, at~the  operating temperature.
In \figref{fig:RSJboloxray}a, the~measured TES resistance and reactance of a Ti/Au   $50\times 50 \, \upmu\mathrm{m}^2$ TES bolometer  biased at $2.4 \, \mathrm{MHz}$ is show as a function of the TES voltage. The~reactance has a periodic dependency on the voltage as predicted by the RSJ model (red line in the graph) \cite{Gottardi_APL14}. 

Using a similar approach, but~then solving the equations numerically, the~peculiar structures observed in the TES reactance of a low resistance ($R_n= 8\, \mathrm{m}\Omega$), Mo/Au TES microcalorimeters with~noise mitigation structures developed at NASA-GSFC, have been explained as well~\cite{Gottardi17}. An~example of the results presented in~\cite{Gottardi17} is reproduced in figure  \figref{fig:RSJboloxray}b, where the TES electrical impedance is shown as a function of bias voltage. The~sharp jumps in the TES reactance $X_\omega$, are related to the Josephson effects in combination with a large TES bias current (detector with high saturation power) and low $R_n$ as predicted in~\cite{McDonald97}.  

% start a new page without indent 4.6cm
%\clearpage
\end{paracol}
\nointerlineskip

\begin{figure}[H]
\widefigure
\includegraphics[width=14.cm]{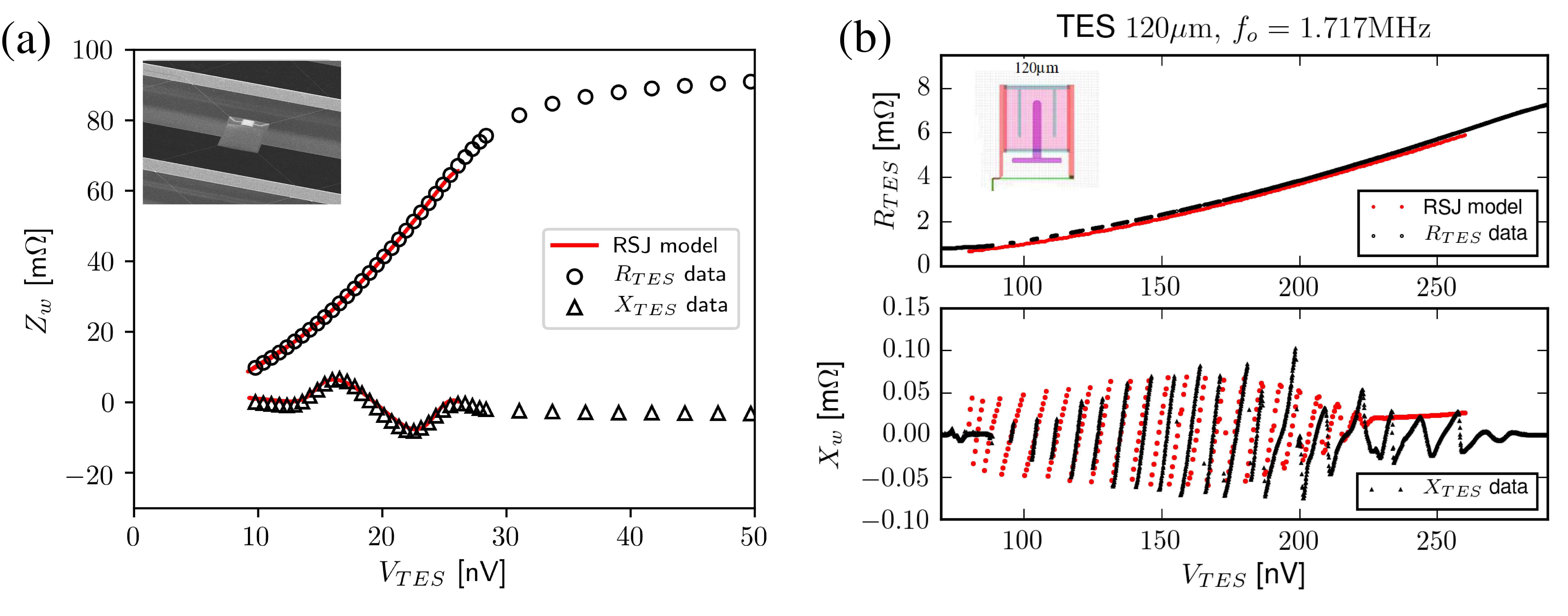}
\caption{\label{fig:RSJboloxray} TES resistance $R_{TES}$ and reactance $X_{TES}$ as a function of the TES voltage for (\textbf{a}) a $50\times 50 \, \upmu\mathrm{m}^2$ bare Ti/Au TES bolometer~\cite{Gottardi_APL14} and (\textbf{b}) a $120\times 120 \, \upmu\mathrm{m}^2$ NASA-GSFC TES microcalorimeters. The~detectors are biased, respectively, at 2.4 and 1.7 MHz. The~red lines show the prediction from the RSJ model. Reprinted from~\cite{Gottardi17}  by permission from IEEE,  IEEE Transactions on Applied Superconductivity. Copyright \copyright  2018, IEEE. }
\end{figure} 

\begin{figure}[H]
\widefigure
\includegraphics[width=18cm]{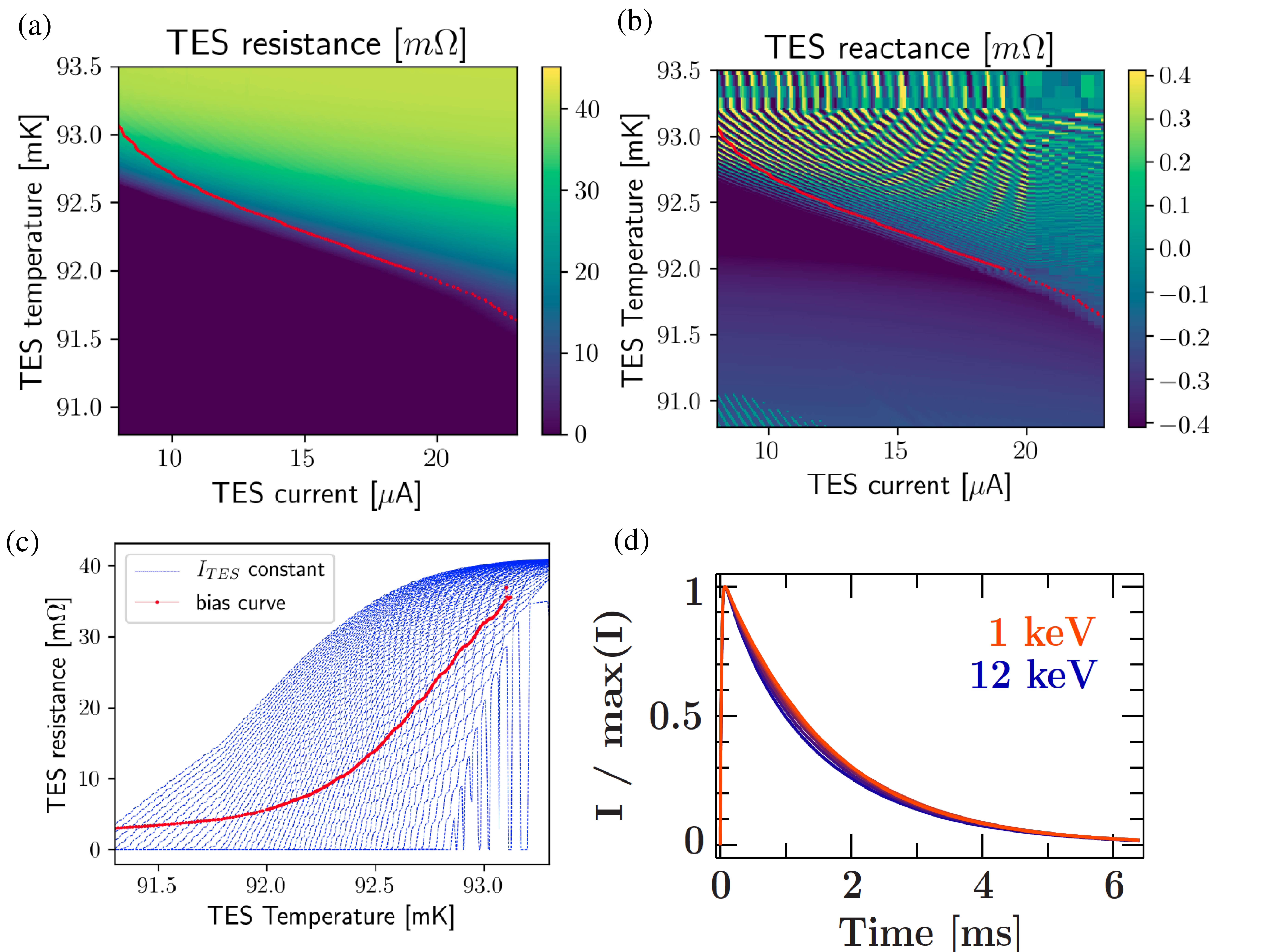}
\caption{\label{fig:RXsim}  Colormaps of the TES resistance $R(I,T)$ (\textbf{a}) and reactance $X(I,T)$ (\textbf{b}) calculated for an NASA-GSFC $100\times 100 \,\upmu\mathrm{m}^2$  TES microcalorimeters biased at $f_{bias}=3.081\, \mathrm{MHz}$.  (\textbf{c}) $R(T)$ curves calculated at a constant TES current. (\textbf{d}) Simulated X-ray pulses from photons of energy from 1--10 keV, for~a TES biased at 1 MHz at $R/R_n=10\%$, credits C. Kirsch, 2019. }
\end{figure}

\begin{paracol}{2}
%\linenumbers
\switchcolumn

Numerical simulations of the resistive transition of a dc biased TES, based on the RSJ model, have been performed by Smith~et~al.~\cite{Smith13}. The~work includes an extensive study on the effect of (self-)magnetic field perpendicular to the TES~surface. 

The coupled differential equations in \Eqref{eq:Langev} have been fully solved numerically, for~an ac biased pixel,  in~the simplified single-thermal body case~\cite{Kirsch2020}. In~the numerical simulations, the~TES impedance is the results of the solution of the RSJ equation and the power balance, as~for the analytical solution described in~\cite{Gottardi_APL14}. Aside from the electric circuit and thermal bath parameters, and~the TES normal resistance $R_n$, the~most important input for the calculation is the $I_c(T)$ curve, which can be evaluated experimentally.  
The equilibrium value of the simulated TES current has been shown to be in excellent agreement with the  experimental IV curve of the modeled pixel. The~strength of this numerical time-domain model is that, beside predicting the steady state of the TES, it can also be used to simulate the TES response to the incoming photons in the non-linear and large signal regime. As~a matter of fact, this TES simulator will become part of the end-to-end (e2e) simulator of the X-IFU  on Athena~\cite{Wilms2016,Lorenz2020}. 
In \figref{fig:RXsim}, we give an example of the simulation of  high normal resistance  TES microcalorimeters developed at NASA-GSFC for the Athena/X-IFU and characterized at SRON under MHz bias. The~devices consist of Mo/Au $100\times 100 \upmu\mathrm{m}^2$ bare TESs, with~a normal resistance $R_n=41 \, \mathrm{m}\Omega$, a~$G_{bath}=83\,\mathrm{pW/K}$ and a saturation power $P_{sat}\sim2\, \mathrm{pW}$ at $T_{bath}=55\, \mathrm{mK}$. They have been characterized under ac-bias at several bias frequencies (from 1.1 up to 4.7 MHz). The~color maps in \mbox{\figref{fig:RXsim}a,b} show, respectively, the TES resistance $R(I,T)$ and reactance $X(I,T)$ as a function of  the current $I$ and temperature for a TES biased at $f_{bias}=3.081\, \mathrm{MHz}$. The~red curve gives the bias trajectory as the solution of the power balance equation.  The~TES impedance has been calculated analytically as in~\cite{Gottardi_APL14,Coffey_2009}. A~set of $R(T)$ curves calculated at a constant TES current are given in \figref{fig:RXsim}c. The~oscillating structures observed experimentally with this type of TESs are clearly visible on each $R-T$ line as well as  on the resulting  bias curve in red. The~oscillation are generally much smaller when the TES is biased at lower frequency, following the prediction from the Josephson~equations. 
% start a new page without indent 4.6cm
\clearpage
\end{paracol}
\nointerlineskip

\begin{paracol}{2}
%\linenumbers
\switchcolumn

The $R(T,I)$ curve calculated from the RSJ model can be used as  input for the
end-to-end simulator under development for Athena/X-IFU to provide  a realistic response of the detector to photon of different energy. An~example of the  simulated X-ray pulses from photons of energy from 1--12 keV, for~a TES biased at 1 MHz at $R/R_n=10\%$ is given in \figref{fig:RXsim}d. Solving numerically the full equations (\Eqref{eq:Langev})   is rather fast~\cite{Kirsch2020}. When the calculations are supported by the graphical process units (GPUs), the simulations of  an array of 3000 pixels can be done within several hours~\cite{Lorenz2020}.  The~preliminary comparison with the experimental data are very promising~\cite{Kirsch2020} and work is in progress  to further validate the simulator and to better understand the energy scale of real TES X-ray calorimeter. When the full noise contributions will be integrated, the end-to-end simulator will become a very powerful tool to assist the detectors development for other TES-based instruments as~well. 

Another interesting technique to guide the detector development is to use optical photons from a laser source to  illuminate the X-ray calorimeter via a coupled optical fiber. Using X-ray microcalorimeter from NASA-GSFC, with~energy resolution of 0.7--2.3 eV at 1--6 keV, F. Jaeckel~et~al.~\cite{Jaeckel2019} have achieved photon number resolution for 405 nm photon pulses with mean photon number up to 130 (corresponding to 400 eV). The~experimental set-up is currently being improved to minimize the thermal crosstalk  and improve on the photon counting capability to higher energy~\cite{Jaeckel2021}. This technique is very promising and relatively simple to implement as a standard, inexpensive, calibration facility in a laboratory. In~combination with the e2e simulator discussed above, it could become a very useful tool to verify the small and large signal TES response, as~well as to understand the effect of the detectors non-linearity on the energy~calibration. 

\subsection{TES~Noise}\label{sect_noise}
The total noise of a TES detector  is a combination of Nyquist--Johnson noise associated with the TES resistance  itself and with the losses in the read-out circuit and bias circuit, and~thermal fluctuation noise  between the TES thermal elements.  The~noise sources can be classified in {\it external} and {\it internal} %can italics be removed?
noise sources. Typical external Johnson noise sources  are the additive white current noise from the SQUID amplifier and the white voltage noise $e_{sh}=\sqrt{4k_BT_{bath}r_{sh}}$ from  the Nyquist--Johnson noise of the shunt resistor $r_{sh}$ in the ac or dc bias circuit. With~a proper choice of the parameters of the bias circuit and the SQUID amplifiers, these noise terms can be made negligible. Other external noise sources like, for~example,  stray light or cosmic photons could generate power fluctuations ($p_{ext,1}$ and $p_{ext,2}$) on the detector thermal elements. A~careful thermal design of the focal plane assembly can reduce these noise sources to a negligible~level.   

In \figref{fig:noise}, we show, as~an example,  a~typical current noise spectra for a Ti/Au TES developed at SRON, ac biased at $2.5\, \mathrm{MHz}$ in the transition at $R/R_n\sim0.2$. The~different noise contributions  are explained here~below. 
\begin{figure}[H]
\includegraphics[width=7.5cm]{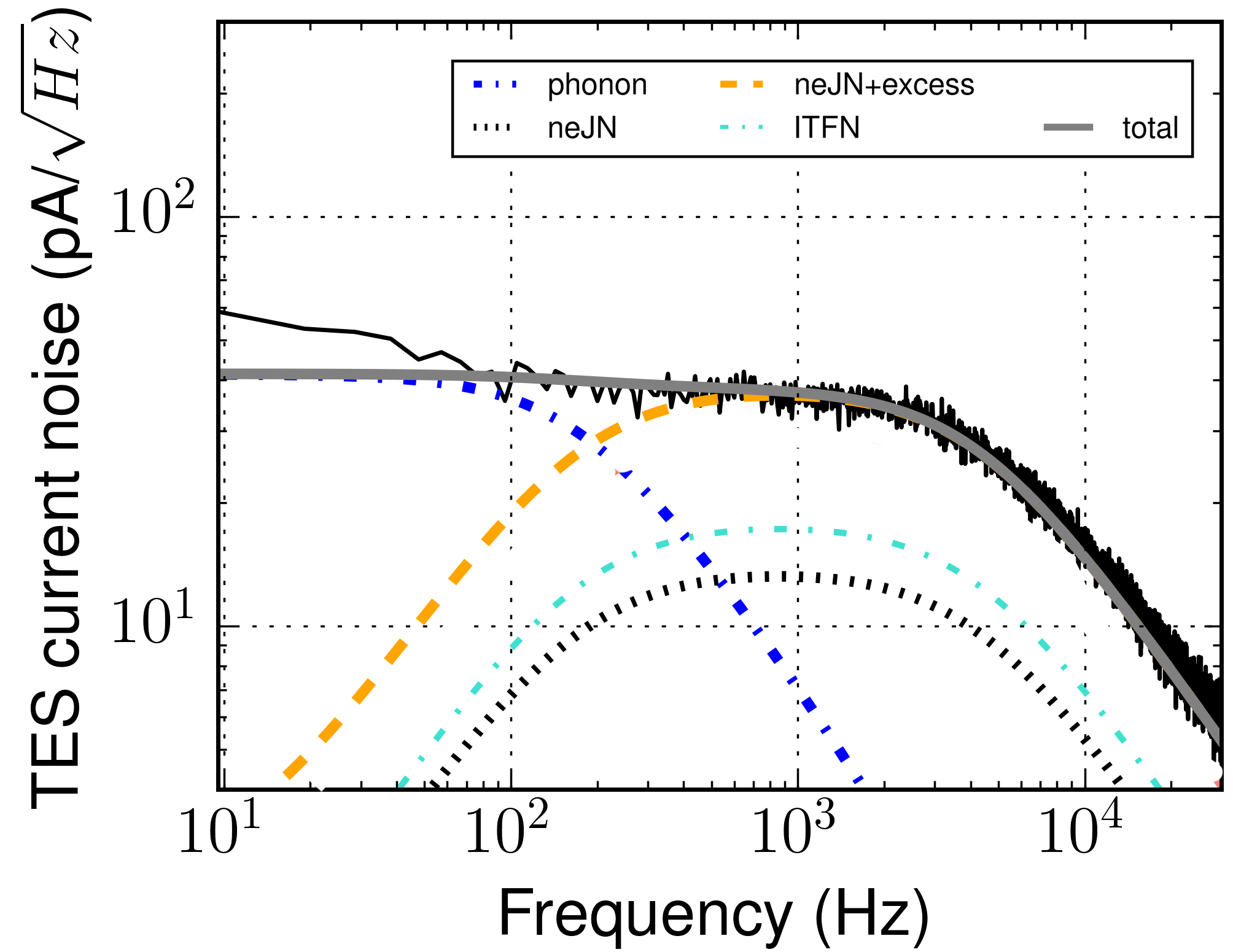}
\caption{\label{fig:noise} Current noise spectra of a Ti/Au $80\times 40\, \upmu\mathrm{m}^2$ TES X-ray calorimeter developed at SRON, biased at $2.5\, \mathrm{MHz}$, at~ $R/R_n\sim0.2$ with $\alpha\sim370$ and $\beta\sim5$  \cite{deWit_2020}.}
\end{figure}

There are three  fundamental thermodynamic internal noise sources of a TES detector, described with a two-body model as in \figref{fig:ETscheme}. 
The first is the thermal fluctuation noise between the TES detector and the heat bath, already discussed in Section \ref{sec:TEStheory}. It is prominent in the detector thermal bandwidth typically below $\sim200\,\mathrm{Hz}$.  It has a white thermal noise given by $p_{bath}=\sqrt{F(T_0,T_{bath})4k_BT^2G_{bath}((T_{bath}/T)^{n+1}+1)(1+M_{ph}^2})$, where  $F(T_0,T_{bath})\sim0.5$, as~discussed in Section \ref{sec:TEStheory}, and~$M_{ph}$ is a parameter introduced {ad-hoc} to account for the excess noise observed  in the experiments. In~the literature, this noise term is typically referred to as {\it phonon} %can italics be removed?
 or thermal fluctuation noise~\cite{IrwinHilton}. 
The noise contribution $p_{2,b}$ to the thermal fluctuation noise from the conductance $G_{2,b}$ is generally neglected since $G_{2,b}\ll G_{bath}$.
The second fundamental noise term is the so called internal fluctuation thermal noise (ITFN), which is generated by thermal fluctuations between distributed heat capacity ($C$ and $C_2$ of \figref{fig:ETscheme}b) internal to the TES-absorber system. It has a  spectral density $ p_{itfn}\equiv p_{12} =\sqrt{4k_BT^2G_{12}}$, where $G_{12}$ is the intrinsic thermal conductance. The~presence of internal thermal fluctuation  noise in TES microcalorimeter was first reported by Hoevers~et~al.~\cite{Hoevers00}. The~origin of the thermal conductance $G_{12}$ can be of different physical nature depending on the specific TES design. In~TES bolometers, for~example, the~parasitic conductances and the distributed heat capacities is typically attributed to the electrical leads patterned onto the thermally isolating nitride legs~\cite{Gildemeister01} or to the SiN legs themselves~\cite{Pourya12}. In~Takei~et~al.~\cite{Takei08}, the~observed dangling heat capacity arose likely from the thin SiN membrane supporting the low $C$ TES calorimeter. In~Goldie~et~al.~\cite{Goldie_2009}, the~internal thermal fluctuation noise is assumed to be generated by a poor thermal conductance of the TES bilayer and $G_{12}$ was estimated from the Wiedeman--Franz relationship $G_{WF}=L_{o}T_0/R_0$, where $L_o=24.5\times 10^{-9}\, \mathrm{W}\Omega\mathrm{K}^{-2}$ is the Lorenz number.
Wakeham~et~al.~\cite{Wake2019} investigated the contribution of the internal thermal fluctuation  noise on TES microcalorimeters with different design and the origin of the finite $G_{12}$ term was thoroughly discussed. 
It can be demonstrated that when  passing through the power-to-current transfer function of the electro-thermal system (see for example~\cite{Maasilta12}), the~ITFN contribution to the current noise has a shape similar  to the Nyquist--Johnson noise described here below, and~it is the highest in the kHz~range.

The third intrinsic noise contribution is the Nyquist--Johnson noise of the TES itself biased in the transition and it is generalized in the form of a voltage noise $e_{int}=\sqrt{4k_BTR\zeta(I)}$. The~response of the TES current to $e_{int}$ is suppressed  at low frequency by the electro-thermal feedback~\cite{Irwin1995} and becomes significant in the detector electrical band at kHz. The~function $\zeta(I)$, introduced in \Eqref{eq:dE}, accounts for the non-linear and non-equilibrium nature of the TES resistance, which is strongly dependent on the TES current. For~linear resistors 
at equilibrium $\zeta(I)=1$. Irwin~\cite{Irwin2006} assumed the TES to be a simple Markovian system with no hidden variables such as internal temperature gradients and fluctuating current paths. Applying the Stratonovich’s nonequilibrium Markovian   
fluctuation–dissipation relations, he calculated the first order, near-equilibrium,  non-linear correction term to the noise to be $\zeta_{ne}(I)=1+2\beta$. A~solution for the higher order terms cannot be found from known TES properties since they contain dissipationally indeterminable parameters~\cite{Irwin2006}. In~this form, the~noise model is known in the literature as non-equilibrium Johnson noise (neJN) and  is extensively used to model the, $\beta$ dependent, Nyquist--Johnson voltage noise observed in the TES. The~broadband unexplained or excess  noise,  typically observed at frequencies larger than the thermal bandwidth of the TES~\cite{Ullom04}, could only partially be explained after the introduction of the correction term $\zeta_{ne}(I)$ and only  at relatively low $\beta$ values~\cite{Takei08,Jethava09,Smith13}. As~shown in~\cite{Maasilta12,Wake2018,Wake2019}, the~characterization of the noise in the electrical bandwidth is complicated by the fact that both the non-equilibrium Johnson noise and the internal thermal fluctuation noise  give a similar contribution  in the measured TES current noise,  after~passing through the system~transadmittance. 
Smith~et~al.~\cite{Smith13}, however, excluded the presence of a significant contribution of the ITFN  in the excess noise observed with their low resistance, large $\beta$, high thermal conductance devices. 

\noindent Noise mitigation normal metal bars and stripes have been proposed and successfully, widely introduced in the TES design of microcalorimeters to  suppress the excess noise~\cite{Ullom04,Jethava09}.

Kozorezov~et~al.~\cite{Kozo12} argued that the observed excess Johnson noise in TES-based detectors could have a natural explanation within the RSJ theory. Following the work of Likharev and Semenov~\cite{LikSem72}, they calculated the power spectral density of the voltage fluctuations across the TES (considered as a resistively shunted junction) $S_V(\omega)$, averaged over the period of the Josephson oscillations. They obtained
\beq{eq:SvRSJ}
S_V(\omega)=\frac{4k_BT}{R_N}\sum_m|Z_{0m}|^2,
 \eeq
where $Z_{mm'}(\omega)$ are  the components of the impedance matrix of the biased TES with the index $m$ standing for the $m_{th}$ harmonic of the Josephson oscillation at $\omega_J$. This approach was used in~\cite{Koch80} to develop the quantum-noise theory of a resistively shunted Josephson junction and it was proved that, to~calculate  the total low frequency voltage fluctuations, one needs to take into account the {\it mixing down} %can the italics be removed?
of high harmonics of the Josephson frequency. 
In~\cite{Kozo12}, it was shown that the spectral density of voltage noise in the RSJ model has a unique analytical structure that cannot be reduced to the expression for the neJN from~\cite{Irwin2006} and that the only source of noise is the  {equilibrium} Johnson normal current noise~\cite{LikSem72}. Moreover, the~RSJ model predicts a significant excess noise, with~respect to the neJN, for~the lower part of the resistive transition, as~generally observed in many experiments. 
In a recent review~\cite{Wessel2019} on the models for the excess Johnson noise in TESs, the~authors derived a simpler expression for \Eqref{eq:SvRSJ} based on the approximations explained in~\cite{Kozo12}. They got  a voltage noise:
\beq{eq:SvRSJbeta}
%\begin{empheq}{align}
\begin{split}
e_{int}=\sqrt{S_V(\omega)}=\sqrt{4k_BTR\zeta_{RSJ}(I)}\\
\mathrm{with}\;\;\; \zeta_{RSJ}(I)=1+\frac{5}{2}\beta+\frac{3}{2}\beta^2,
\end{split}
\eeq
where the $R_N$ in \Eqref{eq:SvRSJ} is replaced by $R$, given the fact that the thermal fluctuations are associated with the real part of the TES impedance at the equilibrium value. In~the same paper, they  compare the measured Johnson noise for a few TES microcalorimeters~\cite{Wake2018} with the prediction \Eqref{eq:SvRSJbeta} and the general form derived by Kogan and Nagaev~\cite{KogNag88} (KN) and  the prediction from the two-fluid model~\cite{Wessel2019,BennetPRB13}. In~the simplified form of the two-fluid model, no noise is mixed-down from the Josephson to low frequency, and~the expected noise is typically underestimated. A~better  agreement with the data is observed with the RSJ and the Kogan--Nagaev models. In~a recent study~\cite{Gottardi2021} on the noise of high-aspect ratio TESs under development at SRON for the MHz bias read-out, a~very good agreement between the observed Johnson noise and the prediction from the RSJ and Kogan--Nagaev has been demonstrated over a large number of TES designs and bias~conditions.

%However, unexplained noise is still present in the new generation of high-$\beta$ TES detectors  fabricated without the normal metal structures~\cite{}, introduced in the past to mitigate the excess Johnson noise~\cite{Ullom04,Lind_NIMA_2004}.

%%%%%%%%%%%%%%%%%%%%%%%%%%%%%%%%%%%%%%%%%%
\section{Large Arrays~Fabrication}\label{sec:LAFab}
%\unskip
 A TES is in  essence  a superconducting metal slice deposited on a substrate. The~earliest TES, fabricated in 1941, was just a thin film of lead evaporated onto a 1.5 cm $\times$ 1 mm glass ribbon~\cite{Andrews41}. A~TES-based X-ray microcalorimeter fabricated nowadays is more complex and consists of three key components: a temperature sensing element (the TES), a~thermal isolation structure, and~an X-ray absorbing layer. In~this section, we will briefly review these components from the aspect of fabrication including materials and microstructures with the emphasis on the technical scalability. An~array of thousands of TESs, realized on a single 4-inch silicon wafer, are under fabrication for the  X-IFU onboard of Athena~\cite{Barret20} (\mbox{Section \ref{sec:athena}}). Moreover, as~it will be reported in \mbox{Section \ref{sec:otherXray}}, beyond~the 2040s, arrays of hundred of thousands of TESs will be required for very large X-ray observatories like, for~example, the~proposed  Lynx space mission~\cite{Gaskin19} or the Cosmic Web Explorer~\cite{Simionescu2019}. In~parallel to these space missions, increasing the size of the array is also mandatory for many other ground-based applications (Section \ref{sec:PP}). We will discuss here  the technical challenges and solutions required to make these large scale arrays a~reality.
\subsection{TES~Bilayer}

The TES itself is the core of an X-ray calorimeter. It is the sensing element that can detect very small  temperature changes. As~shown in \figref{fig:TESabs}, it consists of a superconducting thin film connected between superconducting leads. For~high resolution spectroscopy, the~transition temperature of the film is chosen typically within the range from 50 to 100 mK.  However, other transition temperatures can be selected depending on applications. Various types of films have been proposed and successfully used for almost any photon wavelength to achieve the targeted transition temperatures; a single layer of tungsten~\cite{W_CRESST2018,W_Lita05}, alloys such as Al-Mn~\cite{AlMn2011}, and proximity-coupled bilayers (or multilayers) of Ti/Au~\cite{Nagayoshi19}, Mo/Au~\cite{Fink2017,Parra2013,Mo_Fabrega2009,Cherv2008}, Mo/Cu~\cite{Orlando18}, Ir/Au~\cite{IrAu_Kunieda2006}, and Al/Ti~\cite{AlTi_Lolli2016,AlTi_Kamal2019}. The~geometric size of the superconducting slices is normally tens of microns or larger so wet etch, dry etch, or lift-off processes can be adopted in combination with a standard UV~photolithography.

  The materials for the superconducting leads must have a higher transition temperature and a critical current with respect to the TES film. Niobium is widely used, but~other superconductors such as niobium nitride, molybdenum, and aluminum are also applied. One of the technical challenges for the scalability of the sensing structures is the topology of the wirings on a detector chip. Superconducting broad-side coupled micro-striplines are essential in creating dense-packed arrays. An~example of the structure of the microstripline is given in the schematic of \figref{fig:TESabs}a. It consists of two layers of the superconducting leads, which are electrically isolated by a thin oxide layer. The~state-of-the-art system of wiring lines consists of multiple broad-side coupled lines, embedded within oxide layers by applying chemical mechanical polishing so that planarized microstripline layers can be stacked in a vertical direction. Massachusetts Institute of Technology Lincoln Laboratory has demonstrated the process with six superconducting Nb layers for superconducting electronics~\cite{Huang2020,Devasia2019,Yohannes15}. 
% start a new page without indent 4.6cm
%\clearpage
\end{paracol}
\nointerlineskip

\begin{figure}[H]
\widefigure
\includegraphics[width=15.2cm]{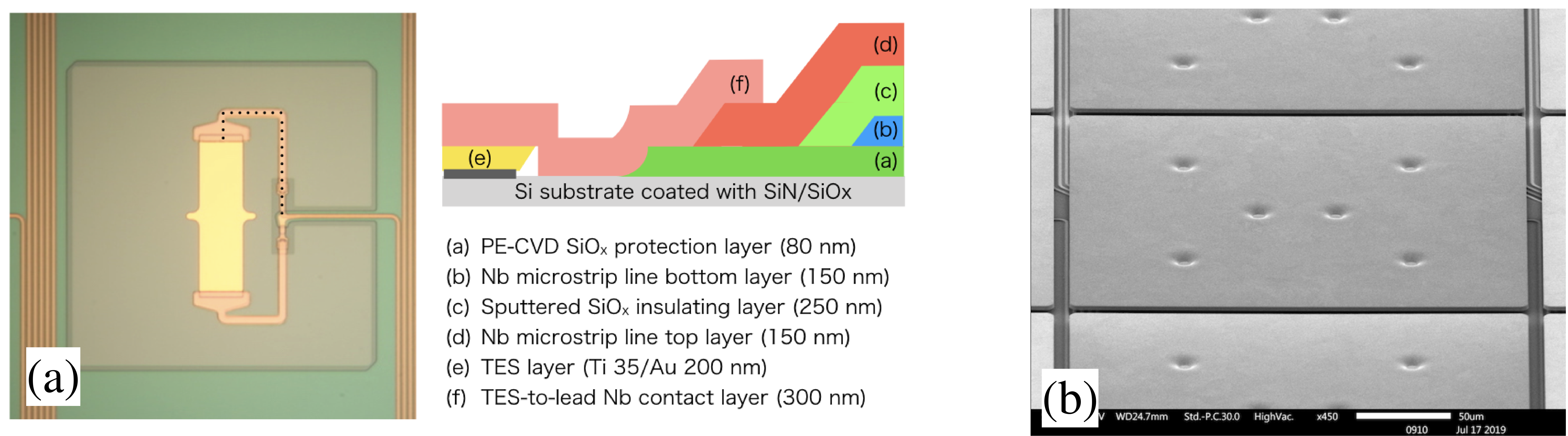}
\caption{(\textbf{a}) Microphotograph of a 100$\times$30 $\upmu$m$^{2}$ Ti/Au TES connected to the Nb superconducting leads and a schematic of a cutting view along with the dotted line. The~separated leads are overlapped near the TES film but electrically separated with an oxide insulating layer, forming a microstripline. The~figure was reprinted from~\cite{Nagayoshi19}  by permission from Springer Nature. Copyright \copyright 2019, Springer Science Business Media. (\textbf{b}) Microphotograph of an Au X-ray absorber. The~absorber is an overhanging structure supported by several stems that can be seen as  dots on top of  the absorber surface. The~two of them located close to the center of the absorbers are thermally connected to the TES thermometer, not visible in the photo.}
\label{fig:TESabs} 
\end{figure}

\begin{paracol}{2}
%\linenumbers
\switchcolumn

\subsection{Thermal Coupling to the~Bath}

A TES film kept at its transition temperature $T_c$ by the electro-thermal feedback, must be weakly connected to the thermal bath stabilized at a temperature $T_{bath}<T_c$. The~thermal link must be properly engineered to be in accordance with  the application requirements.  A~low-stress, silicon-rich $Si_xN_y$ membrane is widely used, with~a few exceptions, to~achieve the thermal isolation because it is mechanically strong and compatible with the  other processes used in TES fabrication. The~thermal conductance can be controlled by changing the thickness of the membrane and its geometric design, by~introducing, for~example, slit holes or other phonon scattering structures. In~ \figref{fig:SiN}a, we show a photograph of an array of ultra-low-noise TES bolometers suspended with 4 diagonal thin and long  legs with dimensions $340\,\upmu\mathrm{m} \times 0.5\,\upmu\mathrm{m} \times 0.25\,\upmu\mathrm{m}$ \cite{Ridder2016}, giving a good example of how the SiN membrane is mechanically stable and flexible in its design. For~X-ray calorimeters, the membrane has a rather simple squared geometry, typically without etched holes or slots.  As~an example, \figref{fig:SiN}b shows a microphotograph of thousands of squared pockets vertically etched through the  Si carrier wafer of 300 $\upmu$m by using a deep-Reactive Ion Etching (RIE) process. It is possible, in~this way,  to~create a uniform large array of the 0.5-$\upmu$m-thick SiN membranes with very high yield. The~scalability towards an array of 100-kilo pockets is not considered a bottleneck~\cite{Bandler2019}.  
\begin{figure}[H]
\includegraphics[width=12cm]{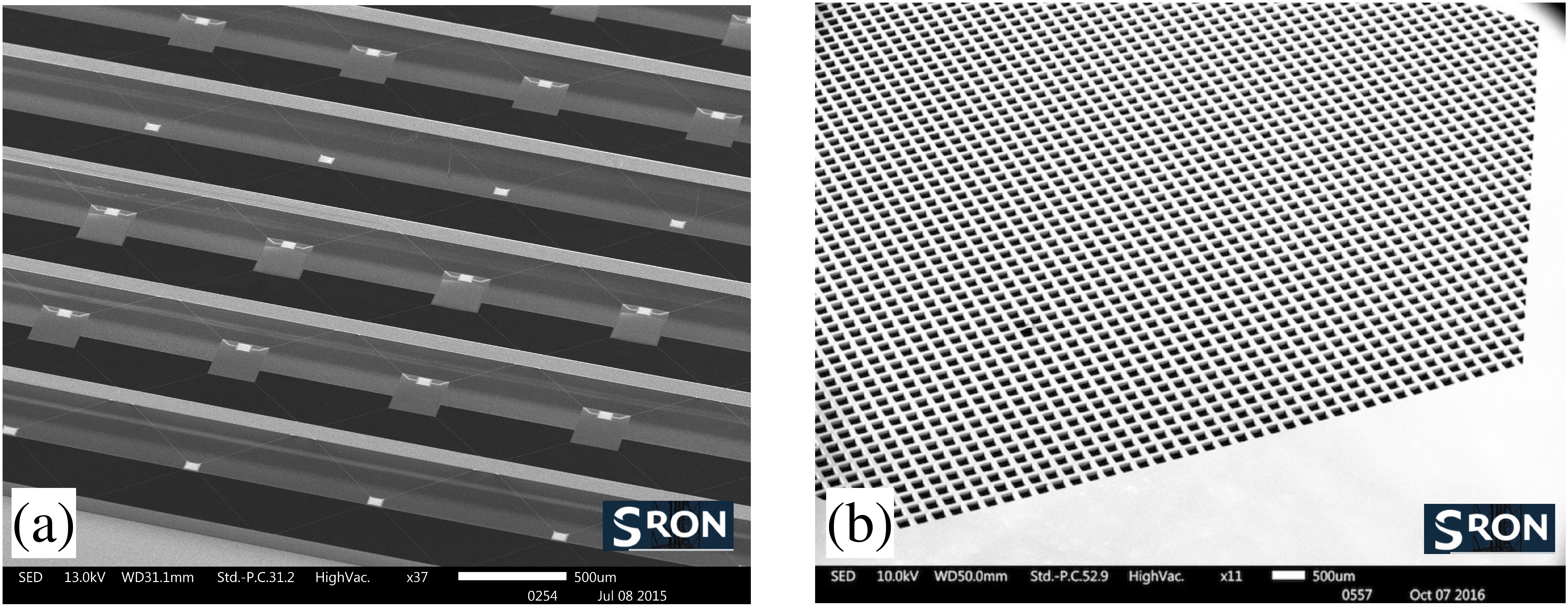}
\caption{(\textbf{a}) SEM micrograph of an array of ultra-low-noise TES bolometers on suspended SiN structures  with diagonal long legs ($340\,\upmu\mathrm{m} \times 0.5\,\upmu\mathrm{m} \times 0.25\,\upmu\mathrm{m}$) design. In~the middle of each detector there is a TES and an absorber on a SiN membrane~\cite{Ridder2016}. (\textbf{b}) Microphotograph of an array of thousands of square pockets etched by using a Si deep-RIE process. The~pocket size is 190$\times$190 $\upmu$m$^{2}$ and the pitch is 250 $\upmu$m.}
\label{fig:SiN} 
\end{figure}
A critical issue however remains the proper thermalization of the Si carrier structure to minimize the cross talk between the nearest neighbor pixels located on each pocket. This could become critical when large arrays need to be built. For~a k-pixel array under development for X-IFU, it was show that the thermal crosstalk between pixels can be made smaller than  
$2\times 10^{-4}$ (well within the requirements derived from the scientific goals of X-IFU) by coating the back-side and sidewalls of the Si grid structure with a layer of Cu or Au with an appropriate thickness~\cite{Miniussi2020}.

\subsection{Absorbing~Layer}
An absorbing layer that is  thermally well connected to the TES is an essential component of an X-ray microcalorimeter.  The~absorbers can be directly attached to a small area of the TES slice, or~can be place over/near the TES films via thermal coupling structures~\cite{Orlando18}. For~imaging detectors, the~absorber is placed directly on top of the TES to maximize the filling factor and hereby defines generally the pixel size. Several absorbers can also be  connected to a single TES via  thermal links of different length, as~it is the case of the emerging class of detectors discussed in Section \ref{sec:hydra}. 

It is important that the absorber has a high intrinsic thermal conductance  to guarantee a fast thermalization, after~a photon hit, and~to minimize potential internal fluctuation noise. 
The absorbers are employed to enhance the quantum efficiency of the devices over the target energy band.  Thus, providing sufficient stopping power over the large detection area is of primary importance. Therefore, materials of high atomic number with good thermal conductivity are desired. In~the soft X-ray field, semi-metals like Bi or normal metal such as Au and Cu have been widely used. In~\figref{fig:TESabs}b, we show a microphotograph of an X-ray absorber made of a 2.3 $\upmu$m thick Au film, which provides a quantum efficiency of 83 \% for a 6 keV photon.
 It is possible to achieve a  quantum efficiency as close as 100 $\%$ by increasing the thickness of Au up to 7 $\upmu$m at the cost of an increase of the heat capacity and a degradation of the detector sensitivity. To~overcome this counter-effect, Bi, which has a very low specific heat at low temperature, can be deposited directly onto a thinner normal metal absorbing layer like~Au. 
 
 The absorbing layers can be deposited by using sputter deposition, evaporating, or electroplating. Any of these can be adopted to form a uniform large array, but~electroplated thin films have been shown to achieve higher Residual-Resistivity Ratio (RRR), usually defined as the ratio of the resistivity of a material at room temperature and at mK,  and better thermal characteristics~\cite{Brown08,Yan17}. Furthermore, electroplating is preferable, especially for the thicker layers, to~minimize the thermal loading during deposition and the stress in the films. On~top of this, evaporating thick layers of Au is far too expensive.
In \figref{fig:BiNIST}, we reproduce the  scanning electron micrographs of evaporated and electroplated bismuth studied at NIST to better understand the non-Gaussian spectral response observed in X-ray absorbers~\cite{Yan17}.
\begin{figure}[H]
%\begin{center}
%\begin{tabular}{c} %% tabular useful for creating an array of images 
\includegraphics[width=13cm]{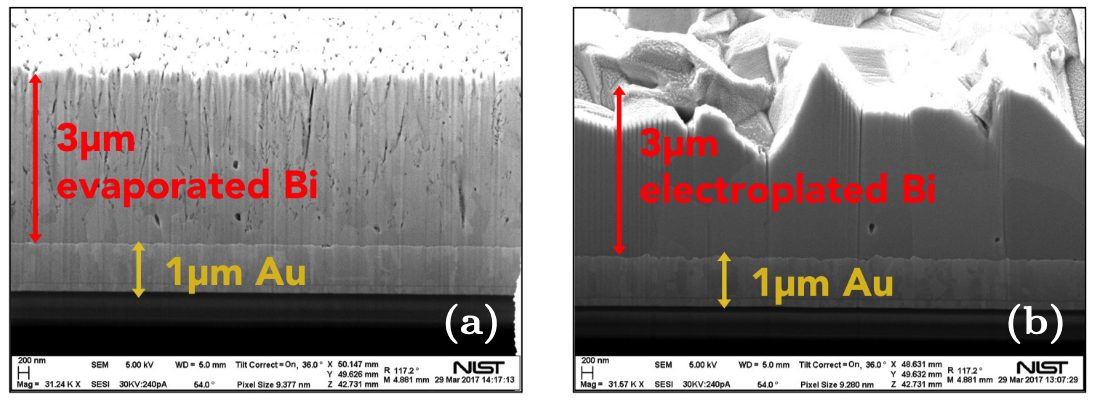}
%\end{tabular}
%\end{center}
\caption[example] 
%>>>> use \label inside caption to get Fig. number with \ref{}
{ \label{fig:BiNIST} Scanning  electron  micrographs  of  the  (\textbf{a}):  evap-Bi and  (\textbf{b}):  elp-Bi  absorber  cross-sections.   The~ evap-Bi  grain appears smaller than that of the elp-Bi, and~shows a columnar structure. Reprinted  from~\cite{Yan17} by permission from AIP Publishing. Copyright \copyright 2017, AIP Publishing.}
\end{figure}

 To achieve the required high QE however, a~careful study needs to be performed on the thickness and the quality of the Au/Bi bilayer. It has been recently shown~\cite{Hummatov2020} that the roughness, the~edge profile, the~angles of incident of the X-ray photons, and the filling factor needs to be known well to obtain an accurate estimate of the QE.
To minimize the effect of the absorption of the stray light  in the detector, it is also desirable to increase the reflectivity of the absorber for long-wavelength infrared radiation. For~an Au/Bi absorber, this can be done for example by adding a Au capping layer on top of the Bi, with~a thin Ti adhesion layer. A~quantum efficiency of 90$\%$  has been reported for TES-X-ray calorimeters designed for Athena/X-IFU~\cite{Hummatov2020} with an increase of reflectivity, measured at room temperature, from~ 45$\%$ to up to 80$\%$  after adding a 40 nm Au capping~layer. 

 In some particle physics experiments that aim to directly detect the neutrino mass using a large array of TES X-ray microcalorimeters (see Section \ref{sec:PP}), a~precise amount of radioactive material like $^{163}$Ho,  has to be embedded in the detector absorber.  A~multi-step micro-fabrication process to achieve this is described in~\cite{Orlando18}. One crucial part of such a process is the co-deposition of gold during the $^{163}$Ho implantation process on the detector~absorber.  

%%%%%%%%%%%%%%%%%%%%%%%%%%%%%%%%%%%%%%%%%%
\section{Single Pixel~Optimization}\label{sec:opt}

In this section, we will review the progress in the pixel optimization for a large array to be used in dc and ac biased multiplexing schemes. Detectors with excellent single pixel energy resolution are shown regularly in many laboratories. The~big challenge however remains to find an optimal pixel design and fabrication process that guarantees  high yield and uniform performance over a large number of pixels.
The underlying physical effects discussed in Section \ref{sec:TEStheory} have different implication on the performance of the TESs according to whether the detector is dc or ac biased, as~it was reported in several works~\cite{Smith13,Gottardi12xray,Taralli2020}.     A~tailored optimization approach is then required, in~view as well of the different~applications.

\subsection{dc~Bias}\label{sec:dcbias}

For many years, the~efforts  to improve the single pixels performance of TES microcalorimenters have been devoted in trying to suppress the poorly understood excess Johnson noise, typically increasing at larger $\beta$. It was empirically observed that adding normal metal stripes on top of the TES bilayer would  mitigate the overall noise~\cite{Ullom04,Lind_NIMA_2004}. Detectors with very good energy resolution have been fabricated following these finding~\cite{Ullom_2015}.
With the increasing demand of larger and larger arrays of pixels, it was soon realized that normal metal structures could strongly affect the shape of the resistive transition in a way that it was not always controllable and reproducible.  Moreover, kinks in the transition have also been regularly observed with those devices. This is partially due to the fact that the proximity effects discussed in Section \ref{sect_WL} are strongly dependent on the device geometry, the~transmissivity of the leads-TES interface  and other parameters of  the SNS structures, such as $T_c$, and~coherence length. In~a series of very detailed experiments, it was shown, by~Smith~et~al.~\cite{Smith13,Smith2014} and in later works~\cite{Wake2018,Zhang2017}, how the normal metal structures added to the TES, including the thermal coupling stems between TES and absorber, could dramatically change the resistive transitions shape and the current flow~\cite{Swetz12}, as~a complex interplay of proximity effects, non-equilibrium superconductivity, and self-induced magnetic field.
An improved  TES design was highly desired, in~particular because the dc bias multiplexing schemes do not allow a single pixel bias optimization.   Due to the common bias configuration, they require an extremely high uniformity over a large array of~pixels.

  The traditional strategy of suppressing $\zeta(I)=(1+2\beta)(1+M^2)$ by adding normal stripes on top of the bilayer has the side effects of reducing $\alpha$ as well, since $M^2$,$\alpha$ and $\beta$ were found  to be correlated~\cite{Ullom_2015,Smith13}.
An alternative approach was needed.
From \mbox{\Eqref{eq:dE}}, it is known that, in~the small signal limit, for~a given  detector $T_{c}$ and heat capacity $C$, to~achieve the best energy resolution one needs to minimize the factor $\sqrt{\zeta(I)}/\alpha$, which depends on the TES geometry and where $\zeta(I)$ generally increases with $\beta$.  
 In a study on the performance of Mo(50nm)/Au(200nm) TESs with different stripes configurations~\cite{Miniussi2018}, it was shown that pixels without stripes could achieve excellent resolution even with a large non-equilibrium Johnson noise in excess. An~excellent energy resolution of $1.58\pm 0.12 \mathrm{eV}$ at $5.9\, \mathrm{keV}$ was reported with a small, $50\times 50\, \upmu\mathrm{m}^2$, bare TES with $T_c=95.6\, \mathrm{mK}$, coupled to a Bi($4.04\, \upmu\mathrm{m}$)/Au($1.49\, \upmu\mathrm{m}$) $240\times 240\,\upmu\mathrm{m}^2$ absorber with heat capacity $C=0.69\, \mathrm{pJ/K}$.  The~stripe-less small device had much larger values of  $\alpha(\sim$1600), $\beta(\sim$86) and $M^2(\sim$6) with respect to the larger size TESs with stripes ($\alpha \sim75$, $\beta\sim1.25$ and $M^2\sim0$) \cite{Yoon2017}, which showed  energy resolution around $2\,\mathrm{eV}$ and were under consideration for~X-IFU.
  
This result was a real breakthrough in the field, because~it suggested that the parameter space for the optimization of TES microcalorimeters for ultra sensitive X-ray spectroscopy, could be much larger than what it  was thought before.  The~major advantages of removing the normal metal stripes from the top of the bilayer and of employing small TESs  are the smaller sensitivity to the external and self-magnetic field, the~ improvement of the uniformity of the TES resistive transition itself and an increased  uniformity within the array. 
Excellent uniformity has been demonstrated with a kilo-pixels array of TES X-ray calorimeters under development for X-IFU (Section \ref{sec:athena}). The~detector design is based on the small,  $50\times 50\, \upmu\mathrm{m}^2$, Mo/Au bare TES described above.  A~combined  energy resolution of $2.16\pm 0.01 \, \mathrm{eV}$ at 5.9 keV, obtained with more than 200 pixels, in~a $8\times 32$ multiplexing configuration, has been reported~\cite{Smith2020}. 

Small pitch ($50\,\upmu\mathrm{m}$) devices based on a similar design, fabricated in an array of \mbox{$60\times 60$ sensors}, with~ $1\,\upmu$ thick, $46\times 46\, \upmu\mathrm{m}^2$, gold  absorbers have been developed at NASA-GSFC for the low energy band (0.2--0.75 keV) of the Ultra High Resolution (UHR) array for the future X-ray space mission Lynx Section \ref{sec:otherXray}. They  measured  2.31  eV  FWHM  at  1.49  keV, limited by  the broadening of the intrinsic linewidth of the Al source,  and~ 0.25  eV  FWHM  at  3 eV using optical photons, from~a  blue laser diode,  delivered  through  an  optical  fiber~\cite{Jaeckel2019}.  The~latter energy resolution result   is  consistent  with  the  estimated  performance  based  on  the  signal  size  and  noise~\cite{Sakai2020}. 

\subsection{ac~Bias}\label{sec:acbias}

TES  microcaloremeters to be read-out under ac bias for the MHz-FDM-based experiments require a different optimization process than the one described above.  The~uniformity of the transition  over a large array is less critical, due to the fact that in FDM each pixel can be biased individually to their optimal point. The~performance of MHz biased TES is deteriorated by two major frequency dependent effects: (i) the  ac losses from eddy currents generated in the normal metal structures and (ii) the ac Josephson effects describe in Section \ref{sec:TEStheory}.
The former induces thermal dissipations inside the TES and absorber~\cite{Gottardi17,Sakai2018}, which effectively decrease $\alpha$ and the detector sensitivity, in~particular at high bias frequency and for low resistance bias points  where the best signal-to-noise is generally achieved. 
The latter is responsible for a relatively large non-linear Josephson inductance in parallel with the TES resistance. At~high bias frequency, it reduces the range for optimal biasing at low TES resistance and generates step-like structures in the TES transition~\cite{Akamatsu2014,Gottardi17,Gottardi18}.

Through  an experimental and theoretical study, the~groups at SRON and  NASA-GSFC have identified  several  optimal TES designs for the ac-bias read-out. They  implemented changes both in the TES geometry and the fabrication process. 
The ac losses  have been minimized to a negligible level by: (a) reducing the Au metal features, like the stripes and the edge banks, in~the proximity of the TES and the leads,  (b) making the non-microstriped loop area formed by the TES and the bias leads smaller, and~(c) increasing the TES aspect ratio (AR = $L\times W$), where the width (W) is reduced relative to its length (L), to~achieve higher value of $R_N$. Moving towards higher and higher TES normal resistances, $R_N$ has also been the strategy to minimize the Josephson effects in ac bias devices, as~it follows from the predictions of  the theoretical models presented in Section \ref{sect_ETEq}. In~TESs with higher $R_N$, operating at the same power, the~gauge invariant phase difference across the Josephson weak link is maximized since $\varphi \propto \sqrt{P_{sat}R}/f_{bias}$ and the TES is less affected by the Josephson effects. Similarly to the dc bias pixel optimization,  a~further strategy was to simplify the geometry of the TES by removing additional metal features such as the Au stripes that are  responsible for a  complex resistive transition. Both these actions are compliant with those also implemented to reduce the ac~losses.

The dependence of the Josephson effect on TES power $P$, $R_N$ and bias frequency 
$f_{bias}$ is phenomenologically demonstrated in \figref{fig:IQvsPRf}.
\begin{figure}[H]
\center
\includegraphics[width=14.cm]{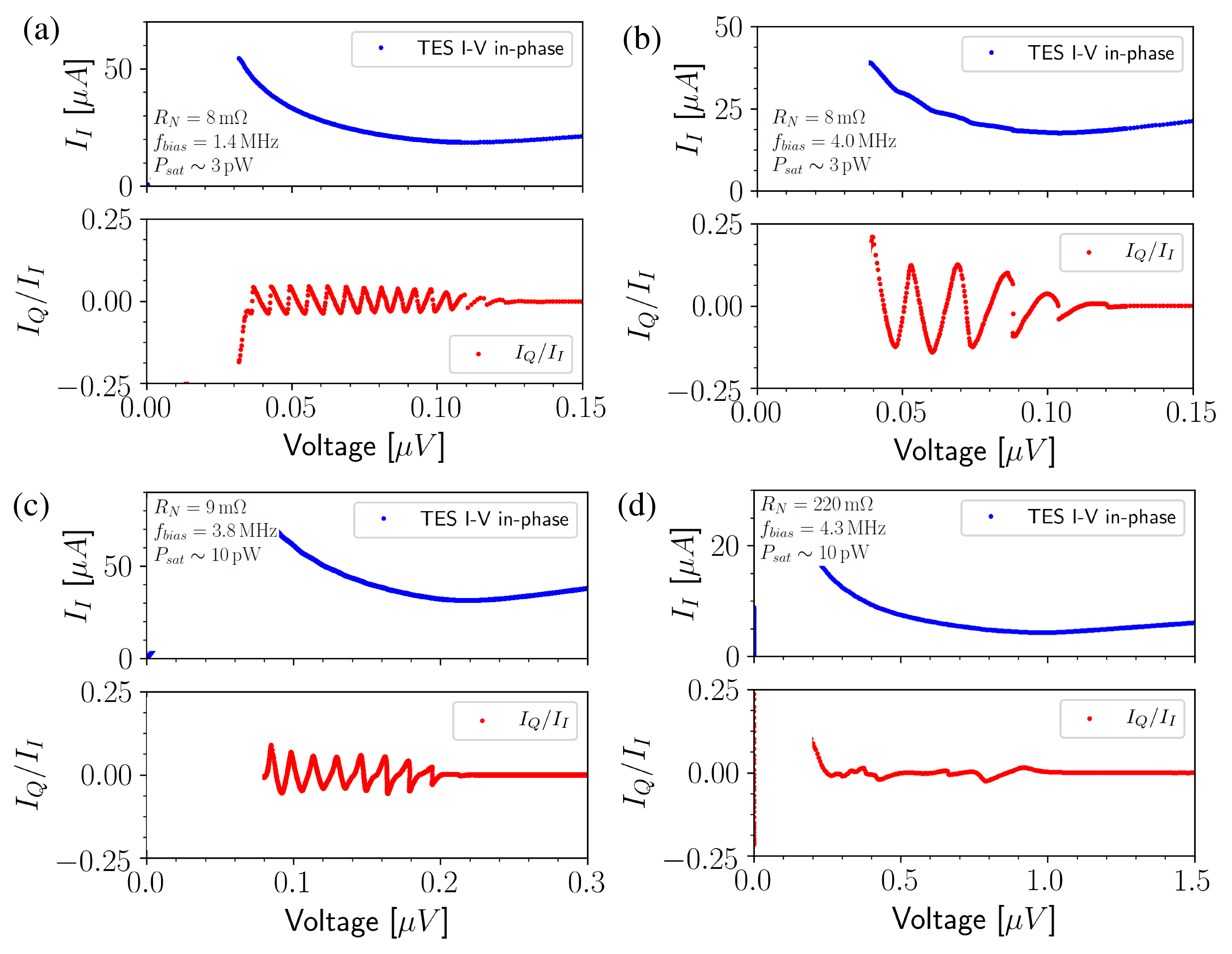}
\caption{\label{fig:IQvsPRf} TES I--V curves for different X-ray microcalorimeter types and MHz bias frequencies. In~the lower graph of each plot the ratio $I_Q/I_I$ of the quadrature or Josephson ($I_Q$) and in-phase $I_I$ current is shown. In~ (\textbf{a}--\textbf{c}), the bias frequency is, respectively, $f_{bias}=\, 1.4, 4.0\, \text{and}\, 3.8\, \mathrm{MHz}$. The~detector is an NASA-GSFC Mo/Au bare TES~\cite{Miniussi2018,Sakai2018}  with $R_N \simeq$ 8--9 $\mathrm{m}\Omega$ and $P_{sat}\sim3\,\mathrm{pW}$(\textbf{a},\textbf{b}) and $P_{sat}\sim10\,\mathrm{pW}$, respectively. In~(\textbf{d}), we consider a Ti/Au, $130\times 140\,\upmu\mathrm{m}$, SRON TES~\cite{Taralli2020} with  $R_N \simeq 220$ $\mathrm{m}\Omega$, $P_{sat}\sim10\,\mathrm{pW}$, and biased at frequency $f_{bias}=4.3\, \mathrm{MHz}$. Reprinted  from~\cite{Gottardi18} by permission from Springer Nature. Copyright \copyright 2018, Springer Science Business~Media.}
\end{figure} 
 The ratio of the out-of-phase Josephson current ($I_Q$) to the in-phase current $I_I$ is shown as a function of TES voltage for low (\figref{fig:IQvsPRf}a--c) and high (d) ohmic devices fabricated, respectively, at NASA-GSFC and SRON. 
  In the top-left and top-right plots, the~low power ($P_{sat}\sim3\, \mathrm{pW}$) and low resistance ($R_N\simeq 8\, \mathrm{m}\Omega$) detectors have been bias at low (1.4 MHz) and high frequency (4.0 MHz). At~high bias frequency, the ratio ($I_Q/I_I$) is increased of more than a factor of 2 at low voltage bias.
   In the two lower graphs, the~detectors operates at a high saturation power ($P_{sat}\sim10\, \mathrm{pW}$) and similar bias frequency ($f_{bias}\sim\, 4\, \mathrm{MHz}$). 
  The two TESs have, respectively, $R_N\simeq 8\, \mathrm{m}\Omega$ and    $R_N\simeq 220\, \mathrm{m}\Omega$. The~high ohmic, high power detectors show almost negligible Josephson effects even at high bias~frequency. 

There are two ways to fabricate devices with high $R_N$. One is to increase the bilayer sheet resistance $R_\square$, by~reducing the thickness of the normal metal film (Au). Despite the good results achieved in squared TESs even after increasing the $R_\square$ from 15 to \mbox{50 $\mathrm{m}\Omega/\square$  \cite{Wake2019}}, this approach is not scalable due to limitation in the fabrication process and, eventually, to~ the increase of the internal thermal fluctuation  noise. A~better way to achieve high $R_N$ is to increase the TES aspect ratio, defined as  AR = $L\times W$, with~$L$ and $W$ the TES length and width, respectively. Higher aspect ratio devices, while increasing $R_N$, offer additional design flexibility to simultaneously reduce the speed of the pixel (by reducing the perimeter) and the TES bias current  (at the same saturation power) measured by the SQUID amplifier. Both the effects are beneficial to increase the multiplexing factor, as~discussed in Section \ref{sec:MHzFDM}. 

After a few promising results obtained at NASA-GSFC, SRON developed high aspect ratio TES microcalorimeters  based  on Ti(35nm)/Au(200nm) bilayer with a $R_\square = 25\,\mathrm{m}\Omega/\square$. In~\figref{fig:IQvsR}, we give  an example of the mitigation of the Josephson effects achievable by using high aspect ratio (AR) TESs. From~left to right, we show the I--V curves and the $I_Q/I_I$ ratio of, respectively, an~NASA-GSFC Mo/Au $75\times 17.5\, \upmu\mathrm{m}^2$ (AR = 4:1, \mbox{$R_N=120\,\mathrm{m}\Omega $}), an~ SRON Ti/Au $80\times 20\, \upmu\mathrm{m}^2$ (AR = 4:1, \mbox{$R_N\simeq\, 100\,\mathrm{m}\Omega$})\cite{deWit_2020}, and an  SRON Ti/Au \mbox{$80\times 10\, \upmu\mathrm{m}^2$} (AR = 8:1, $R_N\simeq 200\,\mathrm{m}\Omega$). All the devices have a $P_{sat}\sim$ 1--2 $\mathrm{pW} $, compatible with the X-IFU requirement. The~NASA-GSFC device is biased at $3.8\,\mathrm{MHz}$, while for the SRON devices, I--V curves for pixels biased at different frequencies (from 1.5 to $4.4 \, \mathrm{MHz}$) are~shown. 

% start a new page without indent 4.6cm
%\clearpage
\end{paracol}
\nointerlineskip

\begin{figure}[H]

\includegraphics[width=15.5cm]{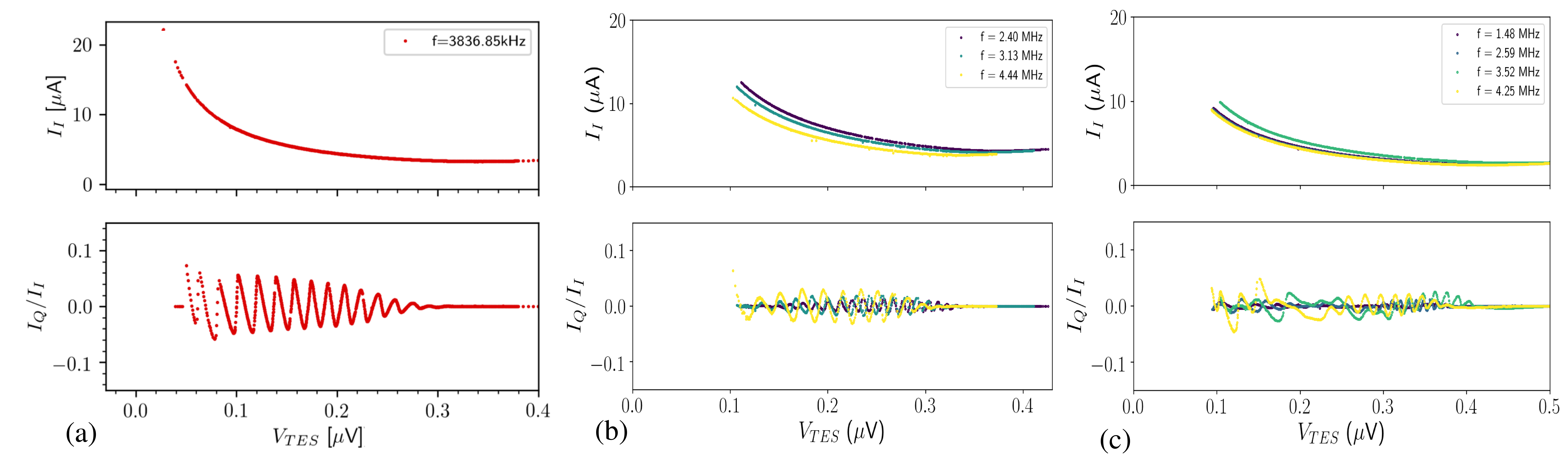}
\widefigure
\caption{\label{fig:IQvsR} TES I--V curves for different X-ray microcalorimeter types and MHz bias frequencies. In~the lower graph of each plot the ratio $I_Q/I_I$ of the quadrature or Josephson ($I_Q$) and in-phase $I_I$ current is shown. The~data are from the following devices: (\textbf{a}) NASA-GSFC Mo/Au $75\times 17.5\, \upmu\mathrm{m}^2$ (\mbox{AR = 4:1}, $R_N=120\,\mathrm{m}\Omega$), ac biased at 3.8 MHz, (\textbf{b})  SRON Ti/Au $80\times 20\, \upmu\mathrm{m}^2$ (\mbox{AR = 4:1}, $R_N\simeq\, 100\,\mathrm{m}\Omega$) \cite{deWit_2020}, ac biased at 2.4--4.4 MHz and (\textbf{c})  SRON Ti/Au $80\times 10\, \upmu\mathrm{m}^2$ (AR = 8:1, $R_N\simeq 200\,\mathrm{m}\Omega$, ac biased at 1.4--4 MHz) \cite{deWit_2020}.} 
\end{figure}
\begin{figure}[H]
\widefigure
\includegraphics[width=15.0cm]{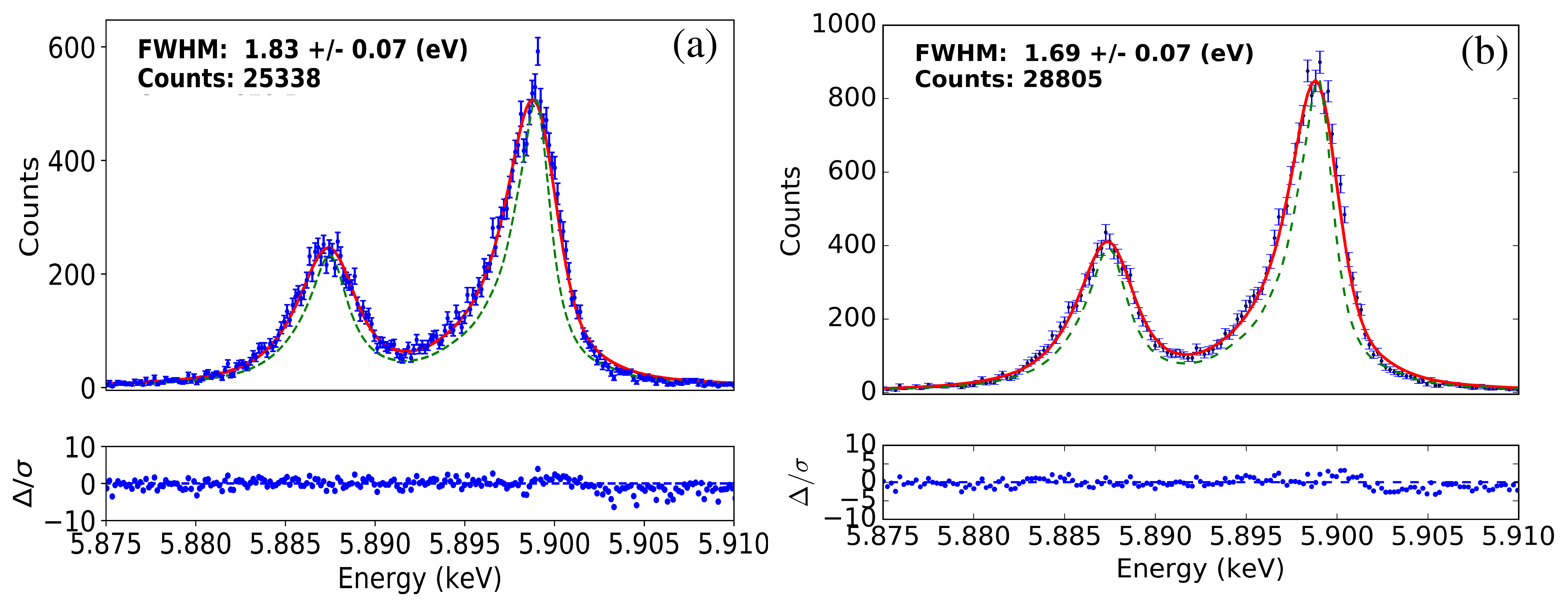}
\caption{\label{fig:Fe55HAR} X-ray spectra of the Fe$^{55}-K\alpha$ complex at 5.9 keV demonstrating excellent energy resolution with SRON high aspect ratio TESs: (\textbf{a}) $80\times 20\, \upmu\mathrm{m}^2$, $R_N=100\,\mathrm{m}\Omega$ and bias frequency 2.5 MHz, and~(\textbf{b}) $80\times 10\, \upmu\mathrm{m}^2$, $R_N=200\,\mathrm{m}\Omega$ and bias frequency 3.5 MHz. The~dashed green line is the natural lineshape of the~complex.}
\end{figure}

\begin{paracol}{2}
%\linenumbers
\switchcolumn

The SRON $80\times 10\, \upmu\mathrm{m}^2$ turns out to be very promising, with~a $I_Q/I_I < \, 0.025$, at~$f_{bias}\sim4\, \mathrm{MHz}$, smaller than the other devices biased at the same frequency. 
Although the weak-link effects are not completely eliminated, these new TES design configuration reduces the size of the oscillatory structures in the transition and allows the access to a wider range of higher signal-to-noise ratio bias points in the~transition.  

\textls[-15]{We have to experimentally investigate the  performance, under~ac bias, of~many high aspect ratio devices fabricated at SRON with high ($T_c\simeq 110\, \mathrm{mK}$) \cite{Taralli2019}, and~low (\mbox{$T_c\simeq 90\, \mathrm{mK}$) \cite{deWit_2020}} critical temperature. The~energy resolution was shown to scale with $T_c$ as expected.  Five different TES designs with low $T_c$ were then further studied. The~geometry of these devices is  rectangular with dimensions (length$\times$ width) $100\times 20\,$, $120\times 20$, $140\times 30$, $80\times 20$, and $80\times 40\, \upmu\mathrm{m}^2$, which correspond to aspect ratios ranging from 2:1 up to 6:1.  The~results on the characterization and uniformity of a kilo-pixels array of $140\times 30\, \upmu\mathrm{m}^2$ is reported in~\cite{Taralli2021,DAndrea2021}.
More details on the TES design and  fabrication are given in~\cite{Nagayoshi19} and in \mbox{Section \ref{sec:LAFab}}.   
The TES bilayer had a $T_c\sim90\,\mathrm{mK}$ and a squared normal resistance of  $26\, \mathrm{m} \Omega/\square$.
The scaling of the devices $T_c$ and the thermal conductance $G_{bath}$ with the TES geometry has been studied carefully.  It is shown that the $T_c$ is decreasing with increasing TES length and decreasing width, in~agreement with the LaiPE and LoPE models described in~\cite{Sadleir11}. 
The thermal conductance, evaluated at the critical temperature, was shown to scale with the perimeter of the device, as expected when the dominant process for the
thermal conductance is two-dimensional radiative transport in the
silicon nitride membrane~\cite{Hoevers2005,Hays2016}. The~thermal characterization of the devices is essential for the development of large uniform arrays tailored for the multiplexing read-out and the instrument requirements.  
The devices showed  excellent energy resolution at $5.9\, \mathrm{keV}$,
with a mean of $2.03\pm0.17\, \mathrm{eV}$ and median of 2.07 eV   over all the tested geometries and bias frequencies from 1 to 4 MHz~\cite{deWit_2020}. The~best results were obtained with the  $120\times 20\, \upmu\mathrm{m}^2$, with~ AR = 6:1, with~energy resolution consistently below 2 eV.
 These results show that, even with  high aspect ratio and very high normal resistance devices, there appears to be no degradation in the performance and in the signal-to-noise ratio. This seems to indicate that the internal thermal fluctuation noise, for~example,  do not simply scale with the TES normal~resistance.}

The good results have motivated the team to explore the performance of extreme aspect ration TESs and more exotic geometries, given the assumption that the Josephson effects should be minimized in high normal resistance devices. Devices based on TiAu TESs have been fabricated at SRON with aspect ratio as high as AR = 15:1, including TES with small width $W=10\,\upmu\mathrm{m}$ and meandering geometries. The~characterization of these new devices is currently on going, but~the first results are very promising. In~\figref{fig:Fe55HAR}, we show the energy spectrum at 5.9 keV of  $80\times 20\, \upmu\mathrm{m}^2$ and  $80\times 10\, \upmu\mathrm{m}^2$ Ti(33)Au(210) TESs with, respectively,  $G_{bath}$ $\sim$ 65 and $50\, \mathrm{pW/K}$, optimized for the MHz frequency-division multiplexing  readout of X-IFU-like instruments. The~$T_c$ of these TESs is $\sim$86.5$\,\mathrm{mK}$, and~they are both coupled to a $240\times 240 \,\upmu\mathrm{m}$ Au ($2.3\,\upmu\mathrm{m}$) absorber with a total heat capacity of $C\sim0.75 \, \mathrm{pJ/K}$ at $T_c$.
Excellent energy resolution at 5.9 keV has been achieved. An~example is given with the two spectra in \figref{fig:Fe55HAR}  taken at about $R/R_N\sim0.1$. An~energy resolution of $dE=1.83\pm 0.07 \, \mathrm{eV}$ and $dE=1.69\pm 0.07 \, \mathrm{eV}$, has been achieved, respectively,  with~the $80\times 20\, \upmu\mathrm{m}^2$
TES biased at $2.5 \, \mathrm{MHz}$ and the $80\times 10\, \upmu\mathrm{m}^2$ TES biased at $3.5 \, \mathrm{MHz}$. The~spectra are the result of many hours of acquisition at a count-rate of about 0.8--1 pulse/s. %abbreviation for second is s. we changed
 They have been taken with the set-up described in~\cite{deWit_2020} hosted in a shared facility where two other experiments were running simultaneously. 
Multiplexing demonstrations for X-IFU-like X-ray  instruments are currently being done  with SRON $8\times 8$ and $32\times 32$ uniform array of the high aspect ratio devices discussed above. As~it will be shown in Section \ref{sec:MHzFDM}, the~first results are very encouraging. It is likely that a further improvement in the energy resolution and multiplexing performance will be shown  in the near future with the new generation of~TESs.
% start a new page without indent 4.6cm
\clearpage
\end{paracol}
\nointerlineskip

\begin{paracol}{2}
%\linenumbers
\switchcolumn

%%%%%%%%%%%%%%%%%%%%%%%%%%%%%%%%%%%%%%%%%%
\section{Multiplexing~Readout}
\label{sec:mux}
Large arrays with thousands of pixels are nowadays routinely fabricated and the demand from future experiments is to scale to higher number of pixels, towards the ambitious target of 100 kilo-pixels or more. A~further improvement of the existing cryogenic  multiplexing read-out  technology is a must, to~reduce the dissipation from the SQUID amplifiers, the~readout, and bias wiring, and~ to minimize the size of the interconnections and the harness complexity. 
It is not our purpose to discuss the details of the existing cryogenic multiplexing systems, which have been well reviewed in~\cite{Ullom_2015}. In~this section, we will give a short overview of the state-of-the-art of the different multiplexing readouts.  
SQUID-based multiplexing techniques for TES bolometers and micro-calorimeters have been  developed and operated at MHz frequencies using only few of the many orthogonal basis sets conceivable~\cite{vdKuur2016}, such as  time-division multiplexing (TDM) \cite{Doriese2016,Doriese2019}, MHz frequency-division multiplexing (MHz-FDM)\cite{vdKuur2016,Vaccaro2021,Akamatsu2020}, and~Walsh code-division multiplexing (CDM) \cite{Morgan2016,Durkin2019,Doriese2019}.
The state-of-the-art of each of these techniques is currently limited by the practical implementation.  The~resources required by  multiplexed signals  will be independent of the applied basis set, when optimally dimensioned and with full signal loading.  The~choice of which  multiplexing readout to use should be then based on the specific system requirements, such as, for~example, cooling power, mass, volume, harness length, and electromagnetic interferences~compliance.  

{Microwave frequencies multiplexing schemes (GHz-FDM), using microwave SQUID multiplexing ($\upmu$w-mux), is  becoming a mature technology   both for transition-edge \linebreak \mbox{sensors~\cite{Mates2017,Bennett2019}} and magnetic microcalorimeters~\cite{Kempf2017}, in~particular for ground-\linebreak base~applications.}

The number of pixels presently multiplexed in each multiplexing  scheme is far below the fundamental limit set by the the information theory~\cite{Irwin2009}. This is partially due to the fact that the cryogenic environment  and the lithographic fabrication 
set stringent constrains that make it difficult to optimally implement any  multiplexing architectures. Much R\&D is  still needed  towards the optimization of the practical implementation of the multiplexing schemes in  the focal-plane design to accommodate the complex cryogenic electronics. 
It is likely that for multiplexing a 100,000-pixels array, a hybrid multiplexing configuration will be required~\cite{Yu_2020}.

\subsection{Time-Division and Code-Division~Multiplexing}

SQUID-based time-division multiplexing (TDM) is, at~the moment,  the~most mature technology to read-out arrays of TES microcalorimeters. In~TDM, the~TESs are dc-biased. Each TES is coupled to its own first-stage SQUID. $N$ rows of first-stage SQUIDs are switched  on sequentially and $M$ columns of SQUIDs are read out in parallel. One single bias line is used for all the $N$ TESs in one column and each TES cannot be individually biased at its optimal bias point. For~this reason, a~highly uniform TES array is~required.  

The latest generation of  TDM architecture  developed at NIST has an overall system bandwidth of $ 6\, \mathrm{MHz}$ bandwidth, which allows to switch the rows faster with transient of $128\,\mathrm{ns}$ and total row time of $160\,\mathrm{ns}$. A~three-stage SQUID configuration is used with a input flux noise of $0.19\, \mu\Phi_0/\sqrt{Hz}$. More details of the state-of-the-art of the TDM architecture, including the description of a practical X-ray spectrometer for beamline science,  can be found in~\cite{Doriese2016,Doriese2017}. 
Recent 40-row TDM experiments  with X-IFU-like TESs make use of  new shunt resistors and the optimized first-stage SQUIDs recently developed at NIST~\cite{Doriese2019}. The~achieved 1x40-row (32 TES rows + 8 repeats of the last row) energy resolution with the 32 TESs was $1.91\pm 0.01\, \mathrm{eV}$ at 1.5 keV, $2.23\pm 0.02\, \mathrm{eV}$ at 5.9 keV, $2.40\pm 0.02\, \mathrm{eV}$ at 6.9 keV, and~$3.44\pm 0.04\, \mathrm{eV}$ at 11.9 keV~\cite{Durkin2019}.
A NIST 8-columns$\times$32-row TDM system with a prototype of a X-IFU kilo-pixel array from NASA-GSFC has shown an exquisite average energy resolution of $2.16\pm 0.006\, \mathrm{eV}$ at 6.9 keV~\cite{Smith2021}. 
The instrument will be soon delivered to LNLN EBIT for astrophysics experiments.  
These results are compliant with the stringent Athena X-IFU~requirements. 

The main disadvantage of TDM is the increase of SQUID noise with the $\sqrt{N}$ due to aliasing, into~the signal bandwidth, of~the Nyquist noise above the rate at which all the $N$ pixels in a column are measured and below the measurement bandwidth. This sets a fundamental limit on the maximum number of pixels measurable with high resolution in multiplexing mode. 
Code Division Multiplexing (CDM) \cite{Stiehl2012} provides a way to  reduce at a  negligible level the SQUID noise aliasing effect and has the big advantage of being compatible with the existing TDM read-out infrastructure. In~ CDM, rather than measuring one TES at a time, all the dc-biased TESs in a column are read out simultaneously at each time step. The~signals from the TESs are encoded with  a Walsh basis and  summed with equal weight, but~different polarity patterns. The~design and  fabrication of the SQUID multiplexing chip is rather complex, since the encoding is achieved by a flux-summing architecture of $N$ different microcalorimeters in $N$ different SQUIDs with different combinations of coupling polarity. CDM is also susceptible  to single-point failure mechanisms  which can result in an unconstrained demodulation matrix. However, potential solutions do exist~\cite{Weber2019}.  Excellent performance has been reported  so far with a 32-channel flux-summed CDM~\cite{Morgan2016} with a mean energy resolution of $2.77\pm 0.02\, \mathrm{eV}$ over 30 working sensors. The~multiplexed noise level and signal slew rate achieved were sufficient to allow the readout of more than 40 pixels per column. Even a larger multiplexing number could be achieved with the recent technological improvement in the fabrication of the shunt resistors chip and the first-stage SQUIDs~\cite{Doriese2019}.

\subsection{MHz-Frequency-Division~Multiplexing}\label{sec:MHzFDM}

The MHz Frequency-division multiplexing (MHz-FDM) scheme is fundamentally different from the other multiplexing architectures from the fact that the TES  itself, and~not the SQUID,  is the modulating element in the multiplexing process. The~detectors are
voltage biased with sinusoidal carriers at different frequencies in the 1 to 5 MHz range and  they operate as amplitude modulators~\cite{JvdK02}. MHz-FDM is under development for the read-out of large array of X-ray calorimeters and infra-red bolometers for future space mission  and used on ground-based CMB experiments~\cite{Hattori2016,Bender2020}.
In MHz-FDM, each TES is coupled to a superconducting high-$Q$ $LC$ bandpass filter~\cite{Bruijn18,Bruijn14,Bruijn12, Rotermund2016}, which has the double function of limiting the information bandwidth and providing a single carrier, stiff ac voltage bias to the TES. The~signals from many TESs are then summed at the input of one  low-temperature first-stage single SQUID current amplifier, amplified further by a second-stage SQUID array, operated at higher temperature, and~finally  demodulated at room temperature by digital electronics. The~SQUID signal is linearized by using a very robust  baseband feedback (BBFB) scheme~\cite{Kiviranta02,Hartog12,Hartog2009} or other digital active nulling  configuration~\cite{deHaan2012}.  
The major advantage of the MHz-FDM architecture is the relative simplicity of the cold electronics, which only requires a limited amount of SQUID amplifiers, and~its robustness to electromagnetic interferences. The~fabrication of high-$Q$ lithographic $LC$ filters has been, for~many years, a~major obstacle to the practical implementation of the MHz-FDM. With the~state-of-the-art of the $LC$-filters fabrication, a~very high yield can be achieved, as~well as  a high accuracy of the frequency definition and a compact design~\cite{Bruijn18}. 
FDM makes as well a better use of the available bandwidth with respect to the other traditional scheme. 

\noindent Moreover, in~an instrument like X-IFU for example, tens of pixels can be readout with a single SQUID amplifier, without~the need of more dynamic range than for the readout of a single pixel~\cite{vdKuur2017}.
 The details of the MHz-FDM system for a X-IFU-like instrument, using for each channel a two-stage SQUID amplifier  developed at VTT~\cite{Kiviranta2011}, can be found in~\cite{vdKuur2016}. The~first-stage SQUID has an input inductance smaller than $3\,\mathrm{nH}$, a~bandwidth $>8\, \mathrm{MHz}$ and a input current noise $<3\,\mathrm{pA}/\sqrt{\mathrm{Hz}}$, for~a total power dissipation smaller than $2\, \mathrm{nW}$.

One of the major issues related to the MHz-FDM readout is the behaviour of a TES under MHz bias, which is strongly influenced by the non-linear Josephson effects describe in Section \ref{sec:TEStheory}. Only recently, the~physical processes involved in an ac bias TES have been  better understood. The~current development undergoing at SRON (see Section \ref{sec:acbias})  is leading to an optimal TES design that will reduce these non-ideal effects to a negligible~level. 

 Moreover, non-linearities in the readout chain, due to, for~example, the~SQUID amplifiers or DACs, can generate intermodulation lines  within the detector bandwidth degrading its performance. To~minimize the effect of  these non-linearities and  to make efficient use of the available readout bandwidth, the~LC resonators are usually designed to be on a regular grid. Despite the high accuracy of the lithographic processes, however, the~observed resonant frequencies have a spread of about  few kHz around the designed values. The~important  recent breakthrough that solves this issue is the experimental demonstration at SRON of the frequency shift algorithm (FSA), a~very elegant approach to electronically tune the resonant frequency of the  cold resonators by means of a feedback on the modulated bias signal~\cite{vdKuur2018}.  The~performance of a new implementation of the FSA, scalable for multi-pixel operation, has been recently shown by Vaccaro~et~al.~\cite{Vaccaro2021}. An~energy resolution of $2.91\pm0.03\, \mathrm{eV}$ at 5.9 keV has been achieved in a summed energy spectrum with  20 pixels, from~a 5x5 array of Ti/Au TESs microcalorimeters with an intrinsic single pixel energy resolution of about 2.7 eV. 
A new experiment has been performed with a larger $LC$ filters array and using an uniform $8\times 8$ array of  $80\times 20\,\upmu\mathrm{m}^2$ X-IFU-like TESs with a measured single pixel energy resolution of $\sim$1.8 $\mathrm{eV}$ (see \figref{fig:Fe55HAR}). An~excellent energy resolution of $2.35\pm 0.03\, \mathrm{eV}$ at 5.9 keV has been demonstrated with 32 pixels multiplexed~\cite{AkaSPIE2020}. 
{At the time of writing this review, a~$2.22\pm 0.02  \, \mathrm{eV}$ at 5.9 keV has been demonstrated with 37 high aspect ratio pixels, with~geometry $80\times 13\,\upmu\mathrm{m}^2$, in~a k-pixels array multiplexed at frequencies from 1 to 5 MHz~\cite{Akamatsu2021}. The~results fulfill the X-IFU requirements.}  
{  
  \noindent This is the first convincing demonstration of MHz-frequency division multiplexing  with TES microcalorimeters. Improvements are still expected in the coming years, after~a fine tuning of the read-out circuit and when employing further optimized detectors currently under fabrication. 
These recent results show no fundamental limitation yet for this multiplexing scheme. The~MHz-FDM is reaching a competitive level of maturity, and~has the potential to achieve  even higher multiplexing capability per SQUID channel.
As a matter of fact, the~current implementation of the current MHz-FDM is far from optimal and increasing the the bandwidth to 6 MHz appears to be feasible. Many new ideas have still to be implemented to efficiently use, for~example, the~limited resources available for space-borne instrumentations~\cite{vdKuur2017,Kiviranta2020classb,kiviranta2018}.  
}

\subsection{GHz-Frequency-Division~Multiplexing}

The multiplexing architectures discussed so far do have a limited ultimate bandwidth around 10 MHz. 
SQUID-based microwave read-out is a promising technology under development  to expand the read-out bandwidth to the GHz range. The~large bandwidth available can be used to multiplex a larger number of TESs or to allow more read-out bandwidth per pixel.  
A microwave SQUID multiplexer ($\upmu$w-mux) performs frequency division multiplexing at GHz (GHz-FDM). Differently from the MHz-FDM, the~TESs are dc-biased and rf-SQUIDs are used  to modulate distinct superconducting microwave resonators coupled to a common microwave feed-line~\cite{IrwLeh2004,Mates2008,Mates2017}. The~signal of thousand of pixels could finally be amplified at 4K by standard  HEMT amplifiers in the, 4- to 8-GHz band, via a single coax cable.  Despite the complexity at many levels of the signal chain, as~for example  the design and fabrication of the cold rf-SQUID and resonators chips~\cite{Mates2019}, the~linearization of the SQUID response~\cite{Mates2017,Richter2021}, the~room temperature electronics, and the packaging at the focal plane assembly stage, the~GHz-FDM will become very competitive in the future. This is particularly true  for ground-based experiments and in applications where a large array of high speed detectors are required~\cite{Kempf2017,Alpert2019}.
The SLEDGEHAMMER instrument with gamma-ray detectors,   has demonstrated the read-out of 128 microwave SQUID multiplexed channels on a single set of coaxial cables~\cite{Mates2017}. With~a similar set-up,  a~82 pixels multiplexing factor has been demonstrated on an array of soft X-ray TESs with an energy resolution of $2.04\pm 0.003\, \mathrm{eV}$ for  the magnesium K$\alpha$ line at 1.25 keV, consistent with the resolution achieved using TDM with the same array~\cite{Bennett2019}. 
A simultaneous readout of 38 high-ohmic Ti/Au SRON TES microcalorimeters has been recently demonstrated with a  low  noise $\upmu$w-mux SQUID amplifier fabricated at AIST in Japan.  With~a readout noise of $0.9\, \mu\Phi_0/\sqrt{\mathrm{Hz}}$,   a~median energy resolution of 3.3 eV at $5.9\,\mathrm{keV}$ has been achieved~\cite{Naka2020} with high normal resistance  and relatively fast detectors. 
The multiplexing factor achievable with microwave SQUIDs is limited by the bandwidth available in the HEMT amplifiers (currently only working between 4--8 GHz) and the high slew rate on the leading edge of X-ray pulses. For~the latter, workarounds are envisioned~\cite{Irwin2018}.

\subsection{Thermal~Multiplexing}
\label{sec:hydra}
The maturity level of the electrical multiplexing architecture discussed above is such that in the next decade they will become a standard technology in the read out of array of  few thousand pixels. Scaling up this technology for the multiplexing of arrays of 100,000 pixels or more will become prohibitive, if~only to electrically connect all the pixels on the wafer.  Multiabsorber TESs~\cite{Smith2019}, also commonly referred to  as `hydras', for~the peculiar shape of these devices, are an alternative approach  to increase the multiplexed number of pixels without a dramatic increase in the number of readout components. The~principle behind the hydra is to couple to a single TES a series of many absorbers, such that each of them acts as a separate imaging element or pixel. The~thermal link between each of the absorbers to the TES is engineered such that it has a slightly different thermal conductance. In~this way, the~TES measures a different temperature excursion for photon hits in the different absorber. The~information of which pixel received the photon can be retrieved from the pulse shape recorded by the electronics. The~Hydras concept and implementation  has been pioneered by the NASA-GSFC team and the most up-to-date theoretical and experimental results have been reported by Smith~et~al.~\cite{Smith2019,Smith2020}.
An intrinsic trade-off between energy resolution and position sensitivity does exist for hydras detectors. It depends upon the ratio
of the internal thermal conductances between absorbers to the external thermal conductance to the bath. In~addition,  the~maximum number of thermally multiplexed absorbers is limited by the ability of the pulse processing algorithm of discriminating the different pulse shapes, which   is  related to the electrical bandwidth available to readout the coupled TES. With~no doubt, the~hydras will become an essential  element of the future hybrid multiplexing architectures required to read out very large arrays of TES~microcalorimeters. 

\section{Future Instruments for Astrophysics and Fundamental Physics~Research}
\label{sec:future}
The current performance of X-ray TES-microcalorimeters and their multiplexing read-out  have reached a level of technological maturity to be employed as baseline detectors at, for~example, synchrotron and free electron laser facilities as very sensitive, science driven, X-ray spectrometers. Those instrument has been reviewed in~\cite{Ullom_2015} and will not be discussed~here.

In this section, we will focus on the development of TES-based instruments for astrophysics space telescopes and fundamental physics experiments. The~state-of-the-art detectors and the ultimate sensitivity is generally required for most of these future ground-breaking scientific~experiments. 

\subsection{X-IFU~Athena}
\label{sec:athena}
One of the most complex TES-based instruments currently under development is the X-ray Integral Field Units (X-IFU) of Athena~\cite{XIFU2018}, the~L-class  ESA’s  X-ray observatory mission, selected in June 2014 to address the Hot and Energetic Universe science theme~\cite{Barret20} and scheduled to be launched  on 2032 at L2 orbit. The~Hot Universe refers to the 
baryons at temperatures above $1056\,\mathrm{K}$, which are believed to account for  half of the total baryonic content of the Universe.  The~best observational 
handle on the energetic universe is through X-ray observations of hot gas 
and accretion around black holes. Athena will be a large observatory offering an unprecedented combination of sensitive X-ray imaging, timing, and high-resolution spectroscopy.
X-IFU~\cite{Barret20} is a cryogenic imaging X-ray spectrometer, with~spatially-resolved high spectra resolution over a $5'$ field-of-view (FoV) that will provide: (i) 3D integral filed spectroscopic mapping of hot cosmic plasma, (ii) weak spectroscopic line detection, (iii) physical characterization of the hot and energetic universe.
X-IFU will host an array of more than 3000 TES pixels with a $T_c\simeq 90\, \mathrm{mK}$, sensitive in the energy range of 0.2--12 $\mathrm{keV}$, with~an unprecedented instrument energy resolution of 2.5~eV at 7 keV. To~achieve the X-IFU requirement, the~intrinsic single pixel energy resolution has to be better than 2 eV, the~quantum efficiency of the absorber $QE>96.7,80 \, \mathrm{and}\, 63\%$ at $1,7,9.5\, \mathrm{keV}$, respectively, and~ pixel filling factor  larger than~0.97. 

 The  prototype kilo pixel array of Mo/Au TESs, developed at NASA-GSFC and read out with the 8-column by 32-row TDM from NIST, has recently shown excellent results that meet the X-IFU requirement with margin.
The detector consists of an  array of $50\times 50\, \upmu\mathrm{m}^2$, low resistance TESs with a single pixel resolution of $<$2 $\mathrm{eV}$ at $6\, \mathrm{keV}$. The~system has demonstrated a  combined exquisite energy resolution, with~more than 200 pixels, of: 1.95~eV for Ti-K$_\alpha$ (4.5 keV), 1.97 eV for Mn-K$\alpha$ (5.9 keV), 2.16 eV for Co-K$_\alpha$ (6.9 keV), 2.33~eV for Cu-K$_\alpha$ (8 keV), 3.26 eV for Br-K$_\alpha$ (11.9 keV) \cite{Smith2020}. The~focus of the team will soon move on demonstrating the performance of a full scale $>\, 3000$ pixels array.
 Thanks to its advanced readiness level, this technology is currently baselined for X-IFU for the mission adoption of Athena expected in~2022.

The X-IFU Focal Plane Assembly (FPA) \cite{Jackson2016,Geoffray2020,vanWeers2020} and the detection chain shall face several implementation challenges related to the interaction between the different components of the read-out (i.e., the main sensor array, the~old electronics stages and the warm electronics) and the thermal, mechanical, and electromagnetic environment. Moreover, the~limited resources available on the  space telescopes  lead to strong constrain on the total allowed mass, volume, electrical power, and cooling power at the  $50\, \mathrm{mK}$ and $2\,\mathrm{K}$  stages. 

 The high-frequency EM-fields generated in the spacecraft call for a careful choice of the read-out technology that needs to be the less sensitive to electromagnetic interferences and capable of driving very long cable harness.    The~high susceptibility of the TES detectors  and the focal plane assembly to DC and low frequency magnetic fields requires a proper magnetic shielding, by~means of $\mu$-metal and Nb shields, to~reduce the DC and the $50\, \mathrm{Hz}$ B-field to level $<1\, \upmu\mathrm{T}$ with a stability $\sim$10 $\mathrm{pT}_{rms}$. Faraday cages needs to be implemented at the FPA, warm electronics, and cryostat level  and the harness needs to be heavily filtered given the high ($\sim$0.4 $\mathrm{nV}/\sqrt{\mathrm{Hz}}$) susceptibility in the DC-MHz range~\cite{Geoffray2020}. In~\figref{fig:XIFUFPA}a, we show  a rendered view of the FPA Demonstration Model  of X-IFU, currently in the commissioning phase at SRON. On~the right side of the figure, we show the pictures of a full scale  X-IFU prototype array from NASA-GSFC with more than 3000 pixels on a pitch of $260\, \upmu\mathrm{m}$. 
The focal plane assembly should also be designed to minimize the infra-red and optical loading on the TES. This should be achieved  by thermal filters in the aperture~\cite{Barbera2018} and Au shielding in the proximity of the array. On~top of that,  the~energy resolution of the detectors can be affected by thermal bath fluctuations from the energy deposited  on the substrate by cosmic rays  from astrophysical sources~\cite{Peille2020,Lotti2018} and by micro-vibrations induced by the mechanical coolers~\cite{Gottardi2019,vanWeers2020}.  A~proper coating and thermalization of the TES array wafer can minimize the impact of the cosmic ray~\cite{Miniussi_cosmicray2020}, while a sophisticated Kevlar-based suspension system is in development  to provide robustness and  stiffness to survive the launch and sufficient thermal and mechanical decoupling from the different cryostat stages~\cite{vanWeers2020}. 
Finally, the~ X-IFU TES array sensitivity  will be  degraded  by  the  particles  background  which  is induced  by  primary  protons  of  both  solar  and  cosmic  rays  origin,  and~ secondary  electrons. 
To prevent that, a~Cryogenic AntiCoincidence (CryoAC) detector is placed in the proximity of  the  X-IFU  TES  array~\cite{Lotti2021}.  The~CryoAC  is  a  4-pixels   detector  made  of  wide  area silicon absorbers sensed by a network of about 120 IrAu TESs connected in parallel. It is located  at a distance  of <1 mm underneath the TES-array. A~set of low-Z material shields in between the TES array and the Niobium shield will minimize the effect of secondaries. In~this configuration, the~X-IFU particle background  will be reduced to a total level  $<$5$\cdot 10^{-3} \mathrm{cts}/\mathrm{cm}^{2}/\mathrm{s}/\mathrm{keV}$ in the 2--10 keV energy band. The~first simultaneous operation of a  CryoAC prototype developed at INAF, with~a NASA-GSFC TES kilo-pixel array read-out in single pixel mode by a MHz-FDM set-up developed at SRON has been recently reported by Macculi~et~al.~\cite{Macculi2020}.

% start a new page without indent 4.6cm
%\clearpage
\end{paracol}
\nointerlineskip

\begin{figure}[H]
\widefigure
%\center
\includegraphics[width=15cm]{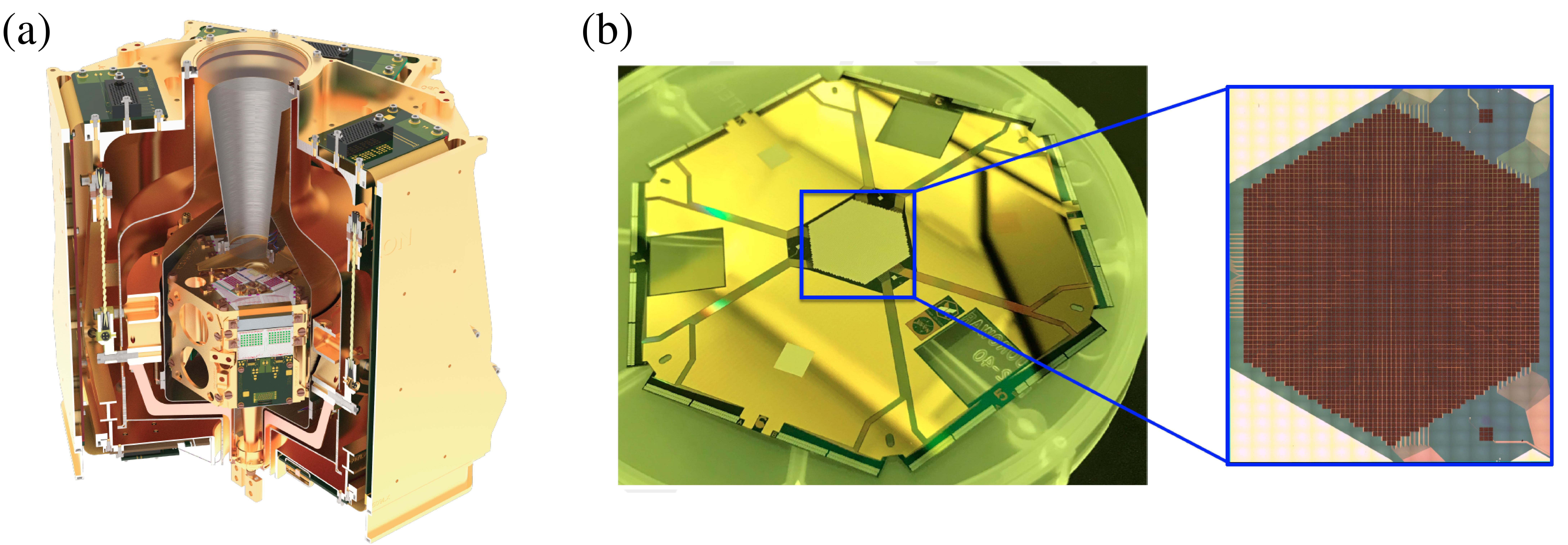}
\caption{\label{fig:XIFUFPA} (\textbf{a}) X-IFU Focal Plane Assembly Demonstration Model. (\textbf{b}) Photograph of a
X-IFU prototype array from NASA-GSFC  with more than  3000 pixels on a pitch of $260\, \upmu\mathrm{m}$. Courtesy of the X-IFU FPA team at SRON and NASA-GSFC.}
\end{figure}

\begin{paracol}{2}
%\linenumbers
\switchcolumn

\subsection{Other Future X-Ray Space~Missions}
\label{sec:otherXray}
After the European mission Athena, other X-ray observatories are currently being proposed in China, Japan, and~USA. 

The Hot Universe Baryon Surveyor (HUBS) \cite{HUBS2020}, proposed in China, is designed to  explore and characterize the circumgalactic medium with a large FoV.  
HUBS will be complementary to Athena. Thanks to the larger FoV ($1'$ 
angular resolution pixels over a $1\, \mathrm{deg}^2$ FoV, while X-IFU has $\sim$5$''$ pixels  over a FoV of $5'$ equivalent diameter), it is optimized to measure the spatial profile of extended warm-hot gas around galaxies and clusters of galaxies, and, potentially, the~filamentary structures in the cosmic web, as~well.  
The array of TES microcalorimeters is
designed to have $60\times 60$ pixels with an area of $1\, \mathrm{mm}^2$ for each pixel.
A $12\times 12$ central array with smaller pixels is foreseen. 
The large Au/Bi absorbers required for this mission  will be an interesting  fabrication and optimization challenge~\cite{Wang2020}.
The X-ray spectrometer  will cover an energy range from 0.2--2 $\mathrm{keV}$ with expected energy resolution of $2\, \mathrm{eV}$ and $0.6\, \mathrm{eV}$ for the outer and inner array, respectively.
The multiplexing read-out will be chosen between TDM, MHz-FDM and GHz-FDM depending on the overall system~requirement. 

Orders of magnitude larger pixels arrays are envisaged for the future proposed large X-ray mission beyond Athena such as Lynx, Super-DIOS, and the Cosmic Web Explorer that will be briefly discussed here below.
The Lynx X-ray Observatory~\cite{Gaskin19}, is  one of four large satellite mission concepts currently being studied by NASA. Lynx will provide a significant increases in sensitivity, FoV with subarcsecond imaging, and~spectral resolution over Chandra and Athena. The~three main science objectives of Lynx are: (i) the observation of the dawn of black holes, (ii) he understanding of  the driving of the galaxy formation and evolutions, (iii) the study of the energetic side of the stellar evolution and the stellar ecosystem.
One of the instruments on board of Lynx is an imaging spectrometer composed of a very large array of TES microcalorimeters~\cite{Bandler2019}. To~meet all the Lynx requirements, the~focal plane array is expected to be divided in different sub-arrays, which will make it unique. Its main array will consist of 86,400, $50\, \upmu \mathrm{m}$ over an area of $1.5\times 1.5\, \mathrm{cm}^2$, with~an energy resolution $dE= 3 \, \mathrm{eV}$ at 0.2--7 $\mathrm{keV}$. At~the innermost central 1 arc-min region, an~array of 12,800, smaller, $25\, \upmu \mathrm{m}$ pixels will provide $dE=2\,\mathrm{eV}$ up to $7\,\mathrm{kev}$. Finally,  there will also be an ultra-high resolution  subarray, optimized to achieve high energy resolution of $0.3\, \mathrm{eV}$ under $1\, \mathrm{keV}$. 
To read-out such a large array of 100,000 pixels will set a challenge not only for the multiplexing, but~for the layout of the wiring on the detector and bias circuit chips as well. NASA-GSFC has developed  a process to combine planarized multi-stack buried wiring layers from Massachusetts Institute of Technology Lincoln Laboratory (MIT/LL)
\cite{Devasia2019,Huang2020} with their TES array design~\cite{Smith2020}.
It will not remain a dream any more to realize a 100k pixels instrument for a future space mission, by~combining  the heritage from X-IFU (Athena), the~TES hydras concept discussed in Section \ref{sec:dcbias}, with~ the state-of-the-art of GHz-FDM~\cite{Bennett2019}, or~a hybrid configuration including the traditional \mbox{multiplexing~read-out}.

In Japan, ISAS/JAXA is considering the Super  Diffuse Intergalactic Oxygen Surveyor (DIOS) space mission~\cite{SuperDIOS2020} for a launch after 2030. The~science case is similar to  HUBS, but~the satellite will perform wide field X-ray spectroscopy  with a comparable  FoV of about 0.5--1 deg, but~with a better angular resolution of about 10--15 arcseconds. To~fulfill these requirement, the~focal plane array will have $\sim$30,000 TES microcalorimeters with expected performance similar to the one under development within the X-IFU program. The~current baseline for the read-out  is GHz-FDM~\cite{Naka2020,Yuki_SPIE2020}

The  Cosmic Web Explorer (CWE) \cite{Simionescu2019} is being proposed in Europe for the future ESA Voyage2050 program to survey and physically characterize  the  vast  majority  of  the  very  faint  warm-hot  diffuse  baryons  in  the  local Universe. It will have an instrument with comparable resolving power $E/\Delta E = 3000$ at 1 keV as the ultra high resolution array  of Lynx, but~will have  a much larger collecting area and angular resolution.
It will eventually be built upon the Athena-XIFU and Lynx technology that is going to be developed in the coming 20~years.

 Table \ref{tab:Xraymissions} summarizes the major properties of the future X-ray missions discussed in this~section.
% start a new page without indent 4.6cm
%\clearpage
\end{paracol}
\nointerlineskip
\newpage

\begin{specialtable}[H]
\widetable
\caption{\label{tab:Xraymissions} Overview of the characteristic of the TES-based instruments planned for the future X-ray space observatories. Athena is currently in the ESA Phase-A study. The~other missions are still in the selection or proposal~phase. }
%\centering
%%% \tablesize{} %% You can specify the fontsize here, e.g.,~\tablesize{\footnotesize}. If commented out \small will be used.
%\begin{tabular}{ccccccc}
\setlength{\cellWidtha}{\columnwidth/7-2\tabcolsep+0.0in}
\setlength{\cellWidthb}{\columnwidth/7-2\tabcolsep+0.0in}
\setlength{\cellWidthc}{\columnwidth/7-2\tabcolsep+0.0in}
\setlength{\cellWidthd}{\columnwidth/7-2\tabcolsep+0.0in}
\setlength{\cellWidthe}{\columnwidth/7-2\tabcolsep+0.0in}
\setlength{\cellWidthf}{\columnwidth/7-2\tabcolsep+0.0in}
\setlength{\cellWidthg}{\columnwidth/7-2\tabcolsep+0.0in}
\scalebox{1}[1]{\begin{tabularx}{\columnwidth}{>{\PreserveBackslash\centering}m{\cellWidtha}>{\PreserveBackslash\centering}m{\cellWidthb}>{\PreserveBackslash\centering}m{\cellWidthc}>{\PreserveBackslash\centering}m{\cellWidthd}>{\PreserveBackslash\centering}m{\cellWidthe}>{\PreserveBackslash\centering}m{\cellWidthf}>{\PreserveBackslash\centering}m{\cellWidthg}}

\toprule
\textbf{Mission}	& \textbf{F.o.V.} &\textbf{Angular Resolution} &	\textbf{Number of Pixels} & \textbf{Energy} & \textbf{dE} & \textbf{Eff. Area}\\
\textbf{ (Instrument)}&\textbf{arc min}& \textbf{arc sec}&& \textbf{(eV)} & \textbf{(eV)} & \textbf{(\boldmath$\mathrm{m}^2$)} \textbf{@ 1 keV}\\
\midrule
Athena (XIFU)		& 5& 5& $\sim$3800			&0.2--12 keV &2--2.5 eV&1.4\\
HUBS		&$60$&$\sim$60& $\sim$3600			&0.2--2&0.8--2		& $\sim$0.05\\
Super DIOS		&$>$30&0--15& $\sim$30,000	&0.2--2&<2 eV	& $>$0.1\\
Lynx (LXM)		&1--5&0.5--1& $\sim$100,000			&0.2--7 &0.3--3&0.2--2
\\
CWE		&$60$&$5$& $\sim$1 M			&0.1--3 &0.3&10\\
\bottomrule
\end{tabularx}}
\end{specialtable}

\begin{paracol}{2}
%\linenumbers
\switchcolumn

\vspace{-6pt}

\subsection{Instruments for Particle-Physics And~Cosmology}
 \label{sec:PP}
 While astrophysics has been, historically, the~main driver to develop TESs as sensitive thermometers in  imaging spectrometers, more and more projects in other fields of particle  physics and cosmology~\cite{PDG2020} are already employing or becoming interested in arrays of TESs for their very sensitive instruments. It happens  thanks to the growing matureness of the TES technology, which is moving from the single pixel development phase  to the  design and construction phase  of complex instrumentations with  larger and larger arrays of pixels.
In this section,  we consider the development of TES-based photon and power detectors for particle-physics and cosmology projects, like the ground-base experiments for (i) the direct detection of the neutrino mass, (ii) the detection of solar axions, and~(iii) the search of signature of the primordial gravitational waves  with space~missions.  

(i)  The absolute values of the masses of the three neutrino flavors, discovered with neutrino oscillation experiments, are unknown. An~upper limit of $1.1\, \mathrm{eV}$ to the absolute mass scale has been set in Spring 2019  by the Karlsruhe Tritium Neutrino experiment KATRIN~\cite{KATRIN2019}.    A~precise measurement of the neutrino mass is one of the important ingredients toward the physics beyond the Standard Model. 
 One way to directly probe the neutrino mass scale in the laboratory is  by performing  calorimetric measurement of the energy released in  electron capture decay of  holmium($^{163}\mathrm{Ho}$). The~HOLMES and ECHO projects~\cite{CamNuc2016,Nucciotti2018,Gastaldo2017} employ, respectively, arrays of  low temperature TES X-ray calorimeters and metallic magnetic calorimeters to~detect the spectrum of $^{163}\mathrm{Ho}$ between  0 and $3\,\mathrm{keV}$, with~an energy resolution better than $3\, \mathrm{eV}$. What makes these experiments unique is the fact that the radioactive source is embedded into the absorber. This means that  the total energy of the decay process, except~for the fraction taken away from the neutrino, is entirely released into the detector. This approach minimize the systematic uncertainties from energy calibration and gain instability correction, and~is neutrino-model independent.  
The goal for these middle-scale experiments, employing an array of $\sim$1000 pixels, is to achieve a neutrino mass statistical sensitivity below $2\, \mathrm{eV}$ and develop the technology for a future megapixels detector~\cite{Nucciotti2018}.
 As for  X-IFU, and~the other low temperature instruments for the future space mission, to~meet the challenging requirement set for the neutrino mass experiment,  cutting edge technology is required both for the detector and the read-out. High energy resolution, $\Delta E_{FWHM}\sim1\, \mathrm{eV}$, and  high statistics (exceeding $10^{13}$ decays) are desired for a competitive experiment. One of the most critical limiting factor to the sensitivity comes from the intrinsic background from unresolved pile-up, as~a results of the fact that the  $^{163}\mathrm{Ho}$ decay generates a signal with finite time resolution. The~fraction of unresolved pile-up is  given approximately by the product of the detector time resolution $\tau_{res}$ with the source activity in the absorber. Ideally, a~fast detector with a rise time $\tau_{res}< 1\, \upmu \mathrm{s}$ and low activity is preferred, considering the fact that  the energy resolution is deteriorated by the increase of the heat capacity from the  nuclei implanted into the $Au$ absorber. Too low an activity, on~the other hand, will require a large number of pixels, to~meet the need of high statistics. This will become challenging for the multiplexing read-out, given the need of a large bandwidth per pixel~\cite{Gastaldo2017,Alpert2019}.  The~HOLMES approach is to use TESs coupled to Au absorbers, similar to the ones developed for the soft X-ray spectroscopy~\cite{Orlando18}, with~pulse rise time engineered to be about $10\, \upmu \mathrm{s}$. Using tailored discrimination algorithms~\cite{Alpert2016,Ferri2016}, a~time resolution better than $3\, \upmu \mathrm{s}$ is expected. The~total amount of implanted $^{163}\mathrm{Ho}$ will be about $6.5\cdot10^{16}$ to give an activity of $300\, \mathrm{dec/s/pixel}$. An~energy resolution of $4.1\, \mathrm{eV}$ has been recently reported  for the HOLMES TES calorimeters before the $^{163}\mathrm{Ho}$ being implanted~\cite{Giachero2021}. 
The k-pixel array is read out using GHz-FDM. Given the speed of the detectors, the~separation among two adjacent GHz tones has to be large, $\Delta f =14\, \mathrm{MHz}$, and~36 detectors could be simultaneously read-out per channel~\cite{Alpert2019}.

(ii) A relatively new field where large array of X-ray microcalorimeters could play an important role in the future is the search of axion-like particles. Axions have been originally predicted by  the Peccei–Quinn theory~\cite{PecQuinPRL1977,PecQuinPRD1977} as a possible solution to the strong CP problem. Axion-like particles in general are proposed in several theories beyond the standard model~\cite{JaeckRing2010}. The~Sun could be a strong axions and axion-like particles source. When passing through a large ($\sim$9$\, \mathrm{T}$) magnetic field, solar axions can be converted to X-ray photons detectable by X-ray spectrometers. The~solar axions spectrum is expected to be generated by axion-photon and axion-electron interactions, respectively. The~former has a smooth shape with a peak at about $3\,\mathrm{keV}$, corresponding to the inner solar temperature. The~latter has a maximum at about $1\, \mathrm{keV}$ and shows many energy peaks up to  $7\,\mathrm{keV}$, which depend on  the metal composition of the Sun. An~instrument similar to X-IFU could contribute to the first axions detection. Thanks to the high resolving power and low energy threshold demonstrated by TES-based calorimeters, TES technology could  become crucial, after~discovery, to~characterize the axions electrons process   and to give a key contribution on the understanding of the stellar physics~\cite{Jaeckel_axion2019}.
The International Axion Observatory (IAXO) is a new generation of axion heliscope that will  improve the CAST state-of-the-art sensitivity~\cite{CAST2017} by more than a factor of $10^4$. It will give access to probe a large fraction of the quantum chromodynamic axion models in the  meV  to eV mass band~\cite{Armengaud_2014}.  A~smaller scale version of IAXO, called BabyIAXO, is currently under development~\cite{babyIAXO2020concept}. Metallic magnetic calorimeters-based instrument has been proposed as one of the X-ray instruments for IAXO~\cite{Unger2020}.
Low temperature X-ray calorimeters do have typically very high quantum efficiency, an~essential feature for this application, and~can be fabricated in large arrays. However, the~most challenging requirement for an X-ray detector for IAXO is the ultra-low background set to be  about $ 10^{-7}\, \mathrm{keV}^{-1} \mathrm{cm}^{-2} \mathrm{s}^{-1}$.
Experiments are on going at Heidelberg to assess the contribution on the radiation background of low temperature detectors.  One of the major  observed sources was the flourescence in the material surrounding the detector induced by cosmic muons and radioactive impurities.  A~level of $1.2 \times 10^{-4}\, \mathrm{keV}^{-1} \mathrm{cm}^{-2} \mathrm{s}^{-1}$,  has been  measured with  their first laboratory prototype~\cite{Unger2020}.  It is clear that a proper choice  of low radioactivity  materials and a careful simulation and design of the focal plane assembly  is very important to be compliant with the extremely low background level requirement. 
The analysis of the residual background composition in the X-IFU focal plane assembly~\cite{Lotti2021}, for~example, revealed that a considerable fraction of the  background is induced by secondary electrons coming from the niobium shield surrounding the detector.  Passive low-Z shields to interpose between the detector and the Niobium shield  will help to reduce this component. A~cryogenic anti-coincidence detector mounted in the proximity of the main array, as~currently under testing in the demonstration model for X-IFU~\cite{Macculi2020,vanWeers2020}, is designed to reduce the contribution from high energy particle of about a factor of  30 around $5\, \mathrm{keV}$. 
One should realize, however, that the volume and mass constraints in space applications are very stringent. In~a ground-based experiment like IAXO, the~passive and active background shielding can be designed to be much more efficient. Correlation analysis between a large array of  very sensitive pixels will help as well in reducing the~background.

(iii) TESs have been successfully employed on ground telescopes as very sensitive bolometers in a series of CMB experiments~\cite{Rini2020,BiCEP2020,Suzuki2016}. One interesting overlap between the TES-bolometers technology and the X-ray instruments discussed so far is the  development of the 40--400 GHz instruments on board of the LiteBIRD mission~\cite{LiteBIRD2019,LiteBIRD2020},  recently selected by the Japanese space agency (JAXA) as a mission to be launched in 2027. LiteBIRD is a Lite satellite for the studies of B-mode polarization and Inflation from cosmic background Radiation Detection. It aims to detect the footprint of the primordial gravitational wave on the CMB. If~successful, it will open up a new era in observational cosmology.
The LiteBIRD detectors core technology will be an array of more than 4000 TES bolometers~\cite{Jaehnig2020}, divided between the low-, mid- and high-frequency bands, and~  read-out with MHz-FDM~\cite{deHaan2020}.  The~current baseline foresees TESs with a normal resistance of about $1 \, \Omega$, a~time constant of 0.1--2.2$\, \mathrm{ms}$ and saturation power around 0.7--0.9$\, \mathrm{pW}$. From~the point of view of the read-out, the~LiteBIRD TES-bolometers  are very similar to the high-aspect ratio and high normal resistance  TESs currently under development at SRON for the MHz-FDM read-out of X-IFU-like detectors, discussed in Section \ref{sec:acbias}. Using MHz-FDM read-out similar to the one shown in Section \ref{sec:MHzFDM}, with~demonstrated  low SQUID current noise  of $\sim$5$\mathrm{pA}/\sqrt{\mathrm{Hz}}$, we believe a multiplexing factor of 72 pixels per SQUID chain is~feasible.

%%%%%%%%%%%%%%%%%%%%%%%%%%%%%%%%%%%%%%%%%%
\section{Conclusions}

We have given a review on the recent progress in the development of low temperature detectors with a large array of TES-based microcalorimeters for application in astro-particle physics.
While X-ray TES microcalorimeters have already demonstrated an exquisite resolving power in the energy band from 0.2 to 10 keV,  we believe a continuous optimization of the detector design and its read-out is required to achieve the ultimate performance in the complex instruments discussed~here. 

 A further improvement in the TES design might lead to the demonstration of   energy resolution better than 1.5 eV at 6 keV and below 1 eV at lower energies, within~the existing X-ray microcalorimeters fabrication technology. For~this reason, we have placed emphasis on the underlying detector physics and the deep interaction between the detectors properties  and the multiplexing~configuration. 

 Researchers are striving to achieve a realistic description of the complex physical phenomena observed at the superconducting transition of TESs with extended geometry and complex structures, including the current leads and the coupled normal metal absorber. We have given the reader an overview of the most recent models developed to predict the TES resistive transition and the noise. Those models can be implemented in the generalized system of coupled differential equations for a TES and its read-out circuit. The~equations can be solved numerically to calculate the detector large signal response and realistically  predict the dynamic interaction with the read-out circuit. This approach can be used as a building block for a detector simulator. Thanks to the available computational power of modern computers, the~simulation of complex instruments with a large amount of pixels distributed on many read-out channels will give the researchers and engineers a powerful tool to guide the detector development and the instrument~calibration. 

 The energy resolution of multiplexed pixels  arrays is approaching the best results observed in single pixel configuration. {Driven by the development of the X-IFU instrument of the ESA future large space mission Athena, the~research field of TES X-ray microcalorimeters is entering a very interesting phase with the focus on demonstrating, in~the next decade,  the~performance of a full instrument with more than thousand pixels. The~progress booked by the community over the past  12 years is summarized in  \figref{fig:dEMUX}. We show, for~each multiplexing technology and for X-ray photons with  5.9 keV energy, the~improvement over the years on the multiplexed energy resolution, the~multiplexing factor per channel, the~multiplexing factor per channel normalized to the energy resolution, and the total number of pixels read-out simultaneously in multi-channels experiments. Both the energy resolution and the multiplexing factor for the time division and the MHz-frequency division multiplexing  are not expected to improve dramatically in the next year, being limited by fundamental noise sources in the TES  and by the detector and read-out bandwidth ratio.  On~the contrary,    the~total number of pixels depends on the total number of read-out channels that an instrument can host. It will likely increase by about a factor of two every two years in the next decade, with~the ultimate demonstration provided by the XIFU flight model, which is expected to be tested on ground in 2029.}
 A lot of efforts have to be devoted in the coming years to improve  even further  the efficiency of the multiplexing read-out techniques to achieve the ambitious goal of building future instruments with 100,000-pixels array beyond~2030. 
 \begin{figure}[H]
%\center
\includegraphics[width=13cm]{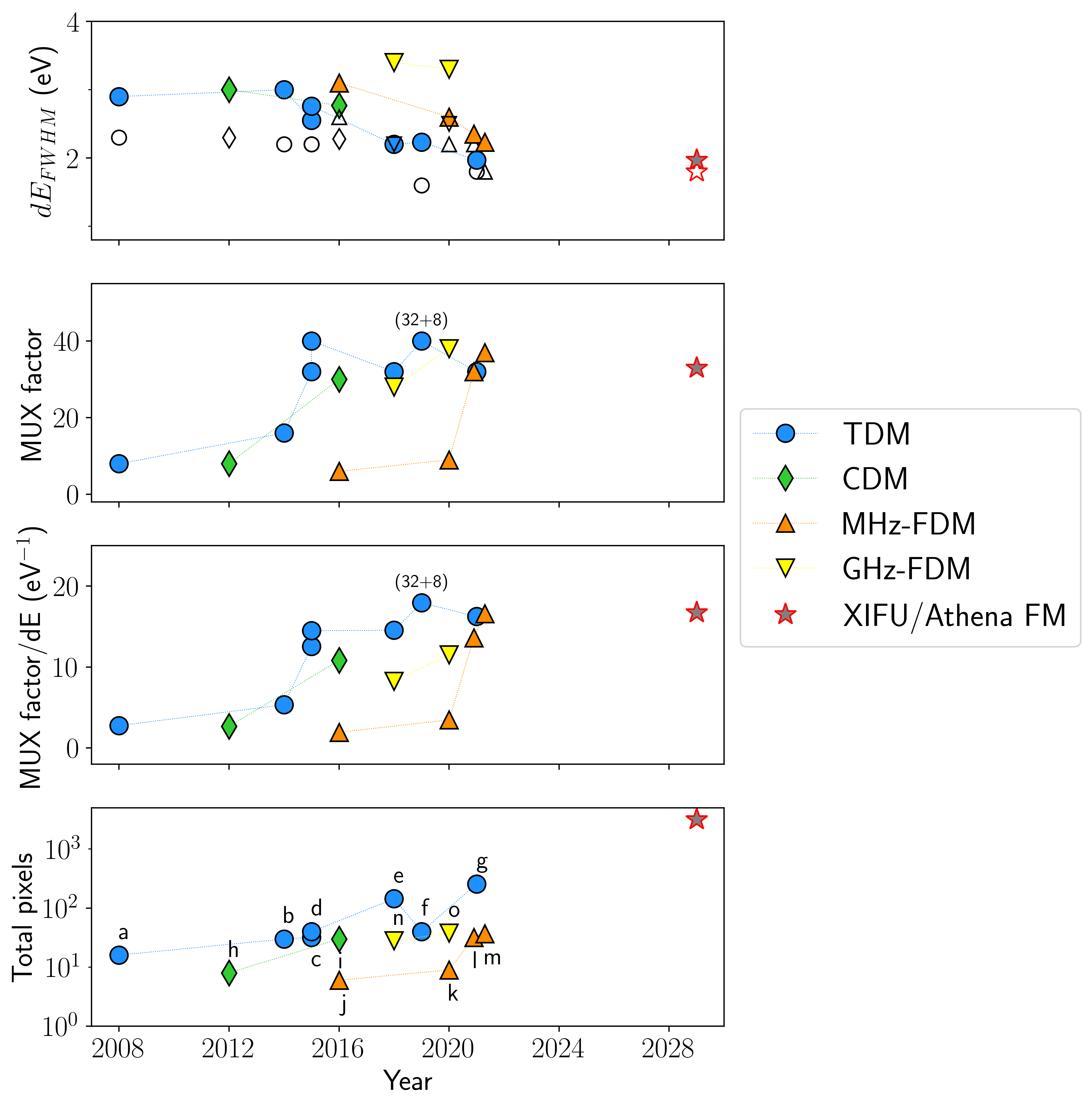}
\caption{\label{fig:dEMUX} Progress on the multiplexing read-out of TES X-ray microcalorimeters over the last decade. From~top to bottom it is shown, for~each multiplexing configuration:  the multiplexed energy resolution, $dE_{FWHM}$, for~photons at 5.9 keV energy, the~multiplexing factor per channel, the~multiplexing factor per channel normalized to the energy resolution, and the total number of pixels read-out simultaneously in multi-channel experiments. The~empty symbols on the top plot correspond to the energy resolution measured in single pixel mode in the corresponding  multiplexing experiments. The~star symbol is the expected demonstration of the readout of 3168 pixels with the XIFU/Athena flight model (FM). The~data points are taken from the following publications, as~annotated in the bottom plot: a = \cite{Kilb2008}, b = \cite{Smith2015}, c = \cite{Doriese2016}, d  =  \cite{Doriese2017}, e = \cite{Ullom2018}, \mbox{f = \cite{Durkin2019}} (32~TESs + 8 repeats of the last row), g = \cite{Smith2021}, h = \cite{Stiehl2012}, i = \cite{Morgan2016}, j = \cite{Akamatsu2016}, k = \cite{Akamatsu2020}, l = \cite{AkaSPIE2020}, \mbox{m = \cite{Akamatsu2021}}, \mbox{n =\cite{Yoon2017}}, o = \cite{Naka2020}.    }
\end{figure}

%\blue{This section is not mandatory, but can be added to the manuscript if the discussion is unusually long or complex.}

%%%%%%%%%%%%%%%%%%%%%%%%%%%%%%%%%%%%%%%%%%
\vspace{6pt} 

%%%%%%%%%%%%%%%%%%%%%%%%%%%%%%%%%%%%%%%%%%
%% optional
%\supplementary{The following are available online at \linksupplementary{s1}, Figure S1: title, Table S1: title, Video S1: title.}


%%%%%%%%%%%%%%%%%%%%%%%%%%%%%%%%%%%%%%%%%%
\authorcontributions{ Conceptualization, writing and editing, L.G.; K.N. took care of the conceptualization and writing of the section on the detector fabrication. All authors have read and agreed to the published version of the~manuscript.}

%%%%%%%%%%%%%%%%%%%%%%%%%%%%%%%%%%%%%%%%%%
\funding{This work is partly funded by European Space Agency
(ESA) under ESA CTP contract ITT AO/1-7947/14/NL/BW and ESA CTP Contract No. 4000130346/20/NL/BW/os, by~the European Union’s Horizon 2020 Program under the AHEAD project (Grant Agreement Number 654215) and is part of the research program Athena (project number 184.034.002), which is partly financed by the Dutch Research Council (NWO).}

\institutionalreview{Not applicable.}%In this section, please add the Institutional Review Board Statement and approval number for studies involving humans or animals. Please note that the Editorial Office might ask you for further information. Please add ``The study was conducted according to the guidelines of the Declaration of Helsinki, and~approved by the Institutional Review Board (or Ethics Committee) of NAME OF INSTITUTE (protocol code XXX and date of approval).'' OR ``Ethical review and approval were waived for this study, due to REASON (please provide a detailed justification).'' OR ``Not applicable'' for studies not involving humans or animals. You might also choose to exclude this statement if the study did not involve humans or animals.}

\informedconsent{Not applicable.}%Any research article describing a study involving humans should contain this statement. Please add ``Informed consent was obtained from all subjects involved in the study.'' OR ``Patient consent was waived due to REASON (please provide a detailed justification).'' OR ``Not applicable'' for studies not involving humans. You might also choose to exclude this statement if the study did not involve~humans.

%Written informed consent for publication must be obtained from participating patients who can be identified (including by the patients themselves). Please state ``Written informed consent has been obtained from the patient(s) to publish this paper'' if applicable.}

\dataavailability{Not applicable.}%In this section, please provide details regarding where data supporting reported results can be found, including links to publicly archived datasets analyzed or generated during the study. Please refer to suggested Data Availability Statements in section ``MDPI Research Data Policies'' at \url{https://www.mdpi.com/ethics}. You might choose to exclude this statement if the study did not report any data.} 

%%%%%%%%%%%%%%%%%%%%%%%%%%%%%%%%%%%%%%%%%%
\acknowledgments{We wish to thank the X-IFU team at SRON for their precious support. In~particular, we are grateful to Hiroki Akamatsu, Marcel Bruin, Martin de Wit, Emanuele Taralli, Davide Vaccaro, Matteo d'Andrea, Marcel Ridder, Jan van der Kuur, and Brian Jackson for the helpful discussions and for their valuable contribution to the work presented here. We acknowledge  the fruitful collaboration over the past years with Mikko Kiviranta at VTT, Steve Smith, Nick Wakeham, Jay Chervenak, Kazu Sakai at  NASA-GSFC,  Douglas Bennet at NIST,  and Alex Kozorezov  at the Lancaster University. We thank Marios Kounalakis from Delft University  for reading and commenting the~manuscript. }

%%%%%%%%%%%%%%%%%%%%%%%%%%%%%%%%%%%%%%%%%%
\conflictsofinterest{The authors declare no conflict of~interest.} 

%%%%%%%%%%%%%%%%%%%%%%%%%%%%%%%%%%%%%%%%%%
%% Only for journal Encyclopedia
%\entrylink{The Link to this entry published on the encyclopedia platform.}

%%%%%%%%%%%%%%%%%%%%%%%%%%%%%%%%%%%%%%%%%%
%% Optional
%\abbreviations{The following abbreviations are used in this manuscript:\\

%\noindent 
%\begin{tabular}{@{}ll}
%SRON & \\
%NASA-GSFC & \\
%NIST & \\
%LD & 
%\end{tabular}}

%%%%%%%%%%%%%%%%%%%%%%%%%%%%%%%%%%%%%%%%%%
%% Optional
%\appendixtitles{no} % Leave argument "no" if all appendix headings stay EMPTY (then no %dot is printed after "Appendix A"). If the appendix sections contain a heading then change the argument to "yes".
%\appendix
%\section{}
%\unskip
%\subsection{}
%The appendix is an optional section that can contain details and data supplemental to the main text. For example, explanations of experimental details that would disrupt the flow of the main text, but nonetheless remain crucial to understanding and reproducing the research shown; figures of replicates for experiments of which representative data is shown in the main text can be added here if brief, or as Supplementary data. Mathematical proofs of results not central to the paper can be added as an appendix.

%\section{}
%All appendix sections must be cited in the main text. In the appendixes, Figures, Tables, etc. should be labeled starting with `A', e.g.,~Figure A1, Figure A2, etc. 
\end{paracol}
%%%%%%%%%%%%%%%%%%%%%%%%%%%%%%%%%%%%%%%%%%
\reftitle{References}%Please check the references carefully


\begin{thebibliography}{-------}
\providecommand{\natexlab}[1]{#1}

\bibitem[{Adams} \em{et~al.}(2021){Adams}, {Baker}, {Bandler}, {Bastidon},
  {Danowski}, {Doriese}, {Eckart}, {Figueroa-Feliciano}, {Fuhrman},
  {Goldfinger}, {Heine}, {Hilton}, {Hubbard}, {Jardin}, {Kelley}, {Kilbourne},
  {Manzagol-Harwood}, {McCammon}, {Okajima}, {Porter}, {Reintsema},
  {Serlemitsos}, {Smith}, and {Wikus}]{Adams2021}
{Adams}, J.S.; {Baker}, R.; {Bandler}, S.R.; {Bastidon}, N.; {Danowski}, M.E.;
  {Doriese}, W.B.; {Eckart}, M.E.; {Figueroa-Feliciano}, E.; {Fuhrman}, J.;
  {Goldfinger}, D.C.; {Heine}, S.N.T.; {Hilton}, G.C.; {Hubbard}, A.J.F.;
  {Jardin}, D.; {Kelley}, R.L.; {Kilbourne}, C.A.; {Manzagol-Harwood}, R.E.;
  {McCammon}, D.; {Okajima}, T.; {Porter}, F.S.; {Reintsema}, C.D.;
  {Serlemitsos}, P.; {Smith}, S.J.; {Wikus}, P.
\newblock {First operation of transition-edge sensors in space with the Micro-X
  sounding rocket}.
\newblock  Society of Photo-Optical Instrumentation Engineers (SPIE) Conference
  Series,  2021, Vol. 11454, {\em Society of Photo-Optical Instrumentation
  Engineers (SPIE) Conference Series}, p. 1145414.
\newblock
  doi:{\changeurlcolor{black}\href{https://doi.org/10.1117/12.2562645}{\detokenize{10.1117/12.2562645}}}.

\bibitem[Barret \em{et~al.}(2020)Barret, Decourchelle, Fabian, Guainazzi,
  Nandra, Smith, and den Herder]{Barret20}
Barret, D.; Decourchelle, A.; Fabian, A.; Guainazzi, M.; Nandra, K.; Smith, R.;
  den Herder, J.W.
\newblock {The Athena space X-ray observatory and the astrophysics of hot
  plasma}.
\newblock {\em Astronomische Nachrichten} {\bf 2020}, {\em 341},~224--235.
\newblock
  doi:{\changeurlcolor{black}\href{https://doi.org/10.1002/asna.202023782}{\detokenize{10.1002/asna.202023782}}}.

\bibitem[Sugai \em{et~al.}(2020)Sugai et~al.]{LiteBIRD2020}
Sugai, H.; others.
\newblock Updated Design of the CMB Polarization Experiment Satellite LiteBIRD.
\newblock {\em J. Low Temp. Phys.} {\bf 2020}, {\em 199},~1107--1117.
\newblock
  doi:{\changeurlcolor{black}\href{https://doi.org/10.1007/s10909-019-02329-w}{\detokenize{10.1007/s10909-019-02329-w}}}.

\bibitem[{Hlozek}(2021)]{Simons2021}
{Hlozek}, R.
\newblock {The Simons Observatory: Science goals and forecasts}.
\newblock  American Astronomical Society Meeting Abstracts,  2021, Vol.~53,
  {\em American Astronomical Society Meeting Abstracts}, p. 214.02.

\bibitem[Thornton(2016)]{ACTPol16}
Thornton, R.J.
\newblock {The Atacama Cosmology Telescope: The Polarization-sensitive ACTPol
  Instrument}.
\newblock {\em The Astrophysical Journal Supplement Series} {\bf 2016}, {\em
  227},~21,  \href{http://xxx.lanl.gov/abs/1605.06569}{{\normalfont
  [arXiv:astro-ph.IM/1605.06569]}}.
\newblock
  doi:{\changeurlcolor{black}\href{https://doi.org/10.3847/1538-4365/227/2/21}{\detokenize{10.3847/1538-4365/227/2/21}}}.

\bibitem[Piat \em{et~al.}(2021)Piat et~al.]{QUIBC2021}
Piat, M.; others.
\newblock {QUBIC IV: Performance of TES Bolometers and Readout Electronics}.
\newblock {\em arXiv e-prints} {\bf 2021}, p. arXiv:2101.06787,
  \href{http://xxx.lanl.gov/abs/2101.06787}{{\normalfont
  [arXiv:astro-ph.IM/2101.06787]}}.

\bibitem[Roelfsema \em{et~al.}(2014)Roelfsema, Giard, Najarro, Wafelbakker,
  Jellema, Jackson, Sibthorpe, Audard, Doi, di~Giorgio, Griffin, Helmich, Kamp,
  Kerschbaum, Meyer, Naylor, Onaka, Poglitch, Spinoglio, van~der Tak, and
  Vandenbussche]{Spica14}
Roelfsema, P.; Giard, M.; Najarro, F.; Wafelbakker, K.; Jellema, W.; Jackson,
  B.; Sibthorpe, B.; Audard, M.; Doi, Y.; di~Giorgio, A.; Griffin, M.; Helmich,
  F.; Kamp, I.; Kerschbaum, F.; Meyer, M.; Naylor, D.; Onaka, T.; Poglitch, A.;
  Spinoglio, L.; van~der Tak, F.; Vandenbussche, B.
\newblock {SAFARI new and improved: extending the capabilities of SPICA's
  imaging spectrometer}.
\newblock  Space Telescopes and Instrumentation 2014: Optical, Infrared, and
  Millimeter Wave; Jr., J.M.O.; Clampin, M.; Fazio, G.G.; MacEwen, H.A., Eds.
  International Society for Optics and Photonics, SPIE,  2014, Vol. 9143, pp.
  419 -- 429.
\newblock
  doi:{\changeurlcolor{black}\href{https://doi.org/10.1117/12.2056449}{\detokenize{10.1117/12.2056449}}}.

\bibitem[Nucciotti \em{et~al.}(2018)Nucciotti, Alpert, Balata, Becker, Bennett,
  Bevilacqua, Biasotti, Ceriale, Ceruti, Corsini, De~Gerone, Dressler,
  Faverzani, Ferri, Fowler, Gallucci, Gard, Gatti, Giachero, Hays-Wehle,
  Heinitz, Hilton, Köster, Lusignoli, Mates, Nisi, Orlando, Parodi, Pessina,
  Puiu, Ragazzi, Reintsema, Ribeiro-Gomez, Schmidt, Schuman, Siccardi, Swetz,
  Ullom, and Vale]{Nucciotti2018}
Nucciotti, A.; Alpert, B.; Balata, M.; Becker, D.; Bennett, D.; Bevilacqua, A.;
  Biasotti, M.; Ceriale, V.; Ceruti, G.; Corsini, D.; De~Gerone, M.; Dressler,
  R.; Faverzani, M.; Ferri, E.; Fowler, J.; Gallucci, G.; Gard, J.; Gatti, F.;
  Giachero, A.; Hays-Wehle, J.; Heinitz, S.; Hilton, G.; Köster, U.;
  Lusignoli, M.; Mates, J.; Nisi, S.; Orlando, A.; Parodi, L.; Pessina, G.;
  Puiu, A.; Ragazzi, S.; Reintsema, C.; Ribeiro-Gomez, M.; Schmidt, D.;
  Schuman, D.; Siccardi, F.; Swetz, D.; Ullom, J.; Vale, L.
\newblock Status of the HOLMES Experiment to Directly Measure the Neutrino
  Mass.
\newblock {\em J. Low Temp. Phys.} {\bf 2018}, {\em 183},~1137--1145.
\newblock
  doi:{\changeurlcolor{black}\href{https://doi.org/10.1007/s10909-018-2025-x}{\detokenize{10.1007/s10909-018-2025-x}}}.

\bibitem[{Doriese} \em{et~al.}(2016){Doriese}, {Morgan}, {Bennett}, {Denison},
  {Fitzgerald}, {Fowler}, {Gard}, {Hays-Wehle}, {Hilton}, {Irwin}, {Joe},
  {Mates}, {O'Neil}, {Reintsema}, {Robbins}, {Schmidt}, {Swetz}, {Tatsuno},
  {Vale}, and {Ullom}]{Doriese2016}
{Doriese}, W.B.; {Morgan}, K.M.; {Bennett}, D.A.; {Denison}, E.V.;
  {Fitzgerald}, C.P.; {Fowler}, J.W.; {Gard}, J.D.; {Hays-Wehle}, J.P.;
  {Hilton}, G.C.; {Irwin}, K.D.; {Joe}, Y.I.; {Mates}, J.A.B.; {O'Neil}, G.C.;
  {Reintsema}, C.D.; {Robbins}, N.O.; {Schmidt}, D.R.; {Swetz}, D.S.;
  {Tatsuno}, H.; {Vale}, L.R.; {Ullom}, J.N.
\newblock {Developments in Time-Division Multiplexing of X-ray Transition-Edge
  Sensors}.
\newblock {\em J. Low Temp. Phys.} {\bf 2016}, {\em 184},~389--395.
\newblock
  doi:{\changeurlcolor{black}\href{https://doi.org/10.1007/s10909-015-1373-z}{\detokenize{10.1007/s10909-015-1373-z}}}.

\bibitem[Morgan \em{et~al.}(2016)Morgan, Alpert, Bennett, Denison, Doriese,
  Fowler, Gard, Hilton, Irwin, Joe, O'Neil, Reintsema, Schmidt, Ullom, and
  Swetz]{Morgan2016}
Morgan, K.M.; Alpert, B.K.; Bennett, D.A.; Denison, E.V.; Doriese, W.B.;
  Fowler, J.W.; Gard, J.D.; Hilton, G.C.; Irwin, K.D.; Joe, Y.I.; O'Neil, G.C.;
  Reintsema, C.D.; Schmidt, D.R.; Ullom, J.N.; Swetz, D.S.
\newblock Code-division-multiplexed readout of large arrays of TES
  microcalorimeters.
\newblock {\em Applied Physics Letters} {\bf 2016}, {\em 109},~112604.
\newblock
  doi:{\changeurlcolor{black}\href{https://doi.org/10.1063/1.4962636}{\detokenize{10.1063/1.4962636}}}.

\bibitem[Mates \em{et~al.}(2017)Mates, Becker, Bennett, Dober, Gard,
  Hays-Wehle, Fowler, Hilton, Reintsema, Schmidt, Swetz, Vale, and
  Ullom]{Mates2017}
Mates, J.A.B.; Becker, D.T.; Bennett, D.A.; Dober, B.J.; Gard, J.D.;
  Hays-Wehle, J.P.; Fowler, J.W.; Hilton, G.C.; Reintsema, C.D.; Schmidt, D.R.;
  Swetz, D.S.; Vale, L.R.; Ullom, J.N.
\newblock Simultaneous readout of 128 X-ray and gamma-ray transition-edge
  microcalorimeters using microwave SQUID multiplexing.
\newblock {\em Applied Physics Letters} {\bf 2017}, {\em 111},~062601.
\newblock
  doi:{\changeurlcolor{black}\href{https://doi.org/10.1063/1.4986222}{\detokenize{10.1063/1.4986222}}}.

\bibitem[van~der Kuur \em{et~al.}(2016)van~der Kuur, Gottardi, Akamatsu, van
  Leeuwen, den Hartog, Haas, Kiviranta, and Jackson]{vdKuur2016}
van~der Kuur, J.; Gottardi, L.; Akamatsu, H.; van Leeuwen, B.J.; den Hartog,
  R.; Haas, D.; Kiviranta, M.; Jackson, B.J.
\newblock {Optimising the multiplex factor of the frequency domain multiplexed
  readout of the TES-based microcalorimeter imaging array for the X-IFU
  instrument on the Athena x-ray observatory}.
\newblock  Space Telescopes and Instrumentation 2016: Ultraviolet to Gamma Ray;
  den Herder, J.W.A.; Takahashi, T.; Bautz, M., Eds. International Society for
  Optics and Photonics, SPIE,  2016, Vol. 9905, pp. 1666 -- 1674.
\newblock
  doi:{\changeurlcolor{black}\href{https://doi.org/10.1117/12.2232830}{\detokenize{10.1117/12.2232830}}}.

\bibitem[McCammon(2005)]{McCammon2005}
McCammon, D., Thermal Equilibrium Calorimeters -- An Introduction.
\newblock In {\em Cryogenic Particle Detection}; Enss, C., Ed.; Springer Berlin
  Heidelberg: Berlin, Heidelberg,  2005; pp. 1--34.
\newblock
  doi:{\changeurlcolor{black}\href{https://doi.org/10.1007/10933596_1}{\detokenize{10.1007/10933596_1}}}.

\bibitem[Irwin and Hilton(2005)]{IrwinHilton}
Irwin, K.; Hilton, G., Transition-Edge Sensors.
\newblock In {\em Cryogenic Particle Detection}; Enss, C., Ed.; Springer Berlin
  Heidelberg: Berlin, Heidelberg,  2005; pp. 63--150.
\newblock
  doi:{\changeurlcolor{black}\href{https://doi.org/10.1007/10933596_3}{\detokenize{10.1007/10933596_3}}}.

\bibitem[Ullom and Bennett(2015)]{Ullom_2015}
Ullom, J.N.; Bennett, D.A.
\newblock Review of superconducting transition-edge sensors for x-ray and
  gamma-ray spectroscopy.
\newblock {\em Superconductor Science and Technology} {\bf 2015}, {\em
  28},~084003.
\newblock
  doi:{\changeurlcolor{black}\href{https://doi.org/10.1088/0953-2048/28/8/084003}{\detokenize{10.1088/0953-2048/28/8/084003}}}.

\bibitem[Barret \em{et~al.}(2018)Barret, Trong, den Herder, Piro, Cappi,
  Houvelin, Kelley, et~al.]{XIFU2018}
Barret, D.; Trong, T.L.; den Herder, J.W.; Piro, L.; Cappi, M.; Houvelin, J.;
  Kelley, R.; others.
\newblock {The ATHENA X-ray Integral Field Unit (X-IFU)}.
\newblock  Space Telescopes and Instrumentation 2018: Ultraviolet to Gamma Ray;
  den Herder, J.W.A.; Nikzad, S.; Nakazawa, K., Eds. International Society for
  Optics and Photonics, SPIE,  2018, Vol. 10699, pp. 324 -- 338.
\newblock
  doi:{\changeurlcolor{black}\href{https://doi.org/10.1117/12.2312409}{\detokenize{10.1117/12.2312409}}}.

\bibitem[Lindeman \em{et~al.}(2004)Lindeman, Bandler, Brekosky, Chervenak,
  Figueroa-Feliciano, Finkbeiner, Li, and Kilbourne]{Lind2004}
Lindeman, M.A.; Bandler, S.; Brekosky, R.P.; Chervenak, J.A.;
  Figueroa-Feliciano, E.; Finkbeiner, F.M.; Li, M.J.; Kilbourne, C.A.
\newblock Impedance measurements and modeling of a transition-edge-sensor
  calorimeter.
\newblock {\em Rev. Sci. Instrum.} {\bf 2004}, {\em 75},~1283--1289.

\bibitem[Irwin(2006)]{Irwin2006}
Irwin, K.
\newblock Thermodynamics of nonlinear bolometers near equilibrium.
\newblock {\em Nucl. Instrum. and Methods Phys. Res. Sect. A} {\bf 2006}, {\em
  559},~718 -- 720.

\bibitem[Irwin(1995)]{Irwin1995}
Irwin, K.D.
\newblock An application of electrothermal feedback for high resolution
  cryogenic particle detection.
\newblock {\em Appl. Phys. Lett.} {\bf 1995}, {\em 66},~1998--2000,
  \href{http://xxx.lanl.gov/abs/https://doi.org/10.1063/1.113674}{{\normalfont
  [https://doi.org/10.1063/1.113674]}}.
\newblock
  doi:{\changeurlcolor{black}\href{https://doi.org/10.1063/1.113674}{\detokenize{10.1063/1.113674}}}.

\bibitem[{Sadleir} \em{et~al.}(2010){Sadleir}, {Smith}, {Bandler}, {Chervenak},
  and {Clem}]{Sadleir10}
{Sadleir}, J.E.; {Smith}, S.J.; {Bandler}, S.R.; {Chervenak}, J.A.; {Clem},
  J.R.
\newblock {Longitudinal Proximity Effects in Superconducting Transition-Edge
  Sensors}.
\newblock {\em Phys. Rev. Lett.} {\bf 2010}, {\em 104},~047003,
  \href{http://xxx.lanl.gov/abs/0910.2451}{{\normalfont
  [arXiv:cond-mat.supr-con/0910.2451]}}.
\newblock
  doi:{\changeurlcolor{black}\href{https://doi.org/10.1103/PhysRevLett.104.047003}{\detokenize{10.1103/PhysRevLett.104.047003}}}.

\bibitem[Sadleir \em{et~al.}(2011)Sadleir, Smith, Robinson, Finkbeiner,
  Chervenak, Bandler, Eckart, and Kilbourne]{Sadleir11}
Sadleir, J.E.; Smith, S.J.; Robinson, I.K.; Finkbeiner, F.M.; Chervenak, J.A.;
  Bandler, S.R.; Eckart, M.E.; Kilbourne, C.A.
\newblock Proximity effects and nonequilibrium superconductivity in
  transition-edge sensors.
\newblock {\em Phys. Rev. B} {\bf 2011}, {\em 84},~184502.
\newblock
  doi:{\changeurlcolor{black}\href{https://doi.org/10.1103/PhysRevB.84.184502}{\detokenize{10.1103/PhysRevB.84.184502}}}.

\bibitem[de~Wit \em{et~al.}(2020)de~Wit, Gottardi, Taralli, Nagayoshi, Ridder,
  Akamatsu, Bruijn, D’Andrea, van~der Kuur, Ravensberg, Vaccaro, Visser, Gao,
  and den Herder]{deWit_2020}
de~Wit, M.; Gottardi, L.; Taralli, E.; Nagayoshi, K.; Ridder, M.L.; Akamatsu,
  H.; Bruijn, M.P.; D’Andrea, M.; van~der Kuur, J.; Ravensberg, K.; Vaccaro,
  D.; Visser, S.; Gao, J.R.; den Herder, J.W.A.
\newblock High aspect ratio transition edge sensors for x-ray spectrometry.
\newblock {\em J. Appl. Phys.} {\bf 2020}, {\em 128},~224501.

\bibitem[{Ridder} \em{et~al.}(2020){Ridder}, {Nagayoshi}, {Bruijn}, {Gottardi},
  {Taralli}, {Khosropanah}, {Akamatsu}, {van der Kuur}, {Ravensberg}, {Visser},
  {Nieuwenhuizen}, {Gao}, and {den Herder}]{Ridder2020}
{Ridder}, M.L.; {Nagayoshi}, K.; {Bruijn}, M.P.; {Gottardi}, L.; {Taralli}, E.;
  {Khosropanah}, P.; {Akamatsu}, H.; {van der Kuur}, J.; {Ravensberg}, K.;
  {Visser}, S.; {Nieuwenhuizen}, A.C.T.; {Gao}, J.R.; {den Herder}, J.W.
\newblock {Study of TES Detector Transition Curve to Optimize the Pixel Design
  for Frequency-Division Multiplexing Readout}.
\newblock {\em J. Low Temp. Phys.} {\bf 2020}, {\em 199},~962--967.
\newblock
  doi:{\changeurlcolor{black}\href{https://doi.org/10.1007/s10909-020-02401-w}{\detokenize{10.1007/s10909-020-02401-w}}}.

\bibitem[Likharev(1979)]{Likharev79}
Likharev, K.K.
\newblock Superconducting weak links.
\newblock {\em Rev. Mod. Phys.} {\bf 1979}, {\em 51},~101--159.
\newblock
  doi:{\changeurlcolor{black}\href{https://doi.org/10.1103/RevModPhys.51.101}{\detokenize{10.1103/RevModPhys.51.101}}}.

\bibitem[Golubov \em{et~al.}(2004)Golubov, Kupriyanov, and Il'ichev]{Golubov04}
Golubov, A.A.; Kupriyanov, M.Y.; Il'ichev, E.
\newblock The current-phase relation in Josephson junctions.
\newblock {\em Rev. Mod. Phys.} {\bf 2004}, {\em 76},~411--469.
\newblock
  doi:{\changeurlcolor{black}\href{https://doi.org/10.1103/RevModPhys.76.411}{\detokenize{10.1103/RevModPhys.76.411}}}.

\bibitem[Kupriyanov \em{et~al.}(1981)Kupriyanov, Likharev, and
  Lukichev]{KupLikLuk81}
Kupriyanov, M.Y.; Likharev, K.K.; Lukichev, V.F.
\newblock Influence of effective electron interaction on the critical current
  of Josephson weak links.
\newblock {\em Physica} {\bf 1981}, {\em 108B},~1001--1002.

\bibitem[Kupriyanov and Lukichev(1982)]{KupLuk82}
Kupriyanov, M.Y.; Lukichev, V.F.
\newblock Influence of the proximity effect in the electrodes on the stationary
  properties of S--N--S Josephson structures.
\newblock {\em Sov. J. Low Temp. Phys. (Engl. Transl.); (United States)} {\bf
  1982}, {\em 8}.

\bibitem[{Kozorezov} \em{et~al.}(2011){Kozorezov}, {Golubov}, {Martin}, {de
  Korte}, {Lindeman}, {Hijmering}, and {Wigmore}]{Kozo_ieee2011}
{Kozorezov}, A.G.; {Golubov}, A.A.; {Martin}, D.D.E.; {de Korte}, P.A.J.;
  {Lindeman}, M.A.; {Hijmering}, R.A.; {Wigmore}, J.K.
\newblock {Microscopic Model of a Transition Edge Sensor as a Weak Link}.
\newblock {\em IEEE Transactions on Applied Superconductivity} {\bf 2011}, {\em
  21},~250--253.
\newblock
  doi:{\changeurlcolor{black}\href{https://doi.org/10.1109/TASC.2010.2094595}{\detokenize{10.1109/TASC.2010.2094595}}}.

\bibitem[Harwin \em{et~al.}(2017)Harwin, Goldie, and Withington]{Harwin_2017}
Harwin, R.C.; Goldie, D.J.; Withington, S.
\newblock Modelling proximity effects in transition edge sensors to investigate
  the influence of lateral metal structures.
\newblock {\em Superconductor Science and Technology} {\bf 2017}, {\em
  30},~084001.
\newblock
  doi:{\changeurlcolor{black}\href{https://doi.org/10.1088/1361-6668/aa73ae}{\detokenize{10.1088/1361-6668/aa73ae}}}.

\bibitem[Harwin \em{et~al.}(2018)Harwin, Goldie, Withington, Khosropanah,
  Gottardi, and Gao]{Harwin_2018}
Harwin, R.C.; Goldie, D.J.; Withington, S.; Khosropanah, P.; Gottardi, L.; Gao,
  J.R.
\newblock {Proximity effect model for x-ray transition edge sensors}.
\newblock  High Energy, Optical, and Infrared Detectors for Astronomy VIII;
  Holland, A.D.; Beletic, J., Eds. International Society for Optics and
  Photonics, SPIE,  2018, Vol. 10709, pp. 350 -- 361.
\newblock
  doi:{\changeurlcolor{black}\href{https://doi.org/10.1117/12.2313359}{\detokenize{10.1117/12.2313359}}}.

\bibitem[{Kozorezov} \em{et~al.}(2011){Kozorezov}, {Golubov}, {Martin}, {de
  Korte}, {Lindeman}, {Hijmering}, {van der Kuur}, {Hoevers}, {Gottardi},
  {Kupriyanov}, and {Wigmore}]{Kozorezov11}
{Kozorezov}, A.; {Golubov}, A.A.; {Martin}, D.D.E.; {de Korte}, P.A.J.;
  {Lindeman}, M.A.; {Hijmering}, R.A.; {van der Kuur}, J.; {Hoevers}, H.F.C.;
  {Gottardi}, L.; {Kupriyanov}, M.Y.; {Wigmore}, J.K.
\newblock {Modelling the resistive state in a transition edge sensor}.
\newblock {\em Appl. Phys. Lett.} {\bf 2011}, {\em 99},~063503.
\newblock
  doi:{\changeurlcolor{black}\href{https://doi.org/10.1063/1.3621829}{\detokenize{10.1063/1.3621829}}}.

\bibitem[Coffey \em{et~al.}(2008)Coffey, Kalmykov, Titov, and
  Cleary]{Coffey_2008}
Coffey, W.T.; Kalmykov, Y.P.; Titov, S.V.; Cleary, L.
\newblock Smoluchowski equation approach for quantum Brownian motion in a
  tilted periodic potential.
\newblock {\em Phys. Rev. E} {\bf 2008}, {\em 78},~031114.
\newblock
  doi:{\changeurlcolor{black}\href{https://doi.org/10.1103/PhysRevE.78.031114}{\detokenize{10.1103/PhysRevE.78.031114}}}.

\bibitem[Ambegaokar and Halperin(1969)]{AmbHal69}
Ambegaokar, V.; Halperin, B.I.
\newblock Voltage Due to Thermal Noise in the dc Josephson Effect.
\newblock {\em Phys. Rev. Lett.} {\bf 1969}, {\em 22},~1364--1366.
\newblock
  doi:{\changeurlcolor{black}\href{https://doi.org/10.1103/PhysRevLett.22.1364}{\detokenize{10.1103/PhysRevLett.22.1364}}}.

\bibitem[{Smith} \em{et~al.}(2013){Smith}, {Adams}, {Bailey}, {Bandler},
  {Busch}, {Chervenak}, {Eckart}, {Finkbeiner}, {Kilbourne}, {Kelley}, {Lee},
  {Porst}, {Porter}, and {Sadleir}]{Smith13}
{Smith}, S.J.; {Adams}, J.S.; {Bailey}, C.N.; {Bandler}, S.R.; {Busch}, S.E.;
  {Chervenak}, J.A.; {Eckart}, M.E.; {Finkbeiner}, F.M.; {Kilbourne}, C.A.;
  {Kelley}, R.L.; {Lee}, S.J.; {Porst}, J.P.; {Porter}, F.S.; {Sadleir}, J.E.
\newblock {Implications of weak-link behavior on the performance of Mo/Au
  bilayer transition-edge sensors}.
\newblock {\em Journal of Applied Physics} {\bf 2013}, {\em
  114},~074513--074513--24.
\newblock
  doi:{\changeurlcolor{black}\href{https://doi.org/10.1063/1.4818917}{\detokenize{10.1063/1.4818917}}}.

\bibitem[{Smith} \em{et~al.}(2012){Smith}, {Adams}, {Bailey}, {Bandler},
  {Chervenak}, {Eckart}, {Finkbeiner}, {Kelley}, {Kilbourne}, {Porter}, and
  {Sadleir}]{Smith12}
{Smith}, S.J.; {Adams}, J.S.; {Bailey}, C.N.; {Bandler}, S.R.; {Chervenak},
  J.A.; {Eckart}, M.E.; {Finkbeiner}, F.M.; {Kelley}, R.L.; {Kilbourne}, C.A.;
  {Porter}, F.S.; {Sadleir}, J.E.
\newblock {Small Pitch Transition-Edge Sensors with Broadband High Spectral
  Resolution for Solar Physics}.
\newblock {\em J. Low Temp. Phys.} {\bf 2012}, {\em 167},~168--175.
\newblock
  doi:{\changeurlcolor{black}\href{https://doi.org/10.1007/s10909-012-0574-y}{\detokenize{10.1007/s10909-012-0574-y}}}.

\bibitem[Gottardi \em{et~al.}(2014)Gottardi, Akamatsu, Bruijn, Gao, den Hartog,
  Hijmering, Hoevers, Khosropanah, Kozorezov, van~der Kuur, van~der Linden, and
  Ridder]{Gottardi14}
Gottardi, L.; Akamatsu, H.; Bruijn, M.; Gao, J.R.; den Hartog, R.; Hijmering,
  R.; Hoevers, H.; Khosropanah, P.; Kozorezov, A.; van~der Kuur, J.; van~der
  Linden, A.; Ridder, M.
\newblock Weak-Link Phenomena in AC-Biased Transition Edge Sensors.
\newblock {\em J. Low Temp. Phys.} {\bf 2014}, {\em 176},~279--284,
  \href{http://xxx.lanl.gov/abs/1604.00680}{{\normalfont [1604.00680]}}.
\newblock
  doi:{\changeurlcolor{black}\href{https://doi.org/10.1007/s10909-014-1093-9}{\detokenize{10.1007/s10909-014-1093-9}}}.

\bibitem[Bennett \em{et~al.}(2013)Bennett, Swetz, Schmidt, and
  Ullom]{BennetPRB13}
Bennett, D.A.; Swetz, D.S.; Schmidt, D.R.; Ullom, J.N.
\newblock Resistance in transition-edge sensors: A comparison of the
  resistively shunted junction and two-fluid models.
\newblock {\em Phys. Rev. B} {\bf 2013}, {\em 87},~020508.
\newblock
  doi:{\changeurlcolor{black}\href{https://doi.org/10.1103/PhysRevB.87.020508}{\detokenize{10.1103/PhysRevB.87.020508}}}.

\bibitem[Bennett \em{et~al.}(2014)Bennett, Schmidt, Swetz, and
  Ullom]{Bennett14}
Bennett, D.A.; Schmidt, D.R.; Swetz, D.S.; Ullom, J.N.
\newblock Phase-slip lines as a resistance mechanism in transition-edge
  sensors.
\newblock {\em Applied Physics Letters} {\bf 2014}, {\em 104},~042602.
\newblock
  doi:{\changeurlcolor{black}\href{https://doi.org/10.1063/1.4863664}{\detokenize{10.1063/1.4863664}}}.

\bibitem[{Irwin} \em{et~al.}(1998){Irwin}, {Hilton}, {Wollman}, and
  {Martinis}]{Irwin1998}
{Irwin}, K.D.; {Hilton}, G.C.; {Wollman}, D.A.; {Martinis}, J.M.
\newblock {Thermal-response time of superconducting transition-edge
  microcalorimeters}.
\newblock {\em Journal of Applied Physics} {\bf 1998}, {\em 83},~3978--3985.
\newblock
  doi:{\changeurlcolor{black}\href{https://doi.org/10.1063/1.367153}{\detokenize{10.1063/1.367153}}}.

\bibitem[{Bennett} \em{et~al.}(2012){Bennett}, {Swetz}, {Horansky}, {Schmidt},
  and {Ullom}]{Bennet2012}
{Bennett}, D.A.; {Swetz}, D.S.; {Horansky}, R.D.; {Schmidt}, D.R.; {Ullom},
  J.N.
\newblock {A Two-Fluid Model for the Transition Shape in Transition-Edge
  Sensors}.
\newblock {\em J. Low Temp. Phys.} {\bf 2012}, {\em 167},~102--107.
\newblock
  doi:{\changeurlcolor{black}\href{https://doi.org/10.1007/s10909-011-0431-4}{\detokenize{10.1007/s10909-011-0431-4}}}.

\bibitem[Morgan \em{et~al.}(2017)Morgan, Pappas, Bennett, Gard, Hays-Wehle,
  Hilton, Reintsema, Schmidt, Ullom, and Swetz]{Morgan2017}
Morgan, K.M.; Pappas, C.G.; Bennett, D.A.; Gard, J.D.; Hays-Wehle, J.P.;
  Hilton, G.C.; Reintsema, C.D.; Schmidt, D.R.; Ullom, J.N.; Swetz, D.S.
\newblock Dependence of transition width on current and critical current in
  transition-edge sensors.
\newblock {\em Applied Physics Letters} {\bf 2017}, {\em 110},~212602.
\newblock
  doi:{\changeurlcolor{black}\href{https://doi.org/10.1063/1.4984065}{\detokenize{10.1063/1.4984065}}}.

\bibitem[{Morgan} \em{et~al.}(2019){Morgan}, {Becker}, {Bennett}, {Doriese},
  {Gard}, {Irwin}, {Lee}, {Li}, {Mates}, {Pappas}, {Schmidt}, {Titus}, {Van
  Winkle}, {Ullom}, {Wessels}, and {Swetz}]{Morgan2019}
{Morgan}, K.M.; {Becker}, D.T.; {Bennett}, D.A.; {Doriese}, W.B.; {Gard}, J.D.;
  {Irwin}, K.D.; {Lee}, S.J.; {Li}, D.; {Mates}, J.A.B.; {Pappas}, C.G.;
  {Schmidt}, D.R.; {Titus}, C.J.; {Van Winkle}, D.D.; {Ullom}, J.N.; {Wessels},
  A.; {Swetz}, D.S.
\newblock Use of Transition Models to Design High Performance TESs for the
  LCLS-II Soft X-Ray Spectrometer.
\newblock {\em IEEE Transactions on Applied Superconductivity} {\bf 2019}, {\em
  29},~1--5.
\newblock
  doi:{\changeurlcolor{black}\href{https://doi.org/10.1109/TASC.2019.2903032}{\detokenize{10.1109/TASC.2019.2903032}}}.

\bibitem[Kupriyanov \em{et~al.}(1975)Kupriyanov, Likharev, and
  Lukichev]{Kupr75}
Kupriyanov, M.Y.; Likharev, K.K.; Lukichev, V.F.
\newblock {\em The $J_{s}(\psi)$ relationship, Abrikosov vortices and Josephson
  vortices in variable thickness bridges}; North-Holland: Netherlands,  1975.

\bibitem[Smith(2014)]{SmithPoster2014}
Smith, S.
\newblock Studies of weak-link phenomena in Mo/Au TESs.
\newblock {\em Poster presentation. TES workshop. Applied Superc. Conference
  2014} {\bf 2014}.

\bibitem[Sadleir(2016)]{Sadleir2016}
Sadleir, J.
\newblock Unexpected nonlinear effects in superconducting Transition-Edge
  Sensors.
\newblock {\em TES workshop. Applied Superc. Conference 2016} {\bf 2016}.

\bibitem[{Zhou} \em{et~al.}(2018){Zhou}, {Ambarish}, {Gruenke}, {Jaeckel},
  {Kripps}, {McCammon}, {Morgan}, {Wulf}, {Zhang}, {Adams}, {Bandler},
  {Chervenak}, {Datesman}, {Eckart}, {Ewin}, {Finkbeiner}, {Kelley},
  {Kilbourne}, {Miniussi}, {Porter}, {Sadleir}, {Sakai}, {Smith}, {Wakeham},
  {Wassell}, and {Yoon}]{Zhou2018}
{Zhou}, Y.; {Ambarish}, C.V.; {Gruenke}, R.; {Jaeckel}, F.T.; {Kripps}, K.L.;
  {McCammon}, D.; {Morgan}, K.M.; {Wulf}, D.; {Zhang}, S.; {Adams}, J.S.;
  {Bandler}, S.R.; {Chervenak}, J.A.; {Datesman}, A.M.; {Eckart}, M.E.; {Ewin},
  A.J.; {Finkbeiner}, F.M.; {Kelley}, R.L.; {Kilbourne}, C.A.; {Miniussi},
  A.R.; {Porter}, F.S.; {Sadleir}, J.E.; {Sakai}, K.; {Smith}, S.J.; {Wakeham},
  N.A.; {Wassell}, E.J.; {Yoon}, W.
\newblock {Mapping TES Temperature Sensitivity and Current Sensitivity as a
  Function of Temperature, Current, and Magnetic Field with IV Curve and
  Complex Admittance Measurements}.
\newblock {\em J. Low Temp. Phys.} {\bf 2018}, {\em 193},~321--327.
\newblock
  doi:{\changeurlcolor{black}\href{https://doi.org/10.1007/s10909-018-1970-8}{\detokenize{10.1007/s10909-018-1970-8}}}.

\bibitem[F{\`{a}}brega \em{et~al.}(2018)F{\`{a}}brega, Cam{\'{o}}n, Pobes,
  Strichovanec, and Gonz{\'{a}}lez-Arrabal]{Fabrega2018}
F{\`{a}}brega, L.; Cam{\'{o}}n, A.; Pobes, C.; Strichovanec, P.;
  Gonz{\'{a}}lez-Arrabal, R.
\newblock Large current-induced broadening of the superconducting transition in
  Mo/Au transition edge sensors.
\newblock {\em Superconductor Science and Technology} {\bf 2018}, {\em
  32},~015006.
\newblock
  doi:{\changeurlcolor{black}\href{https://doi.org/10.1088/1361-6668/aaebf4}{\detokenize{10.1088/1361-6668/aaebf4}}}.

\bibitem[Fraser(2004)]{Fraser2004}
Fraser, G.
\newblock On the nature of the superconducting-to-normal transition in
  transition edge sensors.
\newblock {\em Nuclear Instruments and Methods in Physics Research Section A:
  Accelerators, Spectrometers, Detectors and Associated Equipment} {\bf 2004},
  {\em 523},~234--245.
\newblock
  doi:{\changeurlcolor{black}\href{https://doi.org/https://doi.org/10.1016/j.nima.2003.12.040}{\detokenize{https://doi.org/10.1016/j.nima.2003.12.040}}}.

\bibitem[{Lindeman} \em{et~al.}(2011){Lindeman}, {Dirks}, {van der Kuur}, {de
  Korte}, {den Hartog}, {Gottardi}, {Hijmering}, {Hoevers}, and
  {Khosropanah}]{Lindeman_2011}
{Lindeman}, M.A.; {Dirks}, B.; {van der Kuur}, J.; {de Korte}, P.A.J.; {den
  Hartog}, R.H.; {Gottardi}, L.; {Hijmering}, R.A.; {Hoevers}, H.F.C.;
  {Khosropanah}, P.
\newblock Relationships Between Complex Impedance, Thermal Response, and Noise
  in TES Calorimeters and Bolometers.
\newblock {\em IEEE Transactions on Applied Superconductivity} {\bf 2011}, {\em
  21},~254--257.
\newblock
  doi:{\changeurlcolor{black}\href{https://doi.org/10.1109/TASC.2010.2096173}{\detokenize{10.1109/TASC.2010.2096173}}}.

\bibitem[{Takei} \em{et~al.}(2008){Takei}, {Gottardi}, {Hoevers}, {de Korte},
  {van der Kuur}, {Ridder}, and {Bruijn}]{Takei08}
{Takei}, Y.; {Gottardi}, L.; {Hoevers}, H.F.C.; {de Korte}, P.A.J.; {van der
  Kuur}, J.; {Ridder}, M.L.; {Bruijn}, M.P.
\newblock {Characterization of a High-Performance Ti/Au TES Microcalorimeter
  with a Central Cu Absorber}.
\newblock {\em J. Low Temp. Phys.} {\bf 2008}, {\em 151},~161--166.
\newblock
  doi:{\changeurlcolor{black}\href{https://doi.org/10.1007/s10909-007-9621-5}{\detokenize{10.1007/s10909-007-9621-5}}}.

\bibitem[Kinnunen \em{et~al.}(2012)Kinnunen, Palosaari, and Maasilta]{Kinn12}
Kinnunen, K.M.; Palosaari, M.R.J.; Maasilta, I.J.
\newblock Normal metal-superconductor decoupling as a source of thermal
  fluctuation noise in transition-edge sensors.
\newblock {\em J. Appl. Phys.} {\bf 2012}, {\em 112},~034515,
  \href{http://xxx.lanl.gov/abs/https://doi.org/10.1063/1.4745908}{{\normalfont
  [https://doi.org/10.1063/1.4745908]}}.
\newblock
  doi:{\changeurlcolor{black}\href{https://doi.org/10.1063/1.4745908}{\detokenize{10.1063/1.4745908}}}.

\bibitem[Maasilta(2012)]{Maasilta12}
Maasilta, I.J.
\newblock Complex impedance, responsivity and noise of transition-edge sensors:
  Analytical solutions for two- and three-block thermal models.
\newblock {\em AIP Advances} {\bf 2012}, {\em 2},~042110,
  \href{http://xxx.lanl.gov/abs/https://doi.org/10.1063/1.4759111}{{\normalfont
  [https://doi.org/10.1063/1.4759111]}}.
\newblock
  doi:{\changeurlcolor{black}\href{https://doi.org/10.1063/1.4759111}{\detokenize{10.1063/1.4759111}}}.

\bibitem[Goldie \em{et~al.}(2009)Goldie, Audley, Glowacka, Tsaneva, and
  Withington]{Goldie_2009}
Goldie, D.J.; Audley, M.D.; Glowacka, D.M.; Tsaneva, V.N.; Withington, S.
\newblock Thermal models and noise in transition edge sensors.
\newblock {\em Journal of Applied Physics} {\bf 2009}, {\em 105},~074512,
  \href{http://xxx.lanl.gov/abs/https://doi.org/10.1063/1.3097396}{{\normalfont
  [https://doi.org/10.1063/1.3097396]}}.
\newblock
  doi:{\changeurlcolor{black}\href{https://doi.org/10.1063/1.3097396}{\detokenize{10.1063/1.3097396}}}.

\bibitem[Wakeham \em{et~al.}(2019)Wakeham, Adams, Bandler, Beaumont, Chervenak,
  Datesman, Eckart, Finkbeiner, Hummatov, Kelley, Kilbourne, Miniussi, Porter,
  Sadleir, Sakai, Smith, and Wassell]{Wake2019}
Wakeham, N.A.; Adams, J.S.; Bandler, S.R.; Beaumont, S.; Chervenak, J.A.;
  Datesman, A.M.; Eckart, M.E.; Finkbeiner, F.M.; Hummatov, R.; Kelley, R.L.;
  Kilbourne, C.A.; Miniussi, A.R.; Porter, F.S.; Sadleir, J.E.; Sakai, K.;
  Smith, S.J.; Wassell, E.J.
\newblock Thermal fluctuation noise in Mo/Au superconducting transition-edge
  sensor microcalorimeters.
\newblock {\em J. Appl. Phys.} {\bf 2019}, {\em 125},~164503,
  \href{http://xxx.lanl.gov/abs/https://doi.org/10.1063/1.5086045}{{\normalfont
  [https://doi.org/10.1063/1.5086045]}}.
\newblock
  doi:{\changeurlcolor{black}\href{https://doi.org/10.1063/1.5086045}{\detokenize{10.1063/1.5086045}}}.

\bibitem[{Bruijn} \em{et~al.}(2018){Bruijn}, {van der Linden}, {Ferrari},
  {Gottardi}, {van der Kuur}, {den Hartog}, {Akamatsu}, and
  {Jackson}]{Bruijn18}
{Bruijn}, M.P.; {van der Linden}, A.J.; {Ferrari}, L.; {Gottardi}, L.; {van der
  Kuur}, J.; {den Hartog}, R.H.; {Akamatsu}, H.; {Jackson}, B.D.
\newblock {LC Filters for FDM Readout of the X-IFU TES Calorimeter Instrument
  on Athena}.
\newblock {\em J. Low Temp. Phys.} {\bf 2018}, {\em 193},~661--667.
\newblock
  doi:{\changeurlcolor{black}\href{https://doi.org/10.1007/s10909-018-1951-y}{\detokenize{10.1007/s10909-018-1951-y}}}.

\bibitem[{Bruijn} \em{et~al.}(2014){Bruijn}, {Gottardi}, {den Hartog}, {van der
  Kuur}, {van der Linden}, and {Jackson}]{Bruijn14}
{Bruijn}, M.P.; {Gottardi}, L.; {den Hartog}, R.H.; {van der Kuur}, J.; {van
  der Linden}, A.J.; {Jackson}, B.D.
\newblock {Tailoring the High-Q LC Filter Arrays for Readout of Kilo-Pixel TES
  Arrays in the SPICA-SAFARI Instrument}.
\newblock {\em J. Low Temp. Phys.} {\bf 2014}, {\em 176},~421--425.
\newblock
  doi:{\changeurlcolor{black}\href{https://doi.org/10.1007/s10909-013-1003-6}{\detokenize{10.1007/s10909-013-1003-6}}}.

\bibitem[Kiviranta \em{et~al.}(2002)Kiviranta, Seppä, van~der Kuur, and
  de~Korte]{Kiviranta02}
Kiviranta, M.; Seppä, H.; van~der Kuur, J.; de~Korte, P.
\newblock SQUID-based readout schemes for microcalorimeter arrays.
\newblock {\em AIP Conference Proceedings} {\bf 2002}, {\em 605},~295--300.
\newblock
  doi:{\changeurlcolor{black}\href{https://doi.org/10.1063/1.1457649}{\detokenize{10.1063/1.1457649}}}.

\bibitem[{Gottardi} \em{et~al.}(2019){Gottardi}, {van der Kuur}, {Bruijn}, {van
  der Linden}, {Kiviranta}, {Akamatsu}, {den Hartog}, and
  {Ravensberg}]{Gottardi19}
{Gottardi}, L.; {van der Kuur}, J.; {Bruijn}, M.; {van der Linden}, A.;
  {Kiviranta}, M.; {Akamatsu}, H.; {den Hartog}, R.; {Ravensberg}, K.
\newblock {Intrinsic Losses and Noise of High- Q Lithographic MHz LC Resonators
  for Frequency Division Multiplexing}.
\newblock {\em J. Low Temp. Phys.} {\bf 2019}, {\em 194},~370--376.
\newblock
  doi:{\changeurlcolor{black}\href{https://doi.org/10.1007/s10909-018-2085-y}{\detokenize{10.1007/s10909-018-2085-y}}}.

\bibitem[{van der Kuur} \em{et~al.}(2011){van der Kuur}, {Gottardi},
  {Borderias}, {Dirks}, {de Korte}, {Lindeman}, {Khosropanah}, {den Hartog},
  and {Hoevers}]{JvdKuur11}
{van der Kuur}, J.; {Gottardi}, L.; {Borderias}, M.P.; {Dirks}, B.; {de Korte},
  P.; {Lindeman}, M.; {Khosropanah}, P.; {den Hartog}, R.; {Hoevers}, H.
\newblock {Small-Signal Behavior of a TES Under AC Bias}.
\newblock {\em IEEE Trans. Appl. Supercond.} {\bf 2011}, {\em 21},~281--284.
\newblock
  doi:{\changeurlcolor{black}\href{https://doi.org/10.1109/TASC.2010.2099092}{\detokenize{10.1109/TASC.2010.2099092}}}.

\bibitem[Taralli \em{et~al.}(2019)Taralli, Khosropanah, Gottardi, Nagayoshi,
  Ridder, Bruijn, and Gao]{TaralliAIP2019}
Taralli, E.; Khosropanah, P.; Gottardi, L.; Nagayoshi, K.; Ridder, M.L.;
  Bruijn, M.P.; Gao, J.R.
\newblock Complex impedance of TESs under AC bias using FDM readout system.
\newblock {\em AIP Advances} {\bf 2019}, {\em 9},~045324,
  \href{http://xxx.lanl.gov/abs/https://doi.org/10.1063/1.5089739}{{\normalfont
  [https://doi.org/10.1063/1.5089739]}}.
\newblock
  doi:{\changeurlcolor{black}\href{https://doi.org/10.1063/1.5089739}{\detokenize{10.1063/1.5089739}}}.

\bibitem[Gottardi \em{et~al.}(2014)Gottardi, Kozorezov, Akamatsu, van~der Kuur,
  Bruijn, den Hartog, Hijmering, Khosropanah, Lambert, van~der Linden, Ridder,
  Suzuki, and Gao]{Gottardi_APL14}
Gottardi, L.; Kozorezov, A.; Akamatsu, H.; van~der Kuur, J.; Bruijn, M.P.; den
  Hartog, R.H.; Hijmering, R.; Khosropanah, P.; Lambert, C.; van~der Linden,
  A.J.; Ridder, M.L.; Suzuki, T.; Gao, J.R.
\newblock Josephson effects in an alternating current biased transition edge
  sensor.
\newblock {\em Appl. Phys. Lett.} {\bf 2014}, {\em 105},~162605,
  \href{http://xxx.lanl.gov/abs/https://doi.org/10.1063/1.4899065}{{\normalfont
  [https://doi.org/10.1063/1.4899065]}}.
\newblock
  doi:{\changeurlcolor{black}\href{https://doi.org/10.1063/1.4899065}{\detokenize{10.1063/1.4899065}}}.

\bibitem[Coffey \em{et~al.}(2009)Coffey, Kalmykov, Titov, and
  Cleary]{Coffey_2009}
Coffey, W.T.; Kalmykov, Y.P.; Titov, S.V.; Cleary, L.
\newblock Nonlinear noninertial response of a quantum Brownian particle in a
  tilted periodic potential to a strong ac force as applied to a point
  Josephson junction.
\newblock {\em Phys. Rev. B} {\bf 2009}, {\em 79},~054507.
\newblock
  doi:{\changeurlcolor{black}\href{https://doi.org/10.1103/PhysRevB.79.054507}{\detokenize{10.1103/PhysRevB.79.054507}}}.

\bibitem[{Coffey} \em{et~al.}(2000){Coffey}, {D{\'e}jardin}, and
  {Kalmykov}]{Coffey00}
{Coffey}, W.T.; {D{\'e}jardin}, J.L.; {Kalmykov}, Y.P.
\newblock {Nonlinear impedance of a microwave-driven Josephson junction with
  noise}.
\newblock {\em Phys. Rev.~B} {\bf 2000}, {\em 62},~3480--3487.
\newblock
  doi:{\changeurlcolor{black}\href{https://doi.org/10.1103/PhysRevB.62.3480}{\detokenize{10.1103/PhysRevB.62.3480}}}.

\bibitem[{Gottardi} \em{et~al.}(2017){Gottardi}, {Akamatsu}, {van der Kuur},
  {Smith}, {Kozorezov}, and {Chervenak}]{Gottardi17}
{Gottardi}, L.; {Akamatsu}, H.; {van der Kuur}, J.; {Smith}, S.J.; {Kozorezov},
  A.; {Chervenak}, J.
\newblock Study of TES-Based Microcalorimeters of Different Size and Geometry
  Under AC Bias.
\newblock {\em IEEE Transactions on Applied Superconductivity} {\bf 2017}, {\em
  27},~1--4.
\newblock
  doi:{\changeurlcolor{black}\href{https://doi.org/10.1109/TASC.2017.2655500}{\detokenize{10.1109/TASC.2017.2655500}}}.

\bibitem[{McDonald} and {Clem}(1997)]{McDonald97}
{McDonald}, J.; {Clem}, J.R.
\newblock {Microwave response and surface impedance of weak links}.
\newblock {\em Phys. Rev.~B} {\bf 1997}, {\em 56},~14723--14732.
\newblock
  doi:{\changeurlcolor{black}\href{https://doi.org/10.1103/PhysRevB.56.14723}{\detokenize{10.1103/PhysRevB.56.14723}}}.

\bibitem[Kirsch \em{et~al.}(2020)Kirsch, Gottardi, Lorenz, Dauser, den Hartog,
  Jackson, Peille, Smith, and Wilms]{Kirsch2020}
Kirsch, C.; Gottardi, L.; Lorenz, M.; Dauser, T.; den Hartog, R.; Jackson, B.;
  Peille, P.; Smith, S.; Wilms, J.
\newblock Time-Domain Modeling of TES Microcalorimeters Under AC Bias.
\newblock {\em J. Low Temp. Phys.} {\bf 2020}, {\em 199},~569--576.

\bibitem[Wilms \em{et~al.}(2016)Wilms, Smith, Peille, Ceballos, Cobo, Dauser,
  Brand, den Hartog, Bandler, de~Plaa, and den Herder]{Wilms2016}
Wilms, J.; Smith, S.J.; Peille, P.; Ceballos, M.T.; Cobo, B.; Dauser, T.;
  Brand, T.; den Hartog, R.H.; Bandler, S.R.; de~Plaa, J.; den Herder, J.W.A.
\newblock {TESSIM: a simulator for the Athena-X-IFU}.
\newblock  Space Telescopes and Instrumentation 2016: Ultraviolet to Gamma Ray;
  den Herder, J.W.A.; Takahashi, T.; Bautz, M., Eds. International Society for
  Optics and Photonics, SPIE,  2016, Vol. 9905, pp. 1795 -- 1801.
\newblock
  doi:{\changeurlcolor{black}\href{https://doi.org/10.1117/12.2234435}{\detokenize{10.1117/12.2234435}}}.

\bibitem[{Lorenz} \em{et~al.}(2020){Lorenz}, {Kirsch}, {Merino-Alonso},
  {Peille}, {Dauser}, {Cucchetti}, {Smith}, and {Wilms}]{Lorenz2020}
{Lorenz}, M.; {Kirsch}, C.; {Merino-Alonso}, P.E.; {Peille}, P.; {Dauser}, T.;
  {Cucchetti}, E.; {Smith}, S.J.; {Wilms}, J.
\newblock {GPU Supported Simulation of Transition-Edge Sensor Arrays}.
\newblock {\em J. Low Temp. Phys.} {\bf 2020}, {\em 200},~277--285.
\newblock
  doi:{\changeurlcolor{black}\href{https://doi.org/10.1007/s10909-020-02450-1}{\detokenize{10.1007/s10909-020-02450-1}}}.

\bibitem[{Jaeckel} \em{et~al.}(2019){Jaeckel}, {Ambarish}, {Christensen},
  {Gruenke}, {Hu}, {McCammon}, {McPheron}, {Meyer}, {Nelms}, {Roy}, {Wulf},
  {Zhang}, {Zhou}, {Adams}, {Bandler}, {Chervenak}, {Datesman}, {Eckart},
  {Ewin}, {Finkbeiner}, {Kelley}, {Kilbourne}, {Miniussi}, {Porter}, {Sadleir},
  {Sakai}, {Smith}, {Wakeham}, {Wassell}, {Yoon}, {Morgan}, {Schmidt}, {Swetz},
  and {Ullom}]{Jaeckel2019}
{Jaeckel}, F.T.; {Ambarish}, C.V.; {Christensen}, N.; {Gruenke}, R.; {Hu}, L.;
  {McCammon}, D.; {McPheron}, M.; {Meyer}, M.; {Nelms}, K.L.; {Roy}, A.;
  {Wulf}, D.; {Zhang}, S.; {Zhou}, Y.; {Adams}, J.S.; {Bandler}, S.R.;
  {Chervenak}, J.A.; {Datesman}, A.M.; {Eckart}, M.E.; {Ewin}, A.J.;
  {Finkbeiner}, F.M.; {Kelley}, R.; {Kilbourne}, C.A.; {Miniussi}, A.R.;
  {Porter}, F.S.; {Sadleir}, J.E.; {Sakai}, K.; {Smith}, S.J.; {Wakeham}, N.;
  {Wassell}, E.; {Yoon}, W.; {Morgan}, K.M.; {Schmidt}, D.R.; {Swetz}, D.S.;
  {Ullom}, J.N.
\newblock Energy Calibration of High-Resolution X-Ray TES Microcalorimeters
  With 3 eV Optical Photons.
\newblock {\em IEEE Transactions on Applied Superconductivity} {\bf 2019}, {\em
  29},~1--4.
\newblock
  doi:{\changeurlcolor{black}\href{https://doi.org/10.1109/TASC.2019.2899856}{\detokenize{10.1109/TASC.2019.2899856}}}.

\bibitem[{Jaeckel} \em{et~al.}(2021){Jaeckel}, {Ambarish}, {Dai}, {Liu},
  {Mccammon}, {Mcpheron}, {Nelms}, {Roy}, {Stueber}, {Bandler}, {Chervenak},
  {Sakai}, and {Smith}]{Jaeckel2021}
{Jaeckel}, F.T.; {Ambarish}, C.V.; {Dai}, H.; {Liu}, S.; {Mccammon}, D.;
  {Mcpheron}, M.; {Nelms}, K.L.; {Roy}, A.; {Stueber}, H.R.; {Bandler}, S.R.;
  {Chervenak}, J.; {Sakai}, K.; {Smith}, S.J.
\newblock Calibration and Testing of Small High-Resolution Transition Edge
  Sensor Microcalorimeters with Optical Photons.
\newblock {\em IEEE Transactions on Applied Superconductivity} {\bf 2021}, pp.
  1--1.
\newblock
  doi:{\changeurlcolor{black}\href{https://doi.org/10.1109/TASC.2021.3053506}{\detokenize{10.1109/TASC.2021.3053506}}}.

\bibitem[Hoevers \em{et~al.}(2000)Hoevers, Bento, Bruijn, Gottardi, Korevaar,
  Mels, and de~Korte]{Hoevers00}
Hoevers, H.F.C.; Bento, A.C.; Bruijn, M.P.; Gottardi, L.; Korevaar, M.A.N.;
  Mels, W.A.; de~Korte, P.A.J.
\newblock Thermal fluctuation noise in a voltage biased superconducting
  transition edge thermometer.
\newblock {\em Appl. Phys. Lett.} {\bf 2000}, {\em 77},~4422--4424,
  \href{http://xxx.lanl.gov/abs/https://doi.org/10.1063/1.1336550}{{\normalfont
  [https://doi.org/10.1063/1.1336550]}}.
\newblock
  doi:{\changeurlcolor{black}\href{https://doi.org/10.1063/1.1336550}{\detokenize{10.1063/1.1336550}}}.

\bibitem[Gildemeister \em{et~al.}(2001)Gildemeister, Lee, and
  Richards]{Gildemeister01}
Gildemeister, J.M.; Lee, A.T.; Richards, P.L.
\newblock Model for excess noise in voltage-biased superconducting bolometers.
\newblock {\em Appl. Opt.} {\bf 2001}, {\em 40},~6229--6235.
\newblock
  doi:{\changeurlcolor{black}\href{https://doi.org/10.1364/AO.40.006229}{\detokenize{10.1364/AO.40.006229}}}.

\bibitem[{Khosropanah} \em{et~al.}(2012){Khosropanah}, {Hijmering}, {Ridder},
  {Lindeman}, {Gottardi}, {Bruijn}, {Kuur}, {Korte}, {Gao}, and
  {Hoevers}]{Pourya12}
{Khosropanah}, P.; {Hijmering}, R.A.; {Ridder}, M.; {Lindeman}, M.A.;
  {Gottardi}, L.; {Bruijn}, M.; {Kuur}, J.; {Korte}, P.A.J.; {Gao}, J.R.;
  {Hoevers}, H.
\newblock {Distributed TES Model for Designing Low Noise Bolometers Approaching
  SAFARI Instrument Requirements}.
\newblock {\em J. Low Temp. Phys.} {\bf 2012}, {\em 167},~188--194.
\newblock
  doi:{\changeurlcolor{black}\href{https://doi.org/10.1007/s10909-012-0550-6}{\detokenize{10.1007/s10909-012-0550-6}}}.

\bibitem[Ullom \em{et~al.}(2004)Ullom, Doriese, Hilton, Beall, Deiker, Irwin,
  Reintsema, Vale, and Xu]{Ullom04}
Ullom, J.; Doriese, W.; Hilton, G.; Beall, J.; Deiker, S.; Irwin, K.;
  Reintsema, C.; Vale, L.; Xu, Y.
\newblock Suppression of excess noise in Transition-Edge Sensors using magnetic
  field and geometry.
\newblock {\em Nucl. Instrum. Methods~A} {\bf 2004}, {\em 520},~333 -- 335.

\bibitem[Jethava \em{et~al.}(2009)Jethava, Ullom, Irwin, Doriese, Beall,
  Hilton, Vale, and Zink]{Jethava09}
Jethava, N.; Ullom, J.N.; Irwin, K.D.; Doriese, W.B.; Beall, J.A.; Hilton,
  G.C.; Vale, L.R.; Zink, B.
\newblock Dependence of Excess Noise on the Partial Derivatives of Resistance
  in Superconducting Transition Edge Sensors.
\newblock {\em AIP Conference Proceedings} {\bf 2009}, {\em 1185},~31--33.
\newblock
  doi:{\changeurlcolor{black}\href{https://doi.org/10.1063/1.3292343}{\detokenize{10.1063/1.3292343}}}.

\bibitem[Wakeham \em{et~al.}(2018)Wakeham, Adams, Bandler, Chervenak, Datesman,
  Eckart, Finkbeiner, Kelley, Kilbourne, Miniussi, Porter, Sadleir, Sakai,
  Smith, Wassell, and Yoon]{Wake2018}
Wakeham, N.A.; Adams, J.S.; Bandler, S.R.; Chervenak, J.A.; Datesman, A.M.;
  Eckart, M.E.; Finkbeiner, F.M.; Kelley, R.L.; Kilbourne, C.A.; Miniussi,
  A.R.; Porter, F.S.; Sadleir, J.E.; Sakai, K.; Smith, S.J.; Wassell, E.J.;
  Yoon, W.
\newblock Effects of Normal Metal Features on Superconducting Transition-Edge
  Sensors.
\newblock {\em J. Low Temp. Phys.} {\bf 2018}, {\em 193},~231--240.
\newblock
  doi:{\changeurlcolor{black}\href{https://doi.org/10.1007/s10909-018-1898-z}{\detokenize{10.1007/s10909-018-1898-z}}}.

\bibitem[Kozorezov \em{et~al.}(2012)Kozorezov, Golubov, Martin, de~Korte,
  Lindeman, Hijmering, van~der Kuur, Hoevers, Gottardi, Kupriyanov, and
  J.K.]{Kozo12}
Kozorezov, A.; Golubov, A.; Martin, D.; de~Korte, P.; Lindeman, M.; Hijmering,
  R.; van~der Kuur, J.; Hoevers, H.; Gottardi, L.; Kupriyanov, M.; J.K., W.
\newblock Electrical Noise in a TES as a Resistively Shunted Conducting
  Junction.
\newblock {\em J. Low Temp. Phys.} {\bf 2012}, {\em 167},~108--113.
\newblock
  doi:{\changeurlcolor{black}\href{https://doi.org/10.1007/s10909-012-0489-7}{\detokenize{10.1007/s10909-012-0489-7}}}.

\bibitem[Likharev and Semenov(1972)]{LikSem72}
Likharev, K.; Semenov, V.
\newblock Fluctuation Spectrum in Superconducting Point Junctions.
\newblock {\em Sov.Phys. JEPT Lett} {\bf 1972}, {\em 15},~442.

\bibitem[Koch \em{et~al.}(1980)Koch, Van~Harlingen, and Clarke]{Koch80}
Koch, R.H.; Van~Harlingen, D.J.; Clarke, J.
\newblock Quantum-Noise Theory for the Resistively Shunted Josephson Junction.
\newblock {\em Phys. Rev. Lett.} {\bf 1980}, {\em 45},~2132--2135.
\newblock
  doi:{\changeurlcolor{black}\href{https://doi.org/10.1103/PhysRevLett.45.2132}{\detokenize{10.1103/PhysRevLett.45.2132}}}.

\bibitem[Wessel \em{et~al.}()Wessel, Morgan, Becker, Gard, Hilton, Mates,
  Reintsema, Schmidt, Ullom, et~al.]{Wessel2019}
Wessel, A.; Morgan, K.; Becker, D.; Gard, J.; Hilton, G.; Mates, J.; Reintsema,
  C.; Schmidt, D.R.~Swetz, D.; Ullom, J.; others.
\newblock {\em arXiv:1907.11343}.

\bibitem[Kogan and Nagaev(1988)]{KogNag88}
Kogan, S.M.; Nagaev, K.
\newblock {\em Zh. Eksp. Teor. Fiz} {\bf 1988}, {\em 94},~262.

\bibitem[Gottardi \em{et~al.}(2021)Gottardi, de~Wit, Kozorezov, Taralli, and
  Nagayashi]{Gottardi2021}
Gottardi, L.; de~Wit, M.; Kozorezov, A.; Taralli, E.; Nagayashi, K.
\newblock Voltage fluctuations in ac biased superconducting transition edge
  sensors.
\newblock {\em Accepted for publication in Phys. Rev.Lett.} {\bf 2021}.

\bibitem[Andrews \em{et~al.}(1941)Andrews, Brucksch, Ziegler, and
  Blanchard]{Andrews41}
Andrews, D.H.; Brucksch, W.F.; Ziegler, W.T.; Blanchard, E.R.
\newblock {Superconducting Films as Radiometric Receivers}.
\newblock {\em Physical Review} {\bf 1941}, {\em 59},~1045--1046.
\newblock
  doi:{\changeurlcolor{black}\href{https://doi.org/10.1103/PhysRev.59.1045.2}{\detokenize{10.1103/PhysRev.59.1045.2}}}.

\bibitem[Gaskin \em{et~al.}(2019)Gaskin, Swartz, Vikhlinin, {\"O}zel, Gelmis,
  Arenberg, Bandler, Bautz, Civitani, Dominguez, Eckart, Falcone,
  Figueroa-Feliciano, Freeman, G{\"u}nther, Havey, Heilmann, Kilaru, Kraft,
  McCarley, McEntaffer, Pareschi, Purcell, Reid, Schattenburg, Schwartz,
  Schwartz, Tananbaum, Tremblay, Zhang, and Zuhone]{Gaskin19}
Gaskin, J.A.; Swartz, D.A.; Vikhlinin, A.; {\"O}zel, F.; Gelmis, K.E.;
  Arenberg, J.W.; Bandler, S.R.; Bautz, M.W.; Civitani, M.M.; Dominguez, A.;
  Eckart, M.E.; Falcone, A.D.; Figueroa-Feliciano, E.; Freeman, M.D.;
  G{\"u}nther, H.M.; Havey, K.A.; Heilmann, R.K.; Kilaru, K.; Kraft, R.P.;
  McCarley, K.S.; McEntaffer, R.L.; Pareschi, G.; Purcell, W.; Reid, P.B.;
  Schattenburg, M.L.; Schwartz, D.A.; Schwartz, E.D.; Tananbaum, H.D.;
  Tremblay, G.R.; Zhang, W.W.; Zuhone, J.A.
\newblock {Lynx X-Ray Observatory: an overview}.
\newblock {\em Journal of Astronomical Telescopes} {\bf 2019}, {\em 5},~021001.
\newblock
  doi:{\changeurlcolor{black}\href{https://doi.org/10.1117/1.JATIS.5.2.021001}{\detokenize{10.1117/1.JATIS.5.2.021001}}}.

\bibitem[{Simionescu} \em{et~al.}(2019){Simionescu}, {Ettori}, {Werner},
  {Nagai}, {Vazza}, {Akamatsu}, {Pinto}, {de Plaa}, {Wijers}, {Nelson},
  {Pointecouteau}, {Pratt}, {Spiga}, {Lau}, {Rossetti}, {Gastaldello}, {Biffi},
  {Bulbul}, {den Herder}, {Eckert}, {Fraternali}, {Mingo}, {Pareschi},
  {Pezzulli}, {Reiprich}, {Schaye}, {Walker}, and {Werk}]{Simionescu2019}
{Simionescu}, A.; {Ettori}, S.; {Werner}, N.; {Nagai}, D.; {Vazza}, F.;
  {Akamatsu}, H.; {Pinto}, C.; {de Plaa}, J.; {Wijers}, N.; {Nelson}, D.;
  {Pointecouteau}, E.; {Pratt}, G.W.; {Spiga}, D.; {Lau}, E.; {Rossetti}, M.;
  {Gastaldello}, F.; {Biffi}, V.; {Bulbul}, E.; {den Herder}, J.W.; {Eckert},
  D.; {Fraternali}, F.; {Mingo}, B.; {Pareschi}, G.; {Pezzulli}, G.;
  {Reiprich}, T.H.; {Schaye}, J.; {Walker}, S.A.; {Werk}, J.
\newblock {Voyage through the Hidden Physics of the Cosmic Web}.
\newblock {\em arXiv e-prints} {\bf 2019}, p. arXiv:1908.01778,
  \href{http://xxx.lanl.gov/abs/1908.01778}{{\normalfont
  [arXiv:astro-ph.CO/1908.01778]}}.

\bibitem[{Rothe} \em{et~al.}(2018){Rothe}, {Angloher}, {Bauer}, {Bento},
  {Bucci}, {Canonica}, {D'Addabbo}, {Defay}, {Erb}, {Feilitzsch}, {Ferreiro
  Iachellini}, {Gorla}, {G{\"u}tlein}, {Hauff}, {Jochum}, {Kiefer}, {Kluck},
  {Kraus}, {Lanfranchi}, {Langenk{\"a}mper}, {Loebell}, {Mancuso}, {Mondragon},
  {M{\"u}nster}, {Pagliarone}, {Petricca}, {Potzel}, {Pr{\"o}bst}, {Puig},
  {Reindl}, {Sch{\"a}ffner}, {Schieck}, {Schipperges}, {Sch{\"o}nert},
  {Seidel}, {Stahlberg}, {Stodolsky}, {Strandhagen}, {Strauss}, {Tanzke},
  {Trinh Thi}, {T{\"u}rko{\u{g}}lu}, {Ulrich}, {Usherov}, {Wawoczny},
  {Willers}, and {W{\"u}strich}]{W_CRESST2018}
{Rothe}, J.; {Angloher}, G.; {Bauer}, P.; {Bento}, A.; {Bucci}, C.; {Canonica},
  L.; {D'Addabbo}, A.; {Defay}, X.; {Erb}, A.; {Feilitzsch}, F.v.; {Ferreiro
  Iachellini}, N.; {Gorla}, P.; {G{\"u}tlein}, A.; {Hauff}, D.; {Jochum}, J.;
  {Kiefer}, M.; {Kluck}, H.; {Kraus}, H.; {Lanfranchi}, J.C.;
  {Langenk{\"a}mper}, A.; {Loebell}, J.; {Mancuso}, M.; {Mondragon}, E.;
  {M{\"u}nster}, A.; {Pagliarone}, C.; {Petricca}, F.; {Potzel}, W.;
  {Pr{\"o}bst}, F.; {Puig}, R.; {Reindl}, F.; {Sch{\"a}ffner}, K.; {Schieck},
  J.; {Schipperges}, V.; {Sch{\"o}nert}, S.; {Seidel}, W.; {Stahlberg}, M.;
  {Stodolsky}, L.; {Strandhagen}, C.; {Strauss}, R.; {Tanzke}, A.; {Trinh Thi},
  H.H.; {T{\"u}rko{\u{g}}lu}, C.; {Ulrich}, A.; {Usherov}, I.; {Wawoczny}, S.;
  {Willers}, M.; {W{\"u}strich}, M.
\newblock {TES-Based Light Detectors for the CRESST Direct Dark Matter Search}.
\newblock {\em J. Low Temp. Phys.} {\bf 2018}, {\em 193},~1160--1166.
\newblock
  doi:{\changeurlcolor{black}\href{https://doi.org/10.1007/s10909-018-1944-x}{\detokenize{10.1007/s10909-018-1944-x}}}.

\bibitem[{Lita} \em{et~al.}(2005){Lita}, {Rosenberg}, {Nam}, {Miller},
  {Balzar}, {Kaatz}, and {Schwall}]{W_Lita05}
{Lita}, A.E.; {Rosenberg}, D.; {Nam}, S.; {Miller}, A.J.; {Balzar}, D.;
  {Kaatz}, L.M.; {Schwall}, R.E.
\newblock Tuning of tungsten thin film superconducting transition temperature
  for fabrication of photon number resolving detectors.
\newblock {\em IEEE Transactions on Applied Superconductivity} {\bf 2005}, {\em
  15},~3528--3531.
\newblock
  doi:{\changeurlcolor{black}\href{https://doi.org/10.1109/TASC.2005.849033}{\detokenize{10.1109/TASC.2005.849033}}}.

\bibitem[{Hubmayr} \em{et~al.}(2011){Hubmayr}, {Austermann}, {Beall}, {Becker},
  {Bennett}, {Benson}, {Bleem}, {Chang}, {Carlstrom}, {Cho}, {Crites}, {Dobbs},
  {Everett}, {George}, {Holzapfel}, {Halverson}, {Henning}, {Hilton}, {Irwin},
  {Li}, {Lowell}, {Lueker}, {McMahon}, {Mehl}, {Meyer}, {Nibarger}, {Niemack},
  {Schmidt}, {Shirokoff}, {Simon}, {Yoon}, and {Young}]{AlMn2011}
{Hubmayr}, J.; {Austermann}, J.E.; {Beall}, J.A.; {Becker}, D.; {Bennett},
  D.A.; {Benson}, B.A.; {Bleem}, L.E.; {Chang}, C.L.; {Carlstrom}, J.E.; {Cho},
  H..; {Crites}, A.T.; {Dobbs}, M.; {Everett}, W.; {George}, E.M.; {Holzapfel},
  W.L.; {Halverson}, N.W.; {Henning}, J.W.; {Hilton}, G.C.; {Irwin}, K.D.;
  {Li}, D.; {Lowell}, P.; {Lueker}, M.; {McMahon}, J.; {Mehl}, J.; {Meyer},
  S.S.; {Nibarger}, J.P.; {Niemack}, M.D.; {Schmidt}, D.R.; {Shirokoff}, E.;
  {Simon}, S.M.; {Yoon}, K.W.; {Young}, E.Y.
\newblock Stability of Al-Mn Transition Edge Sensors for Frequency Domain
  Multiplexing.
\newblock {\em IEEE Transactions on Applied Superconductivity} {\bf 2011}, {\em
  21},~203--206.
\newblock
  doi:{\changeurlcolor{black}\href{https://doi.org/10.1109/TASC.2010.2090630}{\detokenize{10.1109/TASC.2010.2090630}}}.

\bibitem[{Nagayoshi} \em{et~al.}(2020){Nagayoshi}, {Ridder}, {Bruijn},
  {Gottardi}, {Taralli}, {Khosropanah}, {Akamatsu}, {Visser}, and
  {Gao}]{Nagayoshi19}
{Nagayoshi}, K.; {Ridder}, M.L.; {Bruijn}, M.P.; {Gottardi}, L.; {Taralli}, E.;
  {Khosropanah}, P.; {Akamatsu}, H.; {Visser}, S.; {Gao}, J.R.
\newblock {Development of a Ti/Au TES Microcalorimeter Array as a Backup Sensor
  for the Athena/X-IFU Instrument}.
\newblock {\em J. Low Temp. Phys.} {\bf 2020}, {\em 199},~943--948.
\newblock
  doi:{\changeurlcolor{black}\href{https://doi.org/10.1007/s10909-019-02282-8}{\detokenize{10.1007/s10909-019-02282-8}}}.

\bibitem[{Finkbeiner} \em{et~al.}(2017){Finkbeiner}, {Adams}, {Bandler},
  {Betancourt-Martinez}, {Brown}, {Chang}, {Chervenak}, {Chiao}, {Datesman},
  {Eckart}, {Kelley}, {Kilbourne}, {Miniussi}, {Moseley}, {Porter}, {Sadleir},
  {Sakai}, {Smith}, {Wakeham}, {Wassell}, and {Yoon}]{Fink2017}
{Finkbeiner}, F.M.; {Adams}, J.S.; {Bandler}, S.R.; {Betancourt-Martinez},
  G.L.; {Brown}, A.D.; {Chang}, M.P.; {Chervenak}, J.A.; {Chiao}, M.P.;
  {Datesman}, A.M.; {Eckart}, M.E.; {Kelley}, R.L.; {Kilbourne}, C.A.;
  {Miniussi}, A.R.; {Moseley}, S.J.; {Porter}, F.S.; {Sadleir}, J.E.; {Sakai},
  K.; {Smith}, S.J.; {Wakeham}, N.A.; {Wassell}, E.J.; {Yoon}, W.
\newblock {Electron-Beam Deposition of Superconducting Molybdenum Thin Films
  for the Development of Mo/Au TES X-ray Microcalorimeter}.
\newblock {\em IEEE Transactions on Applied Superconductivity} {\bf 2017}, {\em
  27},~2633785.
\newblock
  doi:{\changeurlcolor{black}\href{https://doi.org/10.1109/TASC.2016.2633785}{\detokenize{10.1109/TASC.2016.2633785}}}.

\bibitem[{Parra-Borderias} \em{et~al.}(2013){Parra-Borderias},
  {Fernandez-Martinez}, {Fabrega}, {Camon}, {Gil}, {Costa-Kramer},
  {Gonzalez-Arrabal}, {Sese}, {Bueno}, and {Briones}]{Parra2013}
{Parra-Borderias}, M.; {Fernandez-Martinez}, I.; {Fabrega}, L.; {Camon}, A.;
  {Gil}, O.; {Costa-Kramer}, J.L.; {Gonzalez-Arrabal}, R.; {Sese}, J.; {Bueno},
  J.; {Briones}, F.
\newblock {Characterization of a Mo/Au Thermometer for ATHENA}.
\newblock {\em IEEE Transactions on Applied Superconductivity} {\bf 2013}, {\em
  23},~2300405--2300405.
\newblock
  doi:{\changeurlcolor{black}\href{https://doi.org/10.1109/TASC.2012.2236140}{\detokenize{10.1109/TASC.2012.2236140}}}.

\bibitem[{Fabrega} \em{et~al.}(2009){Fabrega}, {Fernandez-Martinez}, {Gil},
  {Parra-Borderias}, {Camon}, {Costa-Kramer}, {Gonzalez-Arrabal}, {Sese},
  {Briones}, {Santiso}, and {Peiro}]{Mo_Fabrega2009}
{Fabrega}, L.; {Fernandez-Martinez}, I.; {Gil}, O.; {Parra-Borderias}, M.;
  {Camon}, A.; {Costa-Kramer}, J.; {Gonzalez-Arrabal}, R.; {Sese}, J.;
  {Briones}, F.; {Santiso}, J.; {Peiro}, F.
\newblock Mo-Based Proximity Bilayers for TES: Microstructure and Properties.
\newblock {\em IEEE Transactions on Applied Superconductivity} {\bf 2009}, {\em
  19},~460--464.
\newblock
  doi:{\changeurlcolor{black}\href{https://doi.org/10.1109/TASC.2009.2019052}{\detokenize{10.1109/TASC.2009.2019052}}}.

\bibitem[{Chervenak} \em{et~al.}(2008){Chervenak}, {Finkbeiner}, {Bandler},
  {Brekosky}, {Brown}, {Iyomoto}, {Kelley}, {Kilbourne}, {Porter}, {Sadleir},
  and {Smith}]{Cherv2008}
{Chervenak}, J.A.; {Finkbeiner}, F.M.; {Bandler}, S.R.; {Brekosky}, R.;
  {Brown}, A.D.; {Iyomoto}, N.; {Kelley}, R.L.; {Kilbourne}, C.A.; {Porter},
  F.S.; {Sadleir}, J.; {Smith}, S.
\newblock {Materials Development for Auxiliary Components for Large Compact
  Mo/Au TES Arrays}.
\newblock {\em J. Low Temp. Phys.} {\bf 2008}, {\em 151},~255--260.
\newblock
  doi:{\changeurlcolor{black}\href{https://doi.org/10.1007/s10909-007-9636-y}{\detokenize{10.1007/s10909-007-9636-y}}}.

\bibitem[Orlando \em{et~al.}(2018)Orlando, Ceriale, Ceruti, De~Gerone,
  Faverzani, Ferri, Gallucci, Giachero, Nucciotti, Puiu, Schmidt, Swetz, and
  Ullom]{Orlando18}
Orlando, A.; Ceriale, V.; Ceruti, G.; De~Gerone, M.; Faverzani, M.; Ferri, E.;
  Gallucci, G.; Giachero, A.; Nucciotti, A.; Puiu, A.; Schmidt, D.; Swetz, D.;
  Ullom, J.
\newblock Microfabrication of Transition-Edge Sensor Arrays of
  Microcalorimeters with 163Ho for Direct Neutrino Mass Measurements with
  HOLMES.
\newblock {\em Journal of Low Temperature Physics} {\bf 2018}, {\em 193}.
\newblock
  doi:{\changeurlcolor{black}\href{https://doi.org/10.1007/s10909-018-1968-2}{\detokenize{10.1007/s10909-018-1968-2}}}.

\bibitem[Kunieda \em{et~al.}(2006)Kunieda, Takahashi, Zen, Damayanthi, Mori,
  Fujita, Nakazawa, Fukuda, and Ohkubo]{IrAu_Kunieda2006}
Kunieda, Y.; Takahashi, H.; Zen, N.; Damayanthi, R.; Mori, F.; Fujita, K.;
  Nakazawa, M.; Fukuda, D.; Ohkubo, M.
\newblock Characterization of Ir/Au pixel TES.
\newblock {\em Nuclear Instruments and Methods in Physics Research Section A:
  Accelerators, Spectrometers, Detectors and Associated Equipment} {\bf 2006},
  {\em 559},~429--431.
\newblock Proceedings of the 11th International Workshop on Low Temperature
  Detectors,
  doi:{\changeurlcolor{black}\href{https://doi.org/https://doi.org/10.1016/j.nima.2005.12.029}{\detokenize{https://doi.org/10.1016/j.nima.2005.12.029}}}.

\bibitem[Lolli \em{et~al.}(2016)Lolli, Taralli, Portesi, Rajteri, and
  Monticone]{AlTi_Lolli2016}
Lolli, L.; Taralli, E.; Portesi, C.; Rajteri, M.; Monticone, E.
\newblock Aluminum–Titanium Bilayer for Near-Infrared Transition Edge
  Sensors.
\newblock {\em Sensors} {\bf 2016}, {\em 16}.
\newblock
  doi:{\changeurlcolor{black}\href{https://doi.org/10.3390/s16070953}{\detokenize{10.3390/s16070953}}}.

\bibitem[{Ahmad} \em{et~al.}(2019){Ahmad}, {Liu}, {Liu}, {Li}, {Liu}, and
  {Chen}]{AlTi_Kamal2019}
{Ahmad}, K.; {Liu}, J.; {Liu}, Q.; {Li}, G.; {Liu}, J.; {Chen}, W.
\newblock {Fabrication and Characterization of Superconducting Bilayer (Al/Ti)
  Transition-Edge Sensor Bolometer Array}.
\newblock {\em Journal of Electronic Materials} {\bf 2019}, {\em 48},~925--929.
\newblock
  doi:{\changeurlcolor{black}\href{https://doi.org/10.1007/s11664-018-6806-4}{\detokenize{10.1007/s11664-018-6806-4}}}.

\bibitem[Huang \em{et~al.}(2020)Huang, Zhong, Zhang, Koza, Smith, and
  Simmons]{Huang2020}
Huang, R.; Zhong, X.F.; Zhang, B.; Koza, J.; Smith, S.; Simmons, S.
\newblock {Super planarizing material for trench and via arrays}.
\newblock  Advances in Patterning Materials and Processes XXXVII; Gronheid, R.;
  Sanders, D.P., Eds. International Society for Optics and Photonics, SPIE,
  2020, Vol. 11326, pp. 317 -- 323.
\newblock
  doi:{\changeurlcolor{black}\href{https://doi.org/10.1117/12.2552129}{\detokenize{10.1117/12.2552129}}}.

\bibitem[{Devasia} \em{et~al.}(2019){Devasia}, {Balvin}, {Bandler},
  {Bolkhovsky}, {Nagler}, {Ryu}, {Smith}, {Stevenson}, and {Yoon}]{Devasia2019}
{Devasia}, A.M.; {Balvin}, M.A.; {Bandler}, S.R.; {Bolkhovsky}, V.; {Nagler},
  P.C.; {Ryu}, K.; {Smith}, S.J.; {Stevenson}, T.R.; {Yoon}, W.
\newblock Fabrication of Magnetic Calorimeter Arrays With Buried Wiring.
\newblock {\em IEEE Transactions on Applied Superconductivity} {\bf 2019}, {\em
  29},~1--6.
\newblock
  doi:{\changeurlcolor{black}\href{https://doi.org/10.1109/TASC.2019.2902530}{\detokenize{10.1109/TASC.2019.2902530}}}.

\bibitem[Yohannes \em{et~al.}(2015)Yohannes, Hunt, Vivalda, Amparo, Cohen,
  Vernik, and Kirichenko]{Yohannes15}
Yohannes, D.T.; Hunt, R.T.; Vivalda, J.A.; Amparo, D.; Cohen, A.; Vernik, I.V.;
  Kirichenko, A.F.
\newblock {Planarized, Extendible, Multilayer Fabrication Process for
  Superconducting Electronics}.
\newblock {\em IEEE Transactions on Applied Superconductivity} {\bf 2015}, {\em
  25},~2365562.
\newblock
  doi:{\changeurlcolor{black}\href{https://doi.org/10.1109/TASC.2014.2365562}{\detokenize{10.1109/TASC.2014.2365562}}}.

\bibitem[{Ridder} \em{et~al.}(2016){Ridder}, {Khosropanah}, {Hijmering},
  {Suzuki}, {Bruijn}, {Hoevers}, {Gao}, and {Zuiddam}]{Ridder2016}
{Ridder}, M.L.; {Khosropanah}, P.; {Hijmering}, R.A.; {Suzuki}, T.; {Bruijn},
  M.P.; {Hoevers}, H.F.C.; {Gao}, J.R.; {Zuiddam}, M.R.
\newblock {Fabrication of Low-Noise TES Arrays for the SAFARI Instrument on
  SPICA}.
\newblock {\em J. Low Temp. Phys.} {\bf 2016}, {\em 184},~60--65.
\newblock
  doi:{\changeurlcolor{black}\href{https://doi.org/10.1007/s10909-015-1381-z}{\detokenize{10.1007/s10909-015-1381-z}}}.

\bibitem[Bandler \em{et~al.}(2019)Bandler, Chervenak, Datesman, Devasia,
  DiPirro, Sakai, Smith, Stevenson, Yoon, Bennett, Mates, Swetz, Ullom, Irwin,
  Eckart, Figueroa-Feliciano, McCammon, Ryu, Olson, and Zeiger]{Bandler2019}
Bandler, S.R.; Chervenak, J.A.; Datesman, A.M.; Devasia, A.M.; DiPirro, M.J.;
  Sakai, K.; Smith, S.J.; Stevenson, T.R.; Yoon, W.; Bennett, D.A.; Mates, B.;
  Swetz, D.S.; Ullom, J.N.; Irwin, K.D.; Eckart, M.E.; Figueroa-Feliciano, E.;
  McCammon, D.; Ryu, K.K.; Olson, J.R.; Zeiger, B.
\newblock {Lynx x-ray microcalorimeter}.
\newblock {\em Journal of Astronomical Telescopes, Instruments, and Systems}
  {\bf 2019}, {\em 5},~1 -- 29.
\newblock
  doi:{\changeurlcolor{black}\href{https://doi.org/10.1117/1.JATIS.5.2.021017}{\detokenize{10.1117/1.JATIS.5.2.021017}}}.

\bibitem[{Miniussi} \em{et~al.}(2020){Miniussi}, {Adams}, {Bandler},
  {Beaumont}, {Chang}, {Chervenak}, {Finkbeiner}, {Ha}, {Hummatov}, {Kelley},
  {Kilbourne}, {Porter}, {Sadleir}, {Sakai}, {Smith}, {Wakeham}, and
  {Wassell}]{Miniussi2020}
{Miniussi}, A.R.; {Adams}, J.S.; {Bandler}, S.R.; {Beaumont}, S.; {Chang},
  M.P.; {Chervenak}, J.A.; {Finkbeiner}, F.M.; {Ha}, J.Y.; {Hummatov}, R.;
  {Kelley}, R.L.; {Kilbourne}, C.A.; {Porter}, F.S.; {Sadleir}, J.E.; {Sakai},
  K.; {Smith}, S.J.; {Wakeham}, N.A.; {Wassell}, E.J.
\newblock {Thermal Crosstalk Measurements and Simulations for an X-ray
  Microcalorimeter Array}.
\newblock {\em J. Low Temp. Phys.} {\bf 2020}, {\em 199},~663--671.
\newblock
  doi:{\changeurlcolor{black}\href{https://doi.org/10.1007/s10909-019-02312-5}{\detokenize{10.1007/s10909-019-02312-5}}}.

\bibitem[Brown \em{et~al.}(2008)Brown, Bandler, Brekosky, Chervenak,
  Figueroa-Feliciano, Finkbeiner, Iyomoto, Kelley, Kilbourne, Porter, Smith,
  Saab, and Sadleir]{Brown08}
Brown, A.D.; Bandler, S.R.; Brekosky, R.; Chervenak, J.A.; Figueroa-Feliciano,
  E.; Finkbeiner, F.; Iyomoto, N.; Kelley, R.L.; Kilbourne, C.A.; Porter, F.S.;
  Smith, S.; Saab, T.; Sadleir, J.
\newblock {Absorber Materials for Transition-Edge Sensor X-ray
  Microcalorimeters}.
\newblock {\em J Low Temp Phys} {\bf 2008}, {\em 151},~413--417.
\newblock
  doi:{\changeurlcolor{black}\href{https://doi.org/10.1007/s10909-007-9669-2}{\detokenize{10.1007/s10909-007-9669-2}}}.

\bibitem[Yan \em{et~al.}(2017)Yan, Divan, Gades, Kenesei, Madden, Miceli, Park,
  Patel, Quaranta, Sharma, Bennett, Doriese, Fowler, Gard, Hays-Wehle, Morgan,
  Schmidt, Swetz, and Ullom]{Yan17}
Yan, D.; Divan, R.; Gades, L.M.; Kenesei, P.; Madden, T.J.; Miceli, A.; Park,
  J.S.; Patel, U.M.; Quaranta, O.; Sharma, H.; Bennett, D.A.; Doriese, W.B.;
  Fowler, J.W.; Gard, J.D.; Hays-Wehle, J.P.; Morgan, K.M.; Schmidt, D.R.;
  Swetz, D.S.; Ullom, J.N.
\newblock {Eliminating the non-Gaussian spectral response of X-ray absorbers
  for transition-edge sensors}.
\newblock {\em Applied Physics Letters} {\bf 2017}, {\em 111},~192602.
\newblock
  doi:{\changeurlcolor{black}\href{https://doi.org/10.1063/1.5001198}{\detokenize{10.1063/1.5001198}}}.

\bibitem[{Hummatov} \em{et~al.}(2020){Hummatov}, {Adams}, {Bandler}, {Barlis},
  {Beaumont}, {Chang}, {Chervenak}, {Datesman}, {Eckart}, {Finkbeiner}, {Ha},
  {Kelley}, {Kilbourne}, {Miniussi}, {Porter}, {Sadleir}, {Sakai}, {Smith},
  {Wakeham}, {Wassell}, and {Wollack}]{Hummatov2020}
{Hummatov}, R.; {Adams}, J.S.; {Bandler}, S.R.; {Barlis}, A.; {Beaumont}, S.;
  {Chang}, M.P.; {Chervenak}, J.A.; {Datesman}, A.M.; {Eckart}, M.E.;
  {Finkbeiner}, F.M.; {Ha}, J.Y.; {Kelley}, R.L.; {Kilbourne}, C.A.;
  {Miniussi}, A.R.; {Porter}, F.S.; {Sadleir}, J.E.; {Sakai}, K.; {Smith},
  S.J.; {Wakeham}, N.; {Wassell}, E.J.; {Wollack}, E.J.
\newblock {Quantum Efficiency Study and Reflectivity Enhancement of Au/Bi
  Absorbers}.
\newblock {\em J. Low Temp. Phys.} {\bf 2020}, {\em 199},~393--400.
\newblock
  doi:{\changeurlcolor{black}\href{https://doi.org/10.1007/s10909-020-02424-3}{\detokenize{10.1007/s10909-020-02424-3}}}.

\bibitem[Gottardi \em{et~al.}(2012)Gottardi, Adams, Bailey, Bandler, Bruijn,
  Chervenak, Eckart, Finkbeiner, den Hartog, Hoevers, Kelley, Kilbourne,
  de~Korte, van~der Kuur, Lindeman, Porter, Sadlier, and Smith]{Gottardi12xray}
Gottardi, L.; Adams, J.; Bailey, C.; Bandler, S.; Bruijn, M.; Chervenak, J.;
  Eckart, M.; Finkbeiner, F.; den Hartog, R.; Hoevers, H.; Kelley, R.;
  Kilbourne, C.; de~Korte, P.; van~der Kuur, J.; Lindeman, M.; Porter, F.;
  Sadlier, J.; Smith, S.
\newblock Study of the Dependency on Magnetic Field and Bias Voltage of an
  AC-Biased {TES} Microcalorimeter.
\newblock {\em J. Low Temp. Phys.} {\bf 2012}, {\em 167},~214--219,
  \href{http://xxx.lanl.gov/abs/1604.02595}{{\normalfont [1604.02595]}}.
\newblock
  doi:{\changeurlcolor{black}\href{https://doi.org/10.1007/s10909-012-0494-x}{\detokenize{10.1007/s10909-012-0494-x}}}.

\bibitem[{Taralli} \em{et~al.}(2020){Taralli}, {Pobes}, {Khosropanah},
  {Fabrega}, {Cam{\'o}n}, {Gottardi}, {Nagayoshi}, {Ridder}, {Bruijn}, and
  {Gao}]{Taralli2020}
{Taralli}, E.; {Pobes}, C.; {Khosropanah}, P.; {Fabrega}, L.; {Cam{\'o}n}, A.;
  {Gottardi}, L.; {Nagayoshi}, K.; {Ridder}, M.L.; {Bruijn}, M.P.; {Gao}, J.R.
\newblock {AC/DC Characterization of a Ti/Au TES with Au/Bi Absorber for X-ray
  Detection}.
\newblock {\em J. Low Temp. Phys.} {\bf 2020}, {\em 199},~102--109.
\newblock
  doi:{\changeurlcolor{black}\href{https://doi.org/10.1007/s10909-020-02390-w}{\detokenize{10.1007/s10909-020-02390-w}}}.

\bibitem[Lindeman \em{et~al.}(2004)Lindeman, Bandler, Brekosky, Chervenak,
  Figueroa-Feliciano, Finkbeiner, Saab, and Stahle]{Lind_NIMA_2004}
Lindeman, M.A.; Bandler, S.; Brekosky, R.P.; Chervenak, J.A.;
  Figueroa-Feliciano, E.; Finkbeiner, F.M.; Saab, T.; Stahle, C.K.
\newblock Characterization and reduction of noise in Mo/Au transition edge
  sensors.
\newblock {\em Nuclear Instruments and Methods in Physics Research Section A:
  Accelerators, Spectrometers, Detectors and Associated Equipment} {\bf 2004},
  {\em 520},~348 -- 350.
\newblock Proceedings of the 10th International Workshop on Low Temperature
  Detectors,
  doi:{\changeurlcolor{black}\href{https://doi.org/https://doi.org/10.1016/j.nima.2003.11.264}{\detokenize{https://doi.org/10.1016/j.nima.2003.11.264}}}.

\bibitem[{Smith} \em{et~al.}(2014){Smith}, {Adams}, {Bandler}, {Busch},
  {Chervenak}, {Eckart}, {Finkbeiner}, {Kelley}, {Kilbourne}, {Lee}, {Porst},
  {Porter}, and {Sadleir}]{Smith2014}
{Smith}, S.J.; {Adams}, J.S.; {Bandler}, S.R.; {Busch}, S.E.; {Chervenak},
  J.A.; {Eckart}, M.E.; {Finkbeiner}, F.M.; {Kelley}, R.L.; {Kilbourne}, C.A.;
  {Lee}, S.J.; {Porst}, J.P.; {Porter}, F.S.; {Sadleir}, J.E.
\newblock {Characterization of Mo/Au Transition-Edge Sensors with Different
  Geometric Configurations}.
\newblock {\em J. Low Temp. Phys.} {\bf 2014}, {\em 176},~356--362.
\newblock
  doi:{\changeurlcolor{black}\href{https://doi.org/10.1007/s10909-013-1031-2}{\detokenize{10.1007/s10909-013-1031-2}}}.

\bibitem[Zhang \em{et~al.}(2017)Zhang, Eckart, Jaeckel, Kripps, McCammon,
  Morgan, and Zhou]{Zhang2017}
Zhang, S.; Eckart, M.E.; Jaeckel, F.T.; Kripps, K.L.; McCammon, D.; Morgan,
  K.M.; Zhou, Y.
\newblock Mapping of the resistance of a superconducting transition edge sensor
  as a function of temperature, current, and applied magnetic field.
\newblock {\em J. Appl. Phys.} {\bf 2017}, {\em 121},~074503.
\newblock
  doi:{\changeurlcolor{black}\href{https://doi.org/10.1063/1.4976562}{\detokenize{10.1063/1.4976562}}}.

\bibitem[Swetz \em{et~al.}(2012)Swetz, Bennett, Irwin, Schmidt, and
  Ullom]{Swetz12}
Swetz, D.; Bennett, D.; Irwin, K.; Schmidt, D.; Ullom, J.
\newblock Current distribution and transition width in superconducting
  transition-edge sensors.
\newblock {\em Appl. Phys. Lett.} {\bf 2012}, {\em 101},~242603,
  \href{http://xxx.lanl.gov/abs/1212.3537}{{\normalfont [1212.3537]}}.
\newblock
  doi:{\changeurlcolor{black}\href{https://doi.org/10.1063/1.4771984}{\detokenize{10.1063/1.4771984}}}.

\bibitem[Miniussi \em{et~al.}(2018)Miniussi, Adams, Bandler, Chervenak,
  Datesman, Eckart, Ewin, Finkbeiner, Kelley, Kilbourne, Porter, Sadleir,
  Sakai, Smith, Wakeham, Wassell, and Yoon]{Miniussi2018}
Miniussi, A.R.; Adams, J.S.; Bandler, S.R.; Chervenak, J.A.; Datesman, A.M.;
  Eckart, M.E.; Ewin, A.J.; Finkbeiner, F.M.; Kelley, R.L.; Kilbourne, C.A.;
  Porter, F.S.; Sadleir, J.E.; Sakai, K.; Smith, S.J.; Wakeham, N.A.; Wassell,
  E.J.; Yoon, W.
\newblock Performance of an X-ray Microcalorimeter with a 240 $\mu$m Absorber
  and a 50 $\mu$m TES Bilayer.
\newblock {\em J. Low Temp. Phys.} {\bf 2018}, {\em 193},~337--343.

\bibitem[{Yoon} \em{et~al.}(2017){Yoon}, {Adams}, {Bandler},
  {Betancourt-Martinez}, {Chiao}, {Chang}, {Chervenak}, {Datesman}, {Eckart},
  {Ewin}, {Finkbeiner}, {Ha}, {Kelley}, {Kilbourne}, {Miniussi}, {Porter},
  {Sadleir}, {Sakai}, {Smith}, {Wakeham}, and {Wassell}]{Yoon2017}
{Yoon}, W.; {Adams}, J.S.; {Bandler}, S.R.; {Betancourt-Martinez}, G.L.;
  {Chiao}, M.P.; {Chang}, M.; {Chervenak}, J.A.; {Datesman}, A.; {Eckart},
  M.E.; {Ewin}, A.J.; {Finkbeiner}, F.M.; {Ha}, J.Y.; {Kelley}, R.;
  {Kilbourne}, C.A.; {Miniussi}, A.R.; {Porter}, F.S.; {Sadleir}, J.E.;
  {Sakai}, K.; {Smith}, S.J.; {Wakeham}, N.A.; {Wassell}, E.
\newblock Design and Performance of Hybrid Arrays of Mo/Au Bilayer
  Transition-Edge Sensors.
\newblock {\em IEEE Transactions on Applied Superconductivity} {\bf 2017}, {\em
  27},~1--5.
\newblock
  doi:{\changeurlcolor{black}\href{https://doi.org/10.1109/TASC.2017.2655718}{\detokenize{10.1109/TASC.2017.2655718}}}.

\bibitem[Smith \em{et~al.}(2020)Smith, Adams, Bandler, Beaumont, Chervenak,
  Datesman, Finkbeiner, Hummatov, Kelly, Kilbourne, Miniussi, Porter, Sadleir,
  Sakai, Wakeham, Wassell, Witthoeft, and Ryu]{Smith2020}
Smith, S.J.; Adams, J.S.; Bandler, S.R.; Beaumont, S.; Chervenak, J.A.;
  Datesman, A.M.; Finkbeiner, F.M.; Hummatov, R.; Kelly, R.L.; Kilbourne, C.A.;
  Miniussi, A.R.; Porter, F.S.; Sadleir, J.E.; Sakai, K.; Wakeham, N.A.;
  Wassell, E.J.; Witthoeft, M.C.; Ryu, K.
\newblock Toward 100,000-Pixel Microcalorimeter Arrays Using Multi-absorber
  Transition-Edge Sensors.
\newblock {\em J. Low Temp. Phys.} {\bf 2020}, {\em 199},~330--338,
  \href{http://xxx.lanl.gov/abs/1908.02687}{{\normalfont
  [arXiv:astro-ph.IM/1908.02687]}}.
\newblock
  doi:{\changeurlcolor{black}\href{https://doi.org/10.1007/s10909-020-02362-0}{\detokenize{10.1007/s10909-020-02362-0}}}.

\bibitem[Sakai \em{et~al.}(2020)Sakai, Adams, Bandler, Beaumont, Chervenak,
  Datesman, Finkbeiner, Kelley, Kilbourne, Miniussi, Porter, Sadleir, Smith,
  Wakeham, Wassell, Jaeckel, McCammon, Eckart, and Ryu]{Sakai2020}
Sakai, K.; Adams, J.S.; Bandler, S.R.; Beaumont, S.; Chervenak, J.A.; Datesman,
  A.M.; Finkbeiner, F.M.; Kelley, R.L.; Kilbourne, C.A.; Miniussi, A.R.;
  Porter, F.S.; Sadleir, J.E.; Smith, S.J.; Wakeham, N.A.; Wassell, E.J.;
  Jaeckel, F.T.; McCammon, D.; Eckart, M.E.; Ryu, K.
\newblock Demonstration of Fine-Pitch High-Resolution X-ray Transition-Edge
  Sensor Microcalorimeters Optimized for Energies below 1 keV.
\newblock {\em J. Low Temp. Phys.} {\bf 2020}, {\em 199},~949--954,
  \href{http://xxx.lanl.gov/abs/1908.02687}{{\normalfont
  [arXiv:astro-ph.IM/1908.02687]}}.
\newblock
  doi:{\changeurlcolor{black}\href{https://doi.org/10.1007/s10909-020-02362-0}{\detokenize{10.1007/s10909-020-02362-0}}}.

\bibitem[{Sakai} \em{et~al.}(2018){Sakai}, {Adams}, {Bandler}, {Chervenak},
  {Datesman}, {Eckart}, {Finkbeiner}, {Kelley}, {Kilbourne}, {Miniussi},
  {Porter}, {Sadleir}, {Smith}, {Wakeham}, {Wassell}, {Yoon}, {Akamatsu},
  {Bruijn}, {Gottardi}, {Jackson}, {van der Kuur}, {van Leeuwen}, {van der
  Linden}, {van Weers}, and {Kiviranta}]{Sakai2018}
{Sakai}, K.; {Adams}, J.S.; {Bandler}, S.R.; {Chervenak}, J.A.; {Datesman},
  A.M.; {Eckart}, M.E.; {Finkbeiner}, F.M.; {Kelley}, R.L.; {Kilbourne}, C.A.;
  {Miniussi}, A.R.; {Porter}, F.S.; {Sadleir}, J.S.; {Smith}, S.J.; {Wakeham},
  N.A.; {Wassell}, E.J.; {Yoon}, W.; {Akamatsu}, H.; {Bruijn}, M.P.;
  {Gottardi}, L.; {Jackson}, B.D.; {van der Kuur}, J.; {van Leeuwen}, B.J.;
  {van der Linden}, A.J.; {van Weers}, H.J.; {Kiviranta}, M.
\newblock {Study of Dissipative Losses in AC-Biased Mo/Au Bilayer
  Transition-Edge Sensors}.
\newblock {\em J. Low Temp. Phys.} {\bf 2018}, {\em 193},~356--364.
\newblock
  doi:{\changeurlcolor{black}\href{https://doi.org/10.1007/s10909-018-2002-4}{\detokenize{10.1007/s10909-018-2002-4}}}.

\bibitem[{Akamatsu} \em{et~al.}(2014){Akamatsu}, {Gottardi}, {Adams},
  {Bandler}, {Bruijn}, {Chervenak}, {Eckart}, {Finkbeiner}, {den Hartog},
  {Hoevers}, {Kelley}, {Kilbourne}, {van der Kuur}, {van den Linden}, {Porter},
  {Sadleir}, {Smith}, and {Kiviranta}]{Akamatsu2014}
{Akamatsu}, H.; {Gottardi}, L.; {Adams}, J.; {Bandler}, S.; {Bruijn}, M.;
  {Chervenak}, J.; {Eckart}, M.; {Finkbeiner}, F.; {den Hartog}, R.; {Hoevers},
  H.; {Kelley}, R.; {Kilbourne}, C.; {van der Kuur}, J.; {van den Linden},
  A.J.; {Porter}, F.; {Sadleir}, J.; {Smith}, S.; {Kiviranta}, M.
\newblock {Performance of TES X-ray Microcalorimeters with AC Bias Read-Out at
  MHz Frequencies}.
\newblock {\em Journal of Low Temperature Physics} {\bf 2014}, {\em
  176},~591--596.
\newblock
  doi:{\changeurlcolor{black}\href{https://doi.org/10.1007/s10909-014-1130-8}{\detokenize{10.1007/s10909-014-1130-8}}}.

\bibitem[{Gottardi} \em{et~al.}(2018){Gottardi}, {Smith}, {Kozorezov},
  {Akamatsu}, {van der Kuur}, {Bandler}, {Bruijn}, {Chervenak}, {Gao}, {den
  Hartog}, {Jackson}, {Khosropanah}, {Miniussi}, {Nagayoshi}, {Ridder},
  {Sadleir}, {Sakai}, and {Wakeham}]{Gottardi18}
{Gottardi}, L.; {Smith}, S.J.; {Kozorezov}, A.; {Akamatsu}, H.; {van der Kuur},
  J.; {Bandler}, S.R.; {Bruijn}, M.P.; {Chervenak}, J.A.; {Gao}, J.R.; {den
  Hartog}, R.H.; {Jackson}, B.D.; {Khosropanah}, P.; {Miniussi}, A.;
  {Nagayoshi}, K.; {Ridder}, M.; {Sadleir}, J.; {Sakai}, K.; {Wakeham}, N.
\newblock {Josephson Effects in Frequency-Domain Multiplexed TES
  Microcalorimeters and Bolometers}.
\newblock {\em J. Low Temp. Phys.} {\bf 2018}, {\em 193},~209--216.
\newblock
  doi:{\changeurlcolor{black}\href{https://doi.org/10.1007/s10909-018-2006-0}{\detokenize{10.1007/s10909-018-2006-0}}}.

\bibitem[{Taralli} \em{et~al.}(2020){Taralli}, {Gottardi}, {Nagayoshi},
  {Ridder}, {Visser}, {Khosropanah}, {Akamatsu}, {van der Kuur}, {Bruijn}, and
  {Gao}]{Taralli2019}
{Taralli}, E.; {Gottardi}, L.; {Nagayoshi}, K.; {Ridder}, M.; {Visser}, S.;
  {Khosropanah}, P.; {Akamatsu}, H.; {van der Kuur}, J.; {Bruijn}, M.; {Gao},
  J.R.
\newblock {Characterization of High Aspect-Ratio TiAu TES X-ray
  Microcalorimeter Array Under AC Bias}.
\newblock {\em J. Low Temp. Phys.} {\bf 2020}, {\em 199},~80--87.
\newblock
  doi:{\changeurlcolor{black}\href{https://doi.org/10.1007/s10909-019-02254-y}{\detokenize{10.1007/s10909-019-02254-y}}}.

\bibitem[Taralli \em{et~al.}(2021)Taralli, D’Andrea, Gottardi, Nagayoshi,
  Ridder, de~Wit, Vaccaro, Akamatsu, Bruijn, and Gao]{Taralli2021}
Taralli, E.; D’Andrea, M.; Gottardi, L.; Nagayoshi, K.; Ridder, M.L.; de~Wit,
  M.; Vaccaro, D.; Akamatsu, H.; Bruijn, M.P.; Gao, J.R.
\newblock Performance and uniformity of a kilo-pixel array of Ti/Au
  transition-edge sensor microcalorimeters.
\newblock {\em Review of Scientific Instruments} {\bf 2021}, {\em 92},~023101,
  \href{http://xxx.lanl.gov/abs/https://doi.org/10.1063/5.0027750}{{\normalfont
  [https://doi.org/10.1063/5.0027750]}}.
\newblock
  doi:{\changeurlcolor{black}\href{https://doi.org/10.1063/5.0027750}{\detokenize{10.1063/5.0027750}}}.

\bibitem[{D'Andrea} \em{et~al.}(2021){D'Andrea}, {Taralli}, {Akamatsu},
  {Gottardi}, {Nagayoshi}, {Ravensberg}, {Ridder}, {Vaccaro}, {de Vries}, {de
  Wit}, {Bruijn}, {Hoogeveen}, and {Gao}]{DAndrea2021}
{D'Andrea}, M.; {Taralli}, E.; {Akamatsu}, H.; {Gottardi}, L.; {Nagayoshi}, K.;
  {Ravensberg}, K.; {Ridder}, M.L.; {Vaccaro}, D.; {de Vries}, C.P.; {de Wit},
  M.; {Bruijn}, M.P.; {Hoogeveen}, R.W.M.; {Gao}, J.R.
\newblock {Single pixel performance of a 32$\times$32 Ti/Au TES array with
  broadband X-ray spectra}.
\newblock {\em arXiv e-prints} {\bf 2021}, p. arXiv:2102.08103,
  \href{http://xxx.lanl.gov/abs/2102.08103}{{\normalfont
  [arXiv:astro-ph.IM/2102.08103]}}.

\bibitem[Hoevers \em{et~al.}(2005)Hoevers, Ridder, Germeau, Bruijn, de~Korte,
  and Wiegerink]{Hoevers2005}
Hoevers, H.F.C.; Ridder, M.L.; Germeau, A.; Bruijn, M.P.; de~Korte, P.A.J.;
  Wiegerink, R.J.
\newblock Radiative ballistic phonon transport in silicon-nitride membranes at
  low temperatures.
\newblock {\em Applied Physics Letters} {\bf 2005}, {\em 86},~251903,
  \href{http://xxx.lanl.gov/abs/https://doi.org/10.1063/1.1949269}{{\normalfont
  [https://doi.org/10.1063/1.1949269]}}.
\newblock
  doi:{\changeurlcolor{black}\href{https://doi.org/10.1063/1.1949269}{\detokenize{10.1063/1.1949269}}}.

\bibitem[{Hays-Wehle} \em{et~al.}(2016){Hays-Wehle}, {Schmidt}, {Ullom}, and
  {Swetz}]{Hays2016}
{Hays-Wehle}, J.P.; {Schmidt}, D.R.; {Ullom}, J.N.; {Swetz}, D.S.
\newblock {Thermal Conductance Engineering for High-Speed TES
  Microcalorimeters}.
\newblock {\em J. Low Temp. Phys.} {\bf 2016}, {\em 184},~492--497.
\newblock
  doi:{\changeurlcolor{black}\href{https://doi.org/10.1007/s10909-015-1416-5}{\detokenize{10.1007/s10909-015-1416-5}}}.

\bibitem[{Doriese} \em{et~al.}(2019){Doriese}, {Bandler}, {Chaudhuri},
  {Dawson}, {Denison}, {Duff}, {Durkin}, {FitzGerald}, {Fowler}, {Gard},
  {Hilton}, {Irwin}, {Joe}, {Morgan}, {O'Neil}, {Pappas}, {Reintsema},
  {Rudman}, {Smith}, {Stevens}, {Swetz}, {Szypryt}, {Ullom}, {Vale}, {Weber},
  and {Young}]{Doriese2019}
{Doriese}, W.B.; {Bandler}, S.R.; {Chaudhuri}, S.; {Dawson}, C.S.; {Denison},
  E.V.; {Duff}, S.M.; {Durkin}, M.; {FitzGerald}, C.T.; {Fowler}, J.W.; {Gard},
  J.D.; {Hilton}, G.C.; {Irwin}, K.D.; {Joe}, Y.I.; {Morgan}, K.M.; {O'Neil},
  G.C.; {Pappas}, C.G.; {Reintsema}, C.D.; {Rudman}, D.A.; {Smith}, S.J.;
  {Stevens}, R.W.; {Swetz}, D.S.; {Szypryt}, P.; {Ullom}, J.N.; {Vale}, L.R.;
  {Weber}, J.C.; {Young}, B.A.
\newblock Optimization of Time- and Code-Division-Multiplexed Readout for
  Athena X-IFU.
\newblock {\em IEEE Transactions on Applied Superconductivity} {\bf 2019}, {\em
  29},~1--5.
\newblock
  doi:{\changeurlcolor{black}\href{https://doi.org/10.1109/TASC.2019.2905577}{\detokenize{10.1109/TASC.2019.2905577}}}.

\bibitem[{Vaccaro} \em{et~al.}(2021){Vaccaro}, {Akamatsu}, {van der Kuur}, {van
  der Hulst}, {Nieuwenhuizen}, {van Winden}, {Gottardi}, {den Hartog},
  {Bruijn}, {D'Andrea}, {Gao}, {den Herder}, {Hoogeveen}, {Jackson}, {van der
  Linden}, {Nagayoshi}, {Ravensberg}, {Ridder}, {Taralli}, and {de
  Wit}]{Vaccaro2021}
{Vaccaro}, D.; {Akamatsu}, H.; {van der Kuur}, J.; {van der Hulst}, P.;
  {Nieuwenhuizen}, A.C.T.; {van Winden}, P.; {Gottardi}, L.; {den Hartog}, R.;
  {Bruijn}, M.P.; {D'Andrea}, M.; {Gao}, J.R.; {den Herder}, J.W.A.;
  {Hoogeveen}, R.W.M.; {Jackson}, B.; {van der Linden}, A.J.; {Nagayoshi}, K.;
  {Ravensberg}, K.; {Ridder}, M.L.; {Taralli}, E.; {de Wit}, M.
\newblock {Frequency Shift Algorithm: Application to a Frequency-Domain
  Multiplexing Readout of X-ray Transition-Edge Sensor Microcalorimeters}.
\newblock {\em arXiv e-prints} {\bf 2021}, p. arXiv:2102.06092,
  \href{http://xxx.lanl.gov/abs/2102.06092}{{\normalfont
  [arXiv:astro-ph.IM/2102.06092]}}.

\bibitem[{Akamatsu} \em{et~al.}(2020){Akamatsu}, {Gottardi}, {van der Kuur},
  {de Vries}, {Bruijn}, {Chervenak}, {Kiviranta}, {van den Linden}, {Jackson},
  {Miniussi}, {Ravensberg}, {Sakai}, {Smith}, and {Wakeham}]{Akamatsu2020}
{Akamatsu}, H.; {Gottardi}, L.; {van der Kuur}, J.; {de Vries}, C.P.; {Bruijn},
  M.P.; {Chervenak}, J.A.; {Kiviranta}, M.; {van den Linden}, A.J.; {Jackson},
  B.D.; {Miniussi}, A.; {Ravensberg}, K.; {Sakai}, K.; {Smith}, S.J.;
  {Wakeham}, N.
\newblock {Progress in the Development of Frequency-Domain Multiplexing for the
  X-ray Integral Field Unit on Board the Athena Mission}.
\newblock {\em J. Low Temp. Phys.} {\bf 2020}, {\em 199},~737--744,
  \href{http://xxx.lanl.gov/abs/2003.11899}{{\normalfont
  [arXiv:astro-ph.IM/2003.11899]}}.
\newblock
  doi:{\changeurlcolor{black}\href{https://doi.org/10.1007/s10909-020-02351-3}{\detokenize{10.1007/s10909-020-02351-3}}}.

\bibitem[{Durkin} \em{et~al.}(2019){Durkin}, {Adams}, {Bandler}, {Chervenak},
  {Chaudhuri}, {Dawson}, {Denison}, {Doriese}, {Duff}, {Finkbeiner},
  {FitzGerald}, {Fowler}, {Gard}, {Hilton}, {Irwin}, {Joe}, {Kelley},
  {Kilbourne}, {Miniussi}, {Morgan}, {O'Neil}, {Pappas}, {Porter}, {Reintsema},
  {Rudman}, {Sakai}, {Smith}, {Stevens}, {Swetz}, {Szypryt}, {Ullom}, {Vale},
  {Wakeham}, {Weber}, and {Young}]{Durkin2019}
{Durkin}, M.; {Adams}, J.S.; {Bandler}, S.R.; {Chervenak}, J.A.; {Chaudhuri},
  S.; {Dawson}, C.S.; {Denison}, E.V.; {Doriese}, W.B.; {Duff}, S.M.;
  {Finkbeiner}, F.M.; {FitzGerald}, C.T.; {Fowler}, J.W.; {Gard}, J.D.;
  {Hilton}, G.C.; {Irwin}, K.D.; {Joe}, Y.I.; {Kelley}, R.L.; {Kilbourne},
  C.A.; {Miniussi}, A.R.; {Morgan}, K.M.; {O'Neil}, G.C.; {Pappas}, C.G.;
  {Porter}, F.S.; {Reintsema}, C.D.; {Rudman}, D.A.; {Sakai}, K.; {Smith},
  S.J.; {Stevens}, R.W.; {Swetz}, D.S.; {Szypryt}, P.; {Ullom}, J.N.; {Vale},
  L.R.; {Wakeham}, N.A.; {Weber}, J.C.; {Young}, B.A.
\newblock Demonstration of Athena X-IFU Compatible 40-Row
  Time-Division-Multiplexed Readout.
\newblock {\em IEEE Transactions on Applied Superconductivity} {\bf 2019}, {\em
  29},~1--5.
\newblock
  doi:{\changeurlcolor{black}\href{https://doi.org/10.1109/TASC.2019.2904472}{\detokenize{10.1109/TASC.2019.2904472}}}.

\bibitem[Bennett \em{et~al.}(2019)Bennett, Mates, Bandler, Becker, Fowler,
  Gard, Hilton, Irwin, Morgan, Reintsema, Sakai, Schmidt, Smith, Swetz, Ullom,
  Vale, and Wessels]{Bennett2019}
Bennett, D.A.; Mates, J.A.B.; Bandler, S.R.; Becker, D.T.; Fowler, J.W.; Gard,
  J.D.; Hilton, G.C.; Irwin, K.D.; Morgan, K.M.; Reintsema, C.D.; Sakai, K.;
  Schmidt, D.R.; Smith, S.J.; Swetz, D.S.; Ullom, J.N.; Vale, L.R.; Wessels,
  A.L.
\newblock {Microwave SQUID multiplexing for the Lynx x-ray microcalorimeter}.
\newblock {\em Journal of Astronomical Telescopes, Instruments, and Systems}
  {\bf 2019}, {\em 5},~1 -- 10.
\newblock
  doi:{\changeurlcolor{black}\href{https://doi.org/10.1117/1.JATIS.5.2.021007}{\detokenize{10.1117/1.JATIS.5.2.021007}}}.

\bibitem[Kempf \em{et~al.}(2017)Kempf, Wegner, Fleischmann, Gastaldo, Herrmann,
  Papst, Richter, and Enss]{Kempf2017}
Kempf, S.; Wegner, M.; Fleischmann, A.; Gastaldo, L.; Herrmann, F.; Papst, M.;
  Richter, D.; Enss, C.
\newblock Demonstration of a scalable frequency-domain readout of metallic
  magnetic calorimeters by means of a microwave SQUID multiplexer.
\newblock {\em AIP Advances} {\bf 2017}, {\em 7},~015007.
\newblock
  doi:{\changeurlcolor{black}\href{https://doi.org/10.1063/1.4973872}{\detokenize{10.1063/1.4973872}}}.

\bibitem[Irwin(2009)]{Irwin2009}
Irwin, K.D.
\newblock Shannon Limits for Low‐Temperature Detector Readout.
\newblock {\em AIP Conference Proceedings} {\bf 2009}, {\em 1185},~229--236,
  \href{http://xxx.lanl.gov/abs/https://aip.scitation.org/doi/pdf/10.1063/1.3292320}{{\normalfont
  [https://aip.scitation.org/doi/pdf/10.1063/1.3292320]}}.
\newblock
  doi:{\changeurlcolor{black}\href{https://doi.org/10.1063/1.3292320}{\detokenize{10.1063/1.3292320}}}.

\bibitem[Yu \em{et~al.}(2020)Yu, Ames, Chaudhuri, Dawson, Irwin, Kuenstner, Li,
  and Titus]{Yu_2020}
Yu, C.; Ames, A.; Chaudhuri, S.; Dawson, C.; Irwin, K.D.; Kuenstner, S.E.; Li,
  D.; Titus, C.J.
\newblock An impedance-modulated code-division microwave {SQUID} multiplexer.
\newblock {\em Engineering Research Express} {\bf 2020}, {\em 2},~015011.
\newblock
  doi:{\changeurlcolor{black}\href{https://doi.org/10.1088/2631-8695/ab68a4}{\detokenize{10.1088/2631-8695/ab68a4}}}.

\bibitem[Doriese \em{et~al.}(2017)Doriese, Abbamonte, Alpert, Bennett, Denison,
  Fang, Fischer, Fitzgerald, Fowler, Gard, Hays-Wehle, Hilton, Jaye, McChesney,
  Miaja-Avila, Morgan, Joe, O’Neil, Reintsema, Rodolakis, Schmidt, Tatsuno,
  Uhlig, Vale, Ullom, and Swetz]{Doriese2017}
Doriese, W.B.; Abbamonte, P.; Alpert, B.K.; Bennett, D.A.; Denison, E.V.; Fang,
  Y.; Fischer, D.A.; Fitzgerald, C.P.; Fowler, J.W.; Gard, J.D.; Hays-Wehle,
  J.P.; Hilton, G.C.; Jaye, C.; McChesney, J.L.; Miaja-Avila, L.; Morgan, K.M.;
  Joe, Y.I.; O’Neil, G.C.; Reintsema, C.D.; Rodolakis, F.; Schmidt, D.R.;
  Tatsuno, H.; Uhlig, J.; Vale, L.R.; Ullom, J.N.; Swetz, D.S.
\newblock A practical superconducting-microcalorimeter X-ray spectrometer for
  beamline and laboratory science.
\newblock {\em Review of Scientific Instruments} {\bf 2017}, {\em 88},~053108.
\newblock
  doi:{\changeurlcolor{black}\href{https://doi.org/10.1063/1.4983316}{\detokenize{10.1063/1.4983316}}}.

\bibitem[{Smith} \em{et~al.}(2021){Smith}, {Adams}, {Bandler}, {Beaumont},
  {Chervenak}, {Denison}, {Doriese}, {Durkin}, {Finkbeiner}, {Fowler},
  {Hilton}, {Hummatov}, {Irwin}, {Kelley}, {Kilbourne}, {Leutenegger},
  {Miniussi}, {Porter}, {Reintsema}, {Sadleir}, {Sakai}, {Swetz}, {Ullom},
  {Vale}, {Wakeham}, {Wassell}, and {Witthoeft}]{Smith2021}
{Smith}, S.J.; {Adams}, J.S.; {Bandler}, S.R.; {Beaumont}, S.; {Chervenak},
  J.A.; {Denison}, E.V.; {Doriese}, W.B.; {Durkin}, M.; {Finkbeiner}, F.M.;
  {Fowler}, J.W.; {Hilton}, G.C.; {Hummatov}, R.; {Irwin}, K.D.; {Kelley},
  R.L.; {Kilbourne}, C.A.; {Leutenegger}, M.A.; {Miniussi}, A.R.; {Porter},
  F.S.; {Reintsema}, C.D.; {Sadleir}, J.E.; {Sakai}, K.; {Swetz}, D.S.;
  {Ullom}, J.N.; {Vale}, L.R.; {Wakeham}, N.A.; {Wassell}, E.J.; {Witthoeft},
  M.C.
\newblock Performance of a Broad-Band, High-Resolution, Transition-Edge Sensor
  Spectrometer for X-ray Astrophysics.
\newblock {\em IEEE Transactions on Applied Superconductivity} {\bf 2021}, {\em
  31},~1--6.
\newblock
  doi:{\changeurlcolor{black}\href{https://doi.org/10.1109/TASC.2021.3061918}{\detokenize{10.1109/TASC.2021.3061918}}}.

\bibitem[Stiehl \em{et~al.}(2012)Stiehl, Doriese, Fowler, Hilton, Irwin,
  Reintsema, Schmidt, Swetz, Ullom, and Vale]{Stiehl2012}
Stiehl, G.M.; Doriese, W.B.; Fowler, J.W.; Hilton, G.C.; Irwin, K.D.;
  Reintsema, C.D.; Schmidt, D.R.; Swetz, D.S.; Ullom, J.N.; Vale, L.R.
\newblock Code-division multiplexing for x-ray microcalorimeters.
\newblock {\em Applied Physics Letters} {\bf 2012}, {\em 100},~072601.
\newblock
  doi:{\changeurlcolor{black}\href{https://doi.org/10.1063/1.3684807}{\detokenize{10.1063/1.3684807}}}.

\bibitem[Weber \em{et~al.}(2019)Weber, Fowler, Durkin, Morgan, Mates, Bennett,
  Doriese, Schmidt, Hilton, Swetz, and Ullom]{Weber2019}
Weber, J.C.; Fowler, J.W.; Durkin, M.; Morgan, K.M.; Mates, J.A.B.; Bennett,
  D.A.; Doriese, W.B.; Schmidt, D.R.; Hilton, G.C.; Swetz, D.S.; Ullom, J.N.
\newblock Configurable error correction of code-division multiplexed TES
  detectors with a cryotron switch.
\newblock {\em Applied Physics Letters} {\bf 2019}, {\em 114},~232602.
\newblock
  doi:{\changeurlcolor{black}\href{https://doi.org/10.1063/1.5089870}{\detokenize{10.1063/1.5089870}}}.

\bibitem[van~der Kuur \em{et~al.}(2002)van~der Kuur, de~Korte, Hoevers,
  Kiviranta, and Sepp\"a]{JvdK02}
van~der Kuur, J.; de~Korte, P.; Hoevers, H.; Kiviranta, M.; Sepp\"a, H.
\newblock Performance of an {X}-ray microcalorimeter under {AC} biasing.
\newblock {\em Appl. Phys. Lett.} {\bf 2002}, {\em 81},~4467--4469.

\bibitem[{Hattori} \em{et~al.}(2016){Hattori}, {Akiba}, {Arnold}, {Barron},
  {Bender}, {Cukierman}, {de Haan}, {Dobbs}, {Elleflot}, {Hasegawa}, {Hazumi},
  {Holzapfel}, {Hori}, {Keating}, {Kusaka}, {Lee}, {Montgomery}, {Rotermund},
  {Shirley}, {Suzuki}, and {Whitehorn}]{Hattori2016}
{Hattori}, K.; {Akiba}, Y.; {Arnold}, K.; {Barron}, D.; {Bender}, A.N.;
  {Cukierman}, A.; {de Haan}, T.; {Dobbs}, M.; {Elleflot}, T.; {Hasegawa}, M.;
  {Hazumi}, M.; {Holzapfel}, W.; {Hori}, Y.; {Keating}, B.; {Kusaka}, A.;
  {Lee}, A.; {Montgomery}, J.; {Rotermund}, K.; {Shirley}, I.; {Suzuki}, A.;
  {Whitehorn}, N.
\newblock {Development of Readout Electronics for POLARBEAR-2 Cosmic Microwave
  Background Experiment}.
\newblock {\em J. Low Temp. Phys.} {\bf 2016}, {\em 184},~512--518,
  \href{http://xxx.lanl.gov/abs/1512.07663}{{\normalfont
  [arXiv:astro-ph.IM/1512.07663]}}.
\newblock
  doi:{\changeurlcolor{black}\href{https://doi.org/10.1007/s10909-015-1448-x}{\detokenize{10.1007/s10909-015-1448-x}}}.

\bibitem[{Bender} \em{et~al.}(2020){Bender}, {Anderson}, {Avva}, {Ade},
  {Ahmed}, et~al.]{Bender2020}
{Bender}, A.N.; {Anderson}, A.J.; {Avva}, J.S.; {Ade}, P.A.R.; {Ahmed}, Z.;
  others.
\newblock {On-Sky Performance of the SPT-3G Frequency-Domain Multiplexed
  Readout}.
\newblock {\em J. Low Temp. Phys.} {\bf 2020}, {\em 199},~182--191,
  \href{http://xxx.lanl.gov/abs/1907.10947}{{\normalfont
  [arXiv:astro-ph.IM/1907.10947]}}.
\newblock
  doi:{\changeurlcolor{black}\href{https://doi.org/10.1007/s10909-019-02280-w}{\detokenize{10.1007/s10909-019-02280-w}}}.

\bibitem[{Bruijn} \em{et~al.}(2012){Bruijn}, {Gottardi}, {den Hartog},
  {Hoevers}, {Kiviranta}, {de Korte}, and {van der Kuur}]{Bruijn12}
{Bruijn}, M.P.; {Gottardi}, L.; {den Hartog}, R.H.; {Hoevers}, H.F.C.;
  {Kiviranta}, M.; {de Korte}, P.A.J.; {van der Kuur}, J.
\newblock {High-Q LC Filters for FDM Read out of Cryogenic Sensor Arrays}.
\newblock {\em J. Low Temp. Phys.} {\bf 2012}, {\em 167},~695--700.
\newblock
  doi:{\changeurlcolor{black}\href{https://doi.org/10.1007/s10909-011-0422-5}{\detokenize{10.1007/s10909-011-0422-5}}}.

\bibitem[{Rotermund} \em{et~al.}(2016){Rotermund}, {Barch}, {Chapman},
  {Hattori}, {Lee}, {Palaio}, {Shirley}, {Suzuki}, and {Tran}]{Rotermund2016}
{Rotermund}, K.; {Barch}, B.; {Chapman}, S.; {Hattori}, K.; {Lee}, A.;
  {Palaio}, N.; {Shirley}, I.; {Suzuki}, A.; {Tran}, C.
\newblock {Planar Lithographed Superconducting LC Resonators for
  Frequency-Domain Multiplexed Readout Systems}.
\newblock {\em J. Low Temp. Phys.} {\bf 2016}, {\em 184},~486--491.
\newblock
  doi:{\changeurlcolor{black}\href{https://doi.org/10.1007/s10909-016-1554-4}{\detokenize{10.1007/s10909-016-1554-4}}}.

\bibitem[{den Hartog} \em{et~al.}(2012){den Hartog}, {Audley}, {Beyer},
  {Boersma}, {Bruijn}, {Gottardi}, {Hoevers}, {Hou}, {Keizer}, {Khosropanah},
  {Kiviranta}, {de Korte}, {van der Kuur}, {van Leeuwen}, {Nieuwenhuizen}, and
  {van Winden}]{Hartog12}
{den Hartog}, R.; {Audley}, M.D.; {Beyer}, J.; {Boersma}, D.; {Bruijn}, M.;
  {Gottardi}, L.; {Hoevers}, H.; {Hou}, R.; {Keizer}, G.; {Khosropanah}, P.;
  {Kiviranta}, M.; {de Korte}, P.; {van der Kuur}, J.; {van Leeuwen}, B.J.;
  {Nieuwenhuizen}, A.C.T.; {van Winden}, P.
\newblock {Low-Noise Readout of TES Detectors with Baseband Feedback Frequency
  Domain Multiplexing}.
\newblock {\em J. Low Temp. Phys.} {\bf 2012}, {\em 167},~652--657.
\newblock
  doi:{\changeurlcolor{black}\href{https://doi.org/10.1007/s10909-012-0577-8}{\detokenize{10.1007/s10909-012-0577-8}}}.

\bibitem[den Hartog \em{et~al.}(2009)den Hartog, Boersma, Bruijn, Dirks,
  Gottardi, Hoevers, Hou, Kiviranta, de~Korte, van~der Kuur, van Leeuwen,
  Nieuwenhuizen, and Popescu]{Hartog2009}
den Hartog, R.; Boersma, D.; Bruijn, M.; Dirks, B.; Gottardi, L.; Hoevers, H.;
  Hou, R.; Kiviranta, M.; de~Korte, P.; van~der Kuur, J.; van Leeuwen, B.;
  Nieuwenhuizen, A.; Popescu, M.
\newblock Baseband Feedback for Frequency‐Domain‐Multiplexed Readout of TES
  X‐ray Detectors.
\newblock {\em AIP Conference Proceedings} {\bf 2009}, {\em 1185},~261--264.
\newblock
  doi:{\changeurlcolor{black}\href{https://doi.org/10.1063/1.3292328}{\detokenize{10.1063/1.3292328}}}.

\bibitem[de~Haan \em{et~al.}(2012)de~Haan, Smecher, and Dobbs]{deHaan2012}
de~Haan, T.; Smecher, G.; Dobbs, M.
\newblock {Improved performance of TES bolometers using digital feedback}.
\newblock  Millimeter, Submillimeter, and Far-Infrared Detectors and
  Instrumentation for Astronomy VI; Holland, W.S., Ed. International Society
  for Optics and Photonics, SPIE,  2012, Vol. 8452, pp. 113 -- 122.
\newblock
  doi:{\changeurlcolor{black}\href{https://doi.org/10.1117/12.925658}{\detokenize{10.1117/12.925658}}}.

\bibitem[{van der Kuur} \em{et~al.}(2017){van der Kuur}, {Gottardi},
  {Akamatsu}, {van Leeuwen}, {den Hartog}, {Kiviranta}, and
  {Jackson}]{vdKuur2017}
{van der Kuur}, J.; {Gottardi}, L.; {Akamatsu}, H.; {van Leeuwen}, B.J.; {den
  Hartog}, R.; {Kiviranta}, M.; {Jackson}, B.
\newblock Thermal Load Minimization of the Frequency Domain Multiplexed Readout
  for the Athenar X-IFU Instrument.
\newblock {\em IEEE Transactions on Applied Superconductivity} {\bf 2017}, {\em
  27},~1--4.
\newblock
  doi:{\changeurlcolor{black}\href{https://doi.org/10.1109/TASC.2017.2659658}{\detokenize{10.1109/TASC.2017.2659658}}}.

\bibitem[Kiviranta \em{et~al.}(2011)Kiviranta, Gr{\"o}nberg, and
  Sipola]{Kiviranta2011}
Kiviranta, M.; Gr{\"o}nberg, L.; Sipola, H.
\newblock Two-stage locally linearized SQUID readout for frequency domain
  multiplexed calorimeter arrays.
\newblock {\em Superconductor Science and Technology} {\bf 2011}, {\em 24}.
\newblock
  doi:{\changeurlcolor{black}\href{https://doi.org/10.1088/0953-2048/24/4/045003}{\detokenize{10.1088/0953-2048/24/4/045003}}}.

\bibitem[{van der Kuur} \em{et~al.}(2018){van der Kuur}, {Gottardi},
  {Akamatsu}, {Nieuwenhuizen}, {den Hartog}, and {Jackson}]{vdKuur2018}
{van der Kuur}, J.; {Gottardi}, L.; {Akamatsu}, H.; {Nieuwenhuizen}, A.C.T.;
  {den Hartog}, R.; {Jackson}, B.D.
\newblock {Active Tuning of the Resonance Frequencies of LC Bandpass Filters
  for Frequency Domain Multiplexed Readout of TES Detector Arrays}.
\newblock {\em J. Low Temp. Phys.} {\bf 2018}, {\em 193},~626--632.
\newblock
  doi:{\changeurlcolor{black}\href{https://doi.org/10.1007/s10909-018-2055-4}{\detokenize{10.1007/s10909-018-2055-4}}}.

\bibitem[Akamatsu \em{et~al.}(2020)Akamatsu, Vaccaro, Gottardi, van~der Kuur,
  de~Vries, Bandler, Bruijn, Chervenak, D'Andrea, Gao, den Herder, Hoogeveen,
  Kiviranta, van~der Linden, Jackson, Miniussi, Nagayoshi, Ravensberg,
  Ravensberg, Ridder, Sakai, Smith, Taralli, Visser, Wakeham, and
  de~Wit]{AkaSPIE2020}
Akamatsu, H.; Vaccaro, D.; Gottardi, L.; van~der Kuur, J.; de~Vries, C.P.;
  Bandler, S.R.; Bruijn, M.P.; Chervenak, J.A.; D'Andrea, M.; Gao, J.R.; den
  Herder, J.W.A.; Hoogeveen, R.W.M.; Kiviranta, M.; van~der Linden, A.J.;
  Jackson, B.D.; Miniussi, A.R.; Nagayoshi, K.; Ravensberg, K.; Ravensberg, K.;
  Ridder, M.L.; Sakai, K.; Smith, S.J.; Taralli, E.; Visser, S.; Wakeham, N.A.;
  de~Wit, M.
\newblock {Frequency domain multiplexing technology of transition-edge sensors
  for x-ray astronomy}.
\newblock  Space Telescopes and Instrumentation 2020: Ultraviolet to Gamma Ray;
  den Herder, J.W.A.; Nikzad, S.; Nakazawa, K., Eds. International Society for
  Optics and Photonics, SPIE,  2020, Vol. 11444.
\newblock
  doi:{\changeurlcolor{black}\href{https://doi.org/10.1117/12.2561271}{\detokenize{10.1117/12.2561271}}}.

\bibitem[Akamatsu(2021)]{Akamatsu2021}
Akamatsu, H.{\it et al.}.
\newblock {\em In preparation} {\bf 2021}.

\bibitem[Kiviranta(2020)]{Kiviranta2020classb}
Kiviranta, M.
\newblock Class-B cable-driving SQUID amplifier {\bf 2020}.
\newblock  \href{http://xxx.lanl.gov/abs/2011.11741}{{\normalfont
  [arXiv:astro-ph.IM/2011.11741]}}.

\bibitem[Kiviranta \em{et~al.}(2018)Kiviranta, Grönberg, and van~der
  Kuur]{kiviranta2018}
Kiviranta, M.; Grönberg, L.; van~der Kuur, J.
\newblock Two SQUID amplifiers intended to alleviate the summing node
  inductance problem in multiplexed arrays of Transition Edge Sensors {\bf
  2018}.
\newblock  \href{http://xxx.lanl.gov/abs/1810.09122}{{\normalfont
  [arXiv:astro-ph.IM/1810.09122]}}.

\bibitem[Irwin and Lehnert(2004)]{IrwLeh2004}
Irwin, K.D.; Lehnert, K.W.
\newblock Microwave SQUID multiplexer.
\newblock {\em Applied Physics Letters} {\bf 2004}, {\em 85},~2107--2109.
\newblock
  doi:{\changeurlcolor{black}\href{https://doi.org/10.1063/1.1791733}{\detokenize{10.1063/1.1791733}}}.

\bibitem[Mates \em{et~al.}(2008)Mates, Hilton, Irwin, Vale, and
  Lehnert]{Mates2008}
Mates, J.A.B.; Hilton, G.C.; Irwin, K.D.; Vale, L.R.; Lehnert, K.W.
\newblock Demonstration of a multiplexer of dissipationless superconducting
  quantum interference devices.
\newblock {\em Applied Physics Letters} {\bf 2008}, {\em 92},~023514.
\newblock
  doi:{\changeurlcolor{black}\href{https://doi.org/10.1063/1.2803852}{\detokenize{10.1063/1.2803852}}}.

\bibitem[Mates \em{et~al.}(2019)Mates, Becker, Bennett, Dober, Gard, Hilton,
  Swetz, Vale, and Ullom]{Mates2019}
Mates, J.A.B.; Becker, D.T.; Bennett, D.A.; Dober, B.J.; Gard, J.D.; Hilton,
  G.C.; Swetz, D.S.; Vale, L.R.; Ullom, J.N.
\newblock Crosstalk in microwave SQUID multiplexers.
\newblock {\em Applied Physics Letters} {\bf 2019}, {\em 115},~202601.
\newblock
  doi:{\changeurlcolor{black}\href{https://doi.org/10.1063/1.5116573}{\detokenize{10.1063/1.5116573}}}.

\bibitem[Richter \em{et~al.}(2021)Richter, Hoibl, Wolber, Karcher, Fleischmann,
  Enss, Weber, Sander, and Kempf]{Richter2021}
Richter, D.; Hoibl, L.; Wolber, T.; Karcher, N.; Fleischmann, A.; Enss, C.;
  Weber, M.; Sander, O.; Kempf, S.
\newblock Flux ramp modulation based MHz frequency-division dc-SQUID
  multiplexer,  2021,  \href{http://xxx.lanl.gov/abs/2101.06424}{{\normalfont
  [arXiv:physics.ins-det/2101.06424]}}.

\bibitem[Alpert \em{et~al.}(2019)Alpert, Becker, Bennet, Biasotti, Borghesi,
  Gallucci, De~Gerone, Faverzani, Ferri, Fowler, Gard, Giachero, Hays–Wehle,
  Hilton, Mates, Nucciotti, Orlando, Pessina, Puiu, Reintsema, Schmidt, Swetz,
  Ullom, and Vale]{Alpert2019}
Alpert, B.; Becker, D.; Bennet, D.; Biasotti, M.; Borghesi, M.; Gallucci, G.;
  De~Gerone, M.; Faverzani, M.; Ferri, E.; Fowler, J.; Gard, J.; Giachero, A.;
  Hays–Wehle, J.; Hilton, G.; Mates, J.; Nucciotti, A.; Orlando, A.; Pessina,
  G.; Puiu, A.; Reintsema, C.; Schmidt, D.; Swetz, D.; Ullom, J.; Vale, L.
\newblock High-resolution high-speed microwave-multiplexed low temperature
  microcalorimeters for the HOLMES experiment.
\newblock {\em The European Physical Journal C} {\bf 2019}, {\em 79},~304.
\newblock
  doi:{\changeurlcolor{black}\href{https://doi.org/10.1140/epjc/s10052-019-6814-4}{\detokenize{10.1140/epjc/s10052-019-6814-4}}}.

\bibitem[Nakashima \em{et~al.}(2020)Nakashima, Hirayama, Kohjiro, Yamamori,
  Nagasawa, Sato, Yamada, Hayakawa, Yamasaki, Mitsuda, Nagayoshi, Akamatsu,
  Gottardi, Taralli, Bruijn, Ridder, Gao, and den Herder]{Naka2020}
Nakashima, Y.; Hirayama, F.; Kohjiro, S.; Yamamori, H.; Nagasawa, S.; Sato, A.;
  Yamada, S.; Hayakawa, R.; Yamasaki, N.Y.; Mitsuda, K.; Nagayoshi, K.;
  Akamatsu, H.; Gottardi, L.; Taralli, E.; Bruijn, M.P.; Ridder, M.L.; Gao,
  J.R.; den Herder, J.W.A.
\newblock Low-noise microwave SQUID multiplexed readout of 38 x-ray
  transition-edge sensor microcalorimeters.
\newblock {\em Applied Physics Letters} {\bf 2020}, {\em 117},~122601.
\newblock
  doi:{\changeurlcolor{black}\href{https://doi.org/10.1063/5.0016333}{\detokenize{10.1063/5.0016333}}}.

\bibitem[{Irwin} \em{et~al.}(2018){Irwin}, {Chaudhuri}, {Cho}, {Dawson},
  {Kuenstner}, {Li}, {Titus}, and {Young}]{Irwin2018}
{Irwin}, K.D.; {Chaudhuri}, S.; {Cho}, H.M.; {Dawson}, C.; {Kuenstner}, S.;
  {Li}, D.; {Titus}, C.J.; {Young}, B.A.
\newblock {A Spread-Spectrum SQUID Multiplexer}.
\newblock {\em J. Low Temp. Phys.} {\bf 2018}, {\em 193},~476--484,
  \href{http://xxx.lanl.gov/abs/1806.02805}{{\normalfont
  [arXiv:physics.ins-det/1806.02805]}}.
\newblock
  doi:{\changeurlcolor{black}\href{https://doi.org/10.1007/s10909-018-1987-z}{\detokenize{10.1007/s10909-018-1987-z}}}.

\bibitem[Smith \em{et~al.}(2019)Smith, Adams, Bandler, Chervenak, Datesman,
  Eckart, Finkbeiner, Hummatov, Kelley, Kilbourne, Miniussi, Porter, Sadleir,
  Sakai, Wakeham, and Wassell]{Smith2019}
Smith, S.J.; Adams, J.S.; Bandler, S.R.; Chervenak, J.A.; Datesman, A.M.;
  Eckart, M.E.; Finkbeiner, F.M.; Hummatov, R.; Kelley, R.L.; Kilbourne, C.A.;
  Miniussi, A.R.; Porter, F.S.; Sadleir, J.E.; Sakai, K.; Wakeham, N.A.;
  Wassell, E.J.
\newblock {Multiabsorber transition-edge sensors for x-ray astronomy}.
\newblock {\em Journal of Astronomical Telescopes, Instruments, and Systems}
  {\bf 2019}, {\em 5},~1 -- 11.
\newblock
  doi:{\changeurlcolor{black}\href{https://doi.org/10.1117/1.JATIS.5.2.021008}{\detokenize{10.1117/1.JATIS.5.2.021008}}}.

\bibitem[{Jackson} \em{et~al.}(2016){Jackson}, {van Weers}, {van der Kuur},
  {den Hartog}, {Akamatsu}, {Argan}, {Bandler}, {Barbera}, {Barret}, {Bruijn},
  {Chervenak}, {Dercksen}, {Gatti}, {Gottardi}, {Haas}, {den Herder},
  {Kilbourne}, {Kiviranta}, {Lam-Trong}, {van Leeuwen}, {Macculi}, {Piro}, and
  {Smith}]{Jackson2016}
{Jackson}, B.D.; {van Weers}, H.; {van der Kuur}, J.; {den Hartog}, R.;
  {Akamatsu}, H.; {Argan}, A.; {Bandler}, S.R.; {Barbera}, M.; {Barret}, D.;
  {Bruijn}, M.P.; {Chervenak}, J.A.; {Dercksen}, J.; {Gatti}, F.; {Gottardi},
  L.; {Haas}, D.; {den Herder}, J.W.; {Kilbourne}, C.A.; {Kiviranta}, M.;
  {Lam-Trong}, T.; {van Leeuwen}, B.J.; {Macculi}, C.; {Piro}, L.; {Smith},
  S.J.
\newblock {The focal plane assembly for the Athena X-ray Integral Field Unit
  instrument}.
\newblock  Space Telescopes and Instrumentation 2016: Ultraviolet to Gamma Ray;
  {den Herder}, J.W.A.; {Takahashi}, T.; {Bautz}, M., Eds.,  2016, Vol. 9905,
  {\em Society of Photo-Optical Instrumentation Engineers (SPIE) Conference
  Series}, p. 99052I.
\newblock
  doi:{\changeurlcolor{black}\href{https://doi.org/10.1117/12.2232544}{\detokenize{10.1117/12.2232544}}}.

\bibitem[Geoffray \em{et~al.}(2020)Geoffray, Jackson, Bandler, Doriese,
  Kirivanta, Prêle, Ravera, Argan, Barbera, van~der Kuur, van Leeuwen, van
  Weers, Hoogeveen, den herder, Smith, Adams, Chervenak, Durkin, Reintsema,
  Ullom, Parot, Barret, Macculi, Piro, Brachet, and Ledot]{Geoffray2020}
Geoffray, H.; Jackson, B.; Bandler, S.; Doriese, W.B.; Kirivanta, M.; Prêle,
  D.; Ravera, L.; Argan, A.; Barbera, M.; van~der Kuur, J.; van Leeuwen, B.J.;
  van Weers, H.; Hoogeveen, R.; den herder, J.W.; Smith, S.; Adams, J.;
  Chervenak, J.; Durkin, M.; Reintsema, C.; Ullom, J.; Parot, Y.; Barret, D.;
  Macculi, C.; Piro, L.; Brachet, F.; Ledot, A.
\newblock {Conceptual design of the detection chain for the X-IFU on Athena}.
\newblock  Space Telescopes and Instrumentation 2020: Ultraviolet to Gamma Ray;
  den Herder, J.W.A.; Nikzad, S.; Nakazawa, K., Eds. International Society for
  Optics and Photonics, SPIE,  2020, Vol. 11444.
\newblock
  doi:{\changeurlcolor{black}\href{https://doi.org/10.1117/12.2563628}{\detokenize{10.1117/12.2563628}}}.

\bibitem[van Weers \em{et~al.}(2020)van Weers et~al.]{vanWeers2020}
van Weers, H.; others.
\newblock {The X-IFU focal plane assembly development model integration and
  first test results}.
\newblock  Space Telescopes and Instrumentation 2020: Ultraviolet to Gamma Ray;
  den Herder, J.W.A.; Nikzad, S.; Nakazawa, K., Eds. International Society for
  Optics and Photonics, SPIE,  2020, Vol. 11444.
\newblock
  doi:{\changeurlcolor{black}\href{https://doi.org/10.1117/12.2563464}{\detokenize{10.1117/12.2563464}}}.

\bibitem[Barbera \em{et~al.}(2018)Barbera, Cicero, Sciortino, D'Anca, Cicero,
  Parodi, Sciortino, Rauw, Branduardi-Raymont, Varisco, Bonura, Collura,
  Candia, Cicca, Giglio, Buttacavoli, Cuttaia, Villa, Cappi, Trong, Mesnager,
  Peille, den Hartog, den Herder, Jackson, Barret, and Piro]{Barbera2018}
Barbera, M.; Cicero, U.L.; Sciortino, L.; D'Anca, F.; Cicero, G.L.; Parodi, G.;
  Sciortino, S.; Rauw, G.; Branduardi-Raymont, G.; Varisco, S.; Bonura, S.F.;
  Collura, A.; Candia, R.; Cicca, G.D.; Giglio, P.; Buttacavoli, A.; Cuttaia,
  F.; Villa, F.; Cappi, M.; Trong, T.L.; Mesnager, J.M.; Peille, P.; den
  Hartog, R.; den Herder, J.W.; Jackson, B.; Barret, D.; Piro, L.
\newblock {ATHENA X-IFU thermal filters development status toward the end of
  the instrument phase-A}.
\newblock  Space Telescopes and Instrumentation 2018: Ultraviolet to Gamma Ray;
  den Herder, J.W.A.; Nikzad, S.; Nakazawa, K., Eds. International Society for
  Optics and Photonics, SPIE,  2018, Vol. 10699, pp. 406 -- 420.
\newblock
  doi:{\changeurlcolor{black}\href{https://doi.org/10.1117/12.2314450}{\detokenize{10.1117/12.2314450}}}.

\bibitem[Peille \em{et~al.}(2020)Peille, den Hartog, Miniussi, Stever, Bandler,
  Kirsch, Lorenz, Dauser, Wilms, Lotti, Gatti, Macculi, Jackson, and
  Pajot]{Peille2020}
Peille, P.; den Hartog, R.; Miniussi, A.; Stever, S.; Bandler, S.; Kirsch, C.;
  Lorenz, M.; Dauser, T.; Wilms, J.; Lotti, S.; Gatti, F.; Macculi, C.;
  Jackson, B.; Pajot, F.
\newblock {\em J. Low Temp. Phys.} {\bf 2020}, {\em 199},~240--249.

\bibitem[Lotti \em{et~al.}(2018)Lotti, Macculi, D'Andrea, Fioretti, Dondero,
  Mantero, Minervini, Argan, and Piro]{Lotti2018}
Lotti, S.; Macculi, C.; D'Andrea, M.; Fioretti, V.; Dondero, P.; Mantero, A.;
  Minervini, G.; Argan, A.; Piro, L.
\newblock {Estimates for the background of the ATHENA X-IFU instrument: the
  cosmic rays contribution}.
\newblock  Space Telescopes and Instrumentation 2018: Ultraviolet to Gamma Ray;
  den Herder, J.W.A.; Nikzad, S.; Nakazawa, K., Eds. International Society for
  Optics and Photonics, SPIE,  2018, Vol. 10699, pp. 397 -- 405.
\newblock
  doi:{\changeurlcolor{black}\href{https://doi.org/10.1117/12.2313236}{\detokenize{10.1117/12.2313236}}}.

\bibitem[Gottardi \em{et~al.}(2019)Gottardi, van Weers, Dercksen, Akamatsu,
  Bruijn, Gao, Jackson, Khosropanah, van~der Kuur, Ravensberg, and
  Ridder]{Gottardi2019}
Gottardi, L.; van Weers, H.; Dercksen, J.; Akamatsu, H.; Bruijn, M.P.; Gao,
  J.R.; Jackson, B.; Khosropanah, P.; van~der Kuur, J.; Ravensberg, K.; Ridder,
  M.L.
\newblock A six-degree-of-freedom micro-vibration acoustic isolator for
  low-temperature radiation detectors based on superconducting transition-edge
  sensors.
\newblock {\em Review of Scientific Instruments} {\bf 2019}, {\em 90},~055107.
\newblock
  doi:{\changeurlcolor{black}\href{https://doi.org/10.1063/1.5088364}{\detokenize{10.1063/1.5088364}}}.

\bibitem[Miniussi \em{et~al.}(2020)Miniussi, Adams, Bandler, Beaumont, Chang,
  Chervenak, Finkbeiner, Ha, Hummatov, Kelley, Kilbourne, Porter, Sadleir,
  Sakai, Smith, Wakeham, and Wassell]{Miniussi_cosmicray2020}
Miniussi, A.R.; Adams, J.S.; Bandler, S.R.; Beaumont, S.; Chang, M.P.;
  Chervenak, J.A.; Finkbeiner, F.M.; Ha, J.Y.; Hummatov, R.; Kelley, R.L.;
  Kilbourne, C.A.; Porter, F.S.; Sadleir, J.E.; Sakai, K.; Smith, S.J.;
  Wakeham, N.A.; Wassell, E.J.
\newblock Thermal Impact of Cosmic Ray Interaction with an X-Ray
  Microcalorimeter Array.
\newblock {\em J. Low Temp. Phys.} {\bf 2020}, {\em 199},~45--55.
\newblock
  doi:{\changeurlcolor{black}\href{https://doi.org/10.1007/s10909-020-02337-1}{\detokenize{10.1007/s10909-020-02337-1}}}.

\bibitem[{Lotti} \em{et~al.}(2021){Lotti}, {D'Andrea}, {Molendi}, {Macculi},
  {Minervini}, {Fioretti}, {Laurenza}, {Jacquey}, and {Piro}]{Lotti2021}
{Lotti}, S.; {D'Andrea}, M.; {Molendi}, S.; {Macculi}, C.; {Minervini}, G.;
  {Fioretti}, V.; {Laurenza}, M.; {Jacquey}, C.; {Piro}, L.
\newblock {Review of the Particle Background of the Athena X-IFU Instrument}.
\newblock {\em The Astrophysical Journal} {\bf 2021}, {\em 909},~111,
  \href{http://xxx.lanl.gov/abs/2101.02526}{{\normalfont
  [arXiv:astro-ph.IM/2101.02526]}}.
\newblock
  doi:{\changeurlcolor{black}\href{https://doi.org/10.3847/1538-4357/abd94c}{\detokenize{10.3847/1538-4357/abd94c}}}.

\bibitem[Macculi \em{et~al.}(2020)Macculi, Argan, Brienza, D'Andrea, Lotti,
  Minervini, Piro, Biasotti, Barusso, Gatti, Rigano, Chiarello, Torrioli,
  Fiorini, Uslenghi, Cavazzuti, Puccetti, and Volpe]{Macculi2020}
Macculi, C.; Argan, A.; Brienza, D.; D'Andrea, M.; Lotti, S.; Minervini, G.;
  Piro, L.; Biasotti, M.; Barusso, L.F.; Gatti, F.; Rigano, M.; Chiarello, F.;
  Torrioli, G.; Fiorini, M.; Uslenghi, M.; Cavazzuti, E.; Puccetti, S.; Volpe,
  A.
\newblock {The cryogenic anticoincidence detector for ATHENA X-IFU: advancement
  in the project}.
\newblock  Space Telescopes and Instrumentation 2020: Ultraviolet to Gamma Ray;
  den Herder, J.W.A.; Nikzad, S.; Nakazawa, K., Eds. International Society for
  Optics and Photonics, SPIE,  2020, Vol. 11444, pp. 686 -- 697.
\newblock
  doi:{\changeurlcolor{black}\href{https://doi.org/10.1117/12.2563983}{\detokenize{10.1117/12.2563983}}}.

\bibitem[Cui \em{et~al.}(2020)Cui et~al.]{HUBS2020}
Cui, W.; others.
\newblock {HUBS: a dedicated hot circumgalactic medium explorer}.
\newblock  Society of Photo-Optical Instrumentation Engineers (SPIE) Conference
  Series,  2020, Vol. 11444, {\em Society of Photo-Optical Instrumentation
  Engineers (SPIE) Conference Series}, p. 114442S,
  \href{http://xxx.lanl.gov/abs/2101.05587}{{\normalfont
  [arXiv:astro-ph.g/2101.05587]}}.
\newblock
  doi:{\changeurlcolor{black}\href{https://doi.org/10.1117/12.2560871}{\detokenize{10.1117/12.2560871}}}.

\bibitem[{Wang} \em{et~al.}(2020){Wang}, {Bruijn}, {Nagayoshi}, {Ridder},
  {Taralli}, {Gottardi}, {Akamatsu}, {den Herder}, {Gao}, and {Cui}]{Wang2020}
{Wang}, G.; {Bruijn}, M.; {Nagayoshi}, K.; {Ridder}, M.; {Taralli}, E.;
  {Gottardi}, L.; {Akamatsu}, H.; {den Herder}, J.W.; {Gao}, J.R.; {Cui}, W.
\newblock {Developing x-ray microcalorimeters based on TiAu TES for HUBS}.
\newblock  Society of Photo-Optical Instrumentation Engineers (SPIE) Conference
  Series,  2020, Vol. 11444, {\em Society of Photo-Optical Instrumentation
  Engineers (SPIE) Conference Series}, p. 114449J.
\newblock
  doi:{\changeurlcolor{black}\href{https://doi.org/10.1117/12.2562100}{\detokenize{10.1117/12.2562100}}}.

\bibitem[Sato \em{et~al.}(2020)Sato, Ohashi, Ishisaki, Ezoe, Yamada,
  et~al.]{SuperDIOS2020}
Sato, K.; Ohashi, T.; Ishisaki, Y.; Ezoe, Y.; Yamada, S.; others.
\newblock {Super DIOS mission for exploring "dark baryon"}.
\newblock  Space Telescopes and Instrumentation 2020: Ultraviolet to Gamma Ray;
  den Herder, J.W.A.; Nikzad, S.; Nakazawa, K., Eds. International Society for
  Optics and Photonics, SPIE,  2020, Vol. 11444, pp. 960 -- 969.
\newblock
  doi:{\changeurlcolor{black}\href{https://doi.org/10.1117/12.2561681}{\detokenize{10.1117/12.2561681}}}.

\bibitem[Nakashima \em{et~al.}(2020{\natexlab{a}})Nakashima, Hirayama, Kohjiro,
  Yamamori, Nagasawa, Sato, Yamada, Hayakawa, Yamasaki, Mitsuda, Nagayoshi,
  Akamatsu, Gottardi, Taralli, Bruijn, Ridder, Gao, and den
  Herder]{Yuki_APL2020}
Nakashima, Y.; Hirayama, F.; Kohjiro, S.; Yamamori, H.; Nagasawa, S.; Sato, A.;
  Yamada, S.; Hayakawa, R.; Yamasaki, N.Y.; Mitsuda, K.; Nagayoshi, K.;
  Akamatsu, H.; Gottardi, L.; Taralli, E.; Bruijn, M.P.; Ridder, M.L.; Gao,
  J.R.; den Herder, J.W.A.
\newblock Low-noise microwave SQUID multiplexed readout of 38 x-ray
  transition-edge sensor microcalorimeters.
\newblock {\em Applied Physics Letters} {\bf 2020}, {\em 117},~122601.
\newblock
  doi:{\changeurlcolor{black}\href{https://doi.org/10.1063/5.0016333}{\detokenize{10.1063/5.0016333}}}.

\bibitem[Nakashima \em{et~al.}(2020{\natexlab{b}})Nakashima, Hirayama, Kohjiro,
  Yamamori, Nagasawa, Sato, Yamada, Hayakawa, Yamasaki, Mitsuda, Nagayoshi,
  Akamatsu, Gottardi, Taralli, Bruijn, Ridder, Gao, and den
  Herder]{Yuki_SPIE2020}
Nakashima, Y.; Hirayama, F.; Kohjiro, S.; Yamamori, H.; Nagasawa, S.; Sato, A.;
  Yamada, S.; Hayakawa, R.; Yamasaki, N.N.; Mitsuda, K.; Nagayoshi, K.;
  Akamatsu, H.; Gottardi, L.; Taralli, E.; Bruijn, M.P.; Ridder, M.L.; Gao,
  J.R.; den Herder, J.W.
\newblock {Development of microwave multiplexer for the Super DIOS mission: 38
  transition-edge sensor x-ray microcalorimeter readout with microwave
  multiplexing}.
\newblock  X-Ray, Optical, and Infrared Detectors for Astronomy IX; Holland,
  A.D.; Beletic, J., Eds. International Society for Optics and Photonics, SPIE,
   2020, Vol. 11454, pp. 186 -- 194.
\newblock
  doi:{\changeurlcolor{black}\href{https://doi.org/10.1117/12.2560819}{\detokenize{10.1117/12.2560819}}}.

\bibitem[{Particle Data Group} \em{et~al.}(2020){Particle Data Group}, Zyla,
  et~al.]{PDG2020}
{Particle Data Group}.; Zyla, P.A.; others.
\newblock Review of Particle Physics.
\newblock {\em Progress of Theoretical and Experimental Physics} {\bf 2020},
  {\em 2020},
  \href{http://xxx.lanl.gov/abs/https://academic.oup.com/ptep/article-pdf/2020/8/083C01/34673722/ptaa104.pdf}{{\normalfont
  [https://academic.oup.com/ptep/article-pdf/2020/8/083C01/34673722/ptaa104.pdf]}}.
\newblock 083C01,
  doi:{\changeurlcolor{black}\href{https://doi.org/10.1093/ptep/ptaa104}{\detokenize{10.1093/ptep/ptaa104}}}.

\bibitem[Aker \em{et~al.}(2019)Aker et~al.]{KATRIN2019}
Aker, M.; others.
\newblock Improved Upper Limit on the Neutrino Mass from a Direct Kinematic
  Method by KATRIN.
\newblock {\em Phys. Rev. Lett.} {\bf 2019}, {\em 123},~221802.
\newblock
  doi:{\changeurlcolor{black}\href{https://doi.org/10.1103/PhysRevLett.123.221802}{\detokenize{10.1103/PhysRevLett.123.221802}}}.

\bibitem[Camilleri and Nucciotti(2016)]{CamNuc2016}
Camilleri, L.; Nucciotti, A.
\newblock The Use of Low Temperature Detectors for Direct Measurements of the
  Mass of the Electron Neutrino.
\newblock {\em Advances in High Energy Physics} {\bf 2016}, {\em
  2016},~9153024.
\newblock
  doi:{\changeurlcolor{black}\href{https://doi.org/0.1155/2016/9153024}{\detokenize{0.1155/2016/9153024}}}.

\bibitem[Gastaldo \em{et~al.}(2017)Gastaldo et~al.]{Gastaldo2017}
Gastaldo, L.; others.
\newblock The electron capture in 163Ho experiment –ECHo.
\newblock {\em The European Physical Journal Special Topics} {\bf 2017}, {\em
  226},~1623--1694.
\newblock
  doi:{\changeurlcolor{black}\href{https://doi.org/10.1140/epjst/e2017-70071-y}{\detokenize{10.1140/epjst/e2017-70071-y}}}.

\bibitem[Alpert \em{et~al.}(2016)Alpert, Ferri, Bennett, Faverzani, Fowler,
  Giachero, Hays-Wehle, Maino, Nucciotti, Puiu, Swetz, and Ullom]{Alpert2016}
Alpert, B.; Ferri, E.; Bennett, D.; Faverzani, M.; Fowler, J.; Giachero, A.;
  Hays-Wehle, J.; Maino, M.; Nucciotti, A.; Puiu, A.; Swetz, D.; Ullom, J.
\newblock Algorithms for Identification of Nearly-Coincident Events in
  Calorimetric Sensors.
\newblock {\em J. Low Temp. Phys.} {\bf 2016}, {\em 184},~263--273.
\newblock
  doi:{\changeurlcolor{black}\href{https://doi.org/10.1007/s10909-015-1402-y}{\detokenize{10.1007/s10909-015-1402-y}}}.

\bibitem[Ferri \em{et~al.}(2016)Ferri, Alpert, Bennett, Faverzani, Fowler,
  Giachero, Hays-Wehle, Maino, Nucciotti, Puiu, and Ullom]{Ferri2016}
Ferri, E.; Alpert, B.; Bennett, D.; Faverzani, M.; Fowler, J.; Giachero, A.;
  Hays-Wehle, J.; Maino, M.; Nucciotti, A.; Puiu, A.; Ullom, J.
\newblock Pile-Up Discrimination Algorithms for the HOLMES Experiment.
\newblock {\em J. Low Temp. Phys.} {\bf 2016}, {\em 184},~405--411.
\newblock
  doi:{\changeurlcolor{black}\href{https://doi.org/0.1007/s10909-015-1466-8}{\detokenize{0.1007/s10909-015-1466-8}}}.

\bibitem[{Giachero} \em{et~al.}(2021){Giachero}, {Alpert}, {Becker}, {Bennett},
  {Borghesi}, {De Gerone}, {Faverzani}, {Fedkevych}, {Ferri}, {Gallucci},
  {Gard}, {Gatti}, {Hilton}, {Mates}, {Nucciotti}, {Pessina}, {Puiu},
  {Reintsema}, {Schmidt}, {Swetz}, {Ullom}, and {Vale}]{Giachero2021}
{Giachero}, A.; {Alpert}, B.; {Becker}, D.T.; {Bennett}, D.A.; {Borghesi}, M.;
  {De Gerone}, M.; {Faverzani}, M.; {Fedkevych}, M.; {Ferri}, E.; {Gallucci},
  G.; {Gard}, J.D.; {Gatti}, F.; {Hilton}, G.C.; {Mates}, J.A.B.; {Nucciotti},
  A.; {Pessina}, G.; {Puiu}, A.; {Reintsema}, C.D.; {Schmidt}, D.R.; {Swetz},
  D.S.; {Ullom}, J.N.; {Vale}, L.R.
\newblock {Progress in the Development of TES Microcalorimeter Detectors
  Suitable for Neutrino Mass Measurement}.
\newblock {\em IEEE Transactions on Applied Superconductivity} {\bf 2021}, {\em
  31},~3051104.
\newblock
  doi:{\changeurlcolor{black}\href{https://doi.org/10.1109/TASC.2021.3051104}{\detokenize{10.1109/TASC.2021.3051104}}}.

\bibitem[Peccei and Quinn(1977{\natexlab{a}})]{PecQuinPRL1977}
Peccei, R.D.; Quinn, H.R.
\newblock $\mathrm{CP}$ Conservation in the Presence of Pseudoparticles.
\newblock {\em Phys. Rev. Lett.} {\bf 1977}, {\em 38},~1440--1443.
\newblock
  doi:{\changeurlcolor{black}\href{https://doi.org/10.1103/PhysRevLett.38.1440}{\detokenize{10.1103/PhysRevLett.38.1440}}}.

\bibitem[Peccei and Quinn(1977{\natexlab{b}})]{PecQuinPRD1977}
Peccei, R.D.; Quinn, H.R.
\newblock Constraints imposed by $\mathrm{CP}$ conservation in the presence of
  pseudoparticles.
\newblock {\em Phys. Rev. D} {\bf 1977}, {\em 16},~1791--1797.
\newblock
  doi:{\changeurlcolor{black}\href{https://doi.org/10.1103/PhysRevD.16.1791}{\detokenize{10.1103/PhysRevD.16.1791}}}.

\bibitem[Jaeckel and Ringwald(2010)]{JaeckRing2010}
Jaeckel, J.; Ringwald, A.
\newblock The Low-Energy Frontier of Particle Physics.
\newblock {\em Annual Review of Nuclear and Particle Science} {\bf 2010}, {\em
  60},~405--437.
\newblock
  doi:{\changeurlcolor{black}\href{https://doi.org/10.1146/annurev.nucl.012809.104433}{\detokenize{10.1146/annurev.nucl.012809.104433}}}.

\bibitem[Jaeckel and Thormaehlen(2019)]{Jaeckel_axion2019}
Jaeckel, J.; Thormaehlen, L.J.
\newblock Distinguishing axion models with {IAXO}.
\newblock {\em Journal of Cosmology and Astroparticle Physics} {\bf 2019}, {\em
  2019},~039--039.
\newblock
  doi:{\changeurlcolor{black}\href{https://doi.org/10.1088/1475-7516/2019/03/039}{\detokenize{10.1088/1475-7516/2019/03/039}}}.

\bibitem[{CAST collaboration}(2017)]{CAST2017}
{CAST collaboration}.
\newblock New CAST limit on the axion–photon interaction.
\newblock {\em Nat. Phys.} {\bf 2017}, {\em 13},~584--590.
\newblock
  doi:{\changeurlcolor{black}\href{https://doi.org/10.1038/nphys4109}{\detokenize{10.1038/nphys4109}}}.

\bibitem[Armengaud \em{et~al.}(2014)Armengaud et~al.]{Armengaud_2014}
Armengaud, E.; others.
\newblock Conceptual design of the International Axion Observatory ({IAXO}).
\newblock {\em Journal of Instrumentation} {\bf 2014}, {\em 9},~T05002--T05002.
\newblock
  doi:{\changeurlcolor{black}\href{https://doi.org/10.1088/1748-0221/9/05/t05002}{\detokenize{10.1088/1748-0221/9/05/t05002}}}.

\bibitem[Abeln \em{et~al.}(2020)Abeln et~al.]{babyIAXO2020concept}
Abeln, A.; others.
\newblock Conceptual Design of BabyIAXO, the intermediate stage towards the
  International Axion Observatory,  2020,
  \href{http://xxx.lanl.gov/abs/2010.12076}{{\normalfont
  [arXiv:physics.ins-det/2010.12076]}}.

\bibitem[Unger \em{et~al.}(2020)Unger, Abeln, Enss, Fleischmann, Hengstler,
  Kempf, and Gastaldo]{Unger2020}
Unger, D.; Abeln, A.; Enss, C.; Fleischmann, A.; Hengstler, D.; Kempf, S.;
  Gastaldo, L.
\newblock High-resolution for IAXO: MMC-based X-ray Detectors,  2020,
  \href{http://xxx.lanl.gov/abs/2010.15348}{{\normalfont
  [arXiv:physics.ins-det/2010.15348]}}.

\bibitem[Matteo(2020)]{Rini2020}
Matteo, R.
\newblock {Hunting Season for Primordial Gravitational Waves}.
\newblock {\em APS Physics} {\bf 2020}, {\em 13},~164.
\newblock
  doi:{\changeurlcolor{black}\href{https://doi.org/doi:10.1103/Physics.13.164.}{\detokenize{doi:10.1103/Physics.13.164.}}}

\bibitem[Schillaci \em{et~al.}(2020)Schillaci et~al.]{BiCEP2020}
Schillaci, A.; others.
\newblock {Design and Performance of the First BICEP Array Receiver}.
\newblock {\em J. Low Temp. Phys.} {\bf 2020}, {\em 199},~976--984.
\newblock
  doi:{\changeurlcolor{black}\href{https://doi.org/10.1007/s10909-020-02394-6}{\detokenize{10.1007/s10909-020-02394-6}}}.

\bibitem[Suzuki \em{et~al.}(2016)Suzuki et~al.]{Suzuki2016}
Suzuki, A.; others.
\newblock {The Polarbear-2 and the Simons Array Experiments}.
\newblock {\em J. Low Temp. Phys.} {\bf 2016}, {\em 184},~805--810.
\newblock
  doi:{\changeurlcolor{black}\href{https://doi.org/10.1007/s10909-015-1425-4}{\detokenize{10.1007/s10909-015-1425-4}}}.

\bibitem[Suzuki \em{et~al.}(2019)Suzuki et~al.]{LiteBIRD2019}
Suzuki, A.; others.
\newblock {LiteBIRD: A Satellite for the Studies of B-Mode Polarization and
  Inflation from Cosmic Background Radiation Detection}.
\newblock {\em J. Low Temp. Phys.} {\bf 2019}, {\em 194},~443--451.
\newblock
  doi:{\changeurlcolor{black}\href{https://doi.org/10.1007/s10909-019-02150-5}{\detokenize{10.1007/s10909-019-02150-5}}}.

\bibitem[Jaehnig \em{et~al.}(2020)Jaehnig, Arnold, Austermann, Becker, Duff,
  Halverson, Hazumi, Hilton, Hubmayr, Lee, Link, Suzuki, Vissers, Walker, and
  Westbrook]{Jaehnig2020}
Jaehnig, G.C.; Arnold, K.; Austermann, J.; Becker, D.; Duff, S.; Halverson,
  N.W.; Hazumi, M.; Hilton, G.; Hubmayr, J.; Lee, A.T.; Link, M.; Suzuki, A.;
  Vissers, M.; Walker, S.; Westbrook, B.
\newblock Development of Space-Optimized TES Bolometer Arrays for LiteBIRD.
\newblock {\em J. Low Temp. Phys.} {\bf 2020}, {\em 199},~646--656.
\newblock
  doi:{\changeurlcolor{black}\href{https://doi.org/10.1007/s10909-020-02425-2}{\detokenize{10.1007/s10909-020-02425-2}}}.

\bibitem[de~Haan \em{et~al.}(2020)de~Haan, Suzuki, Boyd, Cantor, Coerver,
  Dobbs, Hennings-Yeomans, Holzapfel, Lee, Noble, Smecher, and
  Zhou]{deHaan2020}
de~Haan, T.; Suzuki, A.; Boyd, S.T.P.; Cantor, R.H.; Coerver, A.; Dobbs, M.A.;
  Hennings-Yeomans, R.; Holzapfel, W.L.; Lee, A.T.; Noble, G.I.; Smecher, G.;
  Zhou, J.
\newblock Recent Advances in Frequency-Multiplexed TES Readout: Vastly Reduced
  Parasitics and an Increase in Multiplexing Factor with Sub-Kelvin SQUIDs.
\newblock {\em J. Low Temp. Phys.} {\bf 2020}, {\em 199},~754--761.

\bibitem[{Kilbourne} \em{et~al.}(2008){Kilbourne}, {Doriese}, {Bandler},
  {Brekosky}, {Brown}, {Chervenak}, {Eckart}, {Finkbeiner}, {Hilton}, {Irwin},
  {Iyomoto}, {Kelley}, {Porter}, {Reintsema}, {Smith}, and {Ullom}]{Kilb2008}
{Kilbourne}, C.A.; {Doriese}, W.B.; {Bandler}, S.R.; {Brekosky}, R.P.; {Brown},
  A.D.; {Chervenak}, J.A.; {Eckart}, M.E.; {Finkbeiner}, F.M.; {Hilton}, G.C.;
  {Irwin}, K.D.; {Iyomoto}, N.; {Kelley}, R.L.; {Porter}, F.S.; {Reintsema},
  C.D.; {Smith}, S.J.; {Ullom}, J.N.
\newblock {Multiplexed readout of uniform arrays of TES x-ray microcalorimeters
  suitable for Constellation-X} {\bf 2008}.
\newblock {\em 7011},~701104.
\newblock
  doi:{\changeurlcolor{black}\href{https://doi.org/10.1117/12.790027}{\detokenize{10.1117/12.790027}}}.

\bibitem[{Smith} \em{et~al.}(2015){Smith}, {Adams}, {Bandler},
  {Betancourt-Martinez}, {Chervenak}, {Eckart}, {Finkbeiner}, {Kelley},
  {Kilbourne}, {Lee}, {Porter}, {Sadleir}, and {Wassell}]{Smith2015}
{Smith}, S.J.; {Adams}, J.S.; {Bandler}, S.R.; {Betancourt-Martinez}, G.;
  {Chervenak}, J.A.; {Eckart}, M.E.; {Finkbeiner}, F.M.; {Kelley}, R.L.;
  {Kilbourne}, C.A.; {Lee}, S.J.; {Porter}, F.S.; {Sadleir}, J.E.; {Wassell},
  E.J.
\newblock {Uniformity of Kilo-Pixel Arrays of Transition-Edge Sensors for X-ray
  Astronomy}.
\newblock {\em IEEE Transactions on Applied Superconductivity} {\bf 2015}, {\em
  25},~2369352.
\newblock
  doi:{\changeurlcolor{black}\href{https://doi.org/10.1109/TASC.2014.2369352}{\detokenize{10.1109/TASC.2014.2369352}}}.

\bibitem[Ullom \em{et~al.}(2018)Ullom, Adams, Alpert, Bandler, Bennett,
  Chaudhuri, Chervenak, Denison, Dawson, Doriese, Durkin, Fowler, Gard, Hilton,
  Irwin, Joe, Kilbourne, Mates, Morgan, O'Neil, Reintsema, Schmidt, Smith,
  Swetz, Titus, Vale, and Young]{Ullom2018}
Ullom, J.N.; Adams, J.S.; Alpert, B.K.; Bandler, S.R.; Bennett, D.A.;
  Chaudhuri, S.; Chervenak, J.A.; Denison, E.V.; Dawson, C.; Doriese, W.B.;
  Durkin, M.; Fowler, J.W.; Gard, J.; Hilton, G.C.; Irwin, K.D.; Joe, Y.I.;
  Kilbourne, C.A.; Mates, J.A.; Morgan, K.M.; O'Neil, G.C.; Reintsema, C.D.;
  Schmidt, D.R.; Smith, S.J.; Swetz, D.S.; Titus, C.J.; Vale, L.R.; Young, B.A.
\newblock {Time- and code-division SQUID multiplexing options for ATHENA X-IFU
  (Conference Presentation)}.
\newblock  Space Telescopes and Instrumentation 2018: Ultraviolet to Gamma Ray;
  den Herder, J.W.A.; Nikzad, S.; Nakazawa, K., Eds. International Society for
  Optics and Photonics, SPIE,  2018, Vol. 10699.
\newblock
  doi:{\changeurlcolor{black}\href{https://doi.org/10.1117/12.2314111}{\detokenize{10.1117/12.2314111}}}.

\bibitem[Akamatsu \em{et~al.}(2016)Akamatsu, Gottardi, van~der Kuur, de~Vries,
  Ravensberg, Adams, Bandler, Bruijn, Chervenak, Kilbourne, Kiviranta, van~der
  Linden, Jackson, and Smith]{Akamatsu2016}
Akamatsu, H.; Gottardi, L.; van~der Kuur, J.; de~Vries, C.P.; Ravensberg, K.;
  Adams, J.S.; Bandler, S.R.; Bruijn, M.P.; Chervenak, J.A.; Kilbourne, C.A.;
  Kiviranta, M.; van~der Linden, A.J.; Jackson, B.D.; Smith, S.J.
\newblock {Development of frequency domain multiplexing for the X-ray Integral
  Field unit (X-IFU) on the Athena}.
\newblock  Space Telescopes and Instrumentation 2016: Ultraviolet to Gamma Ray;
  den Herder, J.W.A.; Takahashi, T.; Bautz, M., Eds. International Society for
  Optics and Photonics, SPIE,  2016, Vol. 9905, pp. 1675 -- 1682.
\newblock
  doi:{\changeurlcolor{black}\href{https://doi.org/10.1117/12.2232805}{\detokenize{10.1117/12.2232805}}}.

\end{thebibliography}


\begin{thebibliography}{999}

\end{thebibliography}
\end{document}